\documentclass{aa} 
\usepackage{tablefootnote}
\usepackage{caption}
\usepackage{amsmath}
\usepackage{lineno}
\usepackage[utf8]{inputenc}
\usepackage[title]{appendix}%
\usepackage{graphicx}
\usepackage{xcolor}
\usepackage[version=3]{mhchem}
\usepackage{txfonts}
\begin{document}

   \title{Reconstructing the shock history in the CMZ of NGC 253 with ALCHEMI}
   \titlerunning{Reconstructing the shock history in the CMZ of NGC 253 with ALCHEMI}

        \author{K.-Y. Huang
          \inst{\ref{inst.Leiden}}
          \and S. Viti
          \inst{\ref{inst.Leiden},\ref{inst.UCL}}
          \and J. Holdship
          \inst{\ref{inst.Leiden},\ref{inst.UCL}}
          \and J. G. Mangum \inst{\ref{inst.NRAOCV}}
          \and S. Mart\'in \inst{\ref{inst.ESOChile},\ref{inst.JAO}}
          \and N. Harada \inst{\ref{inst.NAOJ},\ref{inst.ASIAA}, \ref{inst.SOKENDAI}}
          \and S. Muller \inst{\ref{inst.ONSALA}}
          \and K. Sakamoto \inst{\ref{inst.ASIAA}}
          \and K. Tanaka \inst{\ref{inst.KeioUniversity}}
          \and Y. Yoshimura \inst{\ref{inst.UTokio}}
          \and R. Herrero-Illana \inst{\ref{inst.ESOChile},\ref{inst.ICECSIC}}
          \and D. S. Meier \inst{\ref{inst.NMIMT},\ref{inst.NRAOSocorro}}
          \and E. Behrens \inst{\ref{inst.UVA}}
          \and P. P.~van der Werf \inst{\ref{inst.Leiden}}
          \and C. Henkel \inst{\ref{inst.MPIfR},\ref{inst.Abdulaziz},\ref{inst.Xinjiang}}
          \and S. Garc{\'{\i}}a-Burillo \inst{\ref{inst.OAN-IGN}}
          \and V. M. Rivilla \inst{\ref{inst.CAB-INTA}}
          \and K. L. Emig \inst{\ref{inst.NRAOCV}}\thanks{Jansky Fellow of the National Radio Astronomy Observatory}
          \and L. Colzi \inst{\ref{inst.CAB-INTA}}
          \and P. K. Humire \inst{\ref{inst.MPIfR}}
          \and R. Aladro \inst{\ref{inst.MPIfR}}
          \and M. Bouvier \inst{\ref{inst.Leiden}}
}
\institute{
\label{inst.Leiden}Leiden Observatory, Leiden University, PO Box 9513, NL - 2300 RA Leiden, The Netherlands
\label{email}kyhuang@strw.leidenuniv.nl
\and\label{inst.UCL}Department of Physics and Astronomy, University College London, Gower Street, London WC1E6BT, UK
\and\label{inst.NRAOCV}National Radio Astronomy Observatory, 520 Edgemont Road, Charlottesville, VA 22903-2475, USA
\and\label{inst.ESOChile}European Southern Observatory, Alonso de C\'ordova, 3107, Vitacura, Santiago 763-0355, Chile  
\and\label{inst.JAO}Joint ALMA Observatory, Alonso de C\'ordova, 3107, Vitacura, Santiago 763-0355, Chile
\and\label{inst.NAOJ}National Astronomical Observatory of Japan, 2-21-1 Osawa, Mitaka, Tokyo 181-8588, Japan
\and\label{inst.ASIAA}Institute of Astronomy and Astrophysics, Academia Sinica, 11F of AS/NTU Astronomy-Mathematics Building, No.1, Sec. 4, Roosevelt Rd, Taipei 10617, Taiwan
\and\label{inst.SOKENDAI}Department of Astronomy, School of Science, The Graduate University for Advanced Studies (SOKENDAI), 2-21-1 Osawa, Mitaka, Tokyo, 181-1855 Japan
\and\label{inst.ONSALA}Department of Space, Earth and Environment, Chalmers University of Technology, Onsala Space Observatory, SE-43992 Onsala, Sweden
\and\label{inst.KeioUniversity}Department of Physics, Faculty of Science and Technology, Keio University, 3-14-1 Hiyoshi, Yokohama, Kanagawa 223--8522 Japan
\and\label{inst.UTokio}Institute of Astronomy, Graduate School of Science, The University of Tokyo, 2-21-1 Osawa, Mitaka, Tokyo 181-0015, Japan
\and\label{inst.ICECSIC}Institute of Space Sciences (ICE, CSIC), Campus UAB, Carrer de Magrans, E-08193 Barcelona, Spain
\and\label{inst.NMIMT}New Mexico Institute of Mining and Technology, 801 Leroy Place, Socorro, NM 87801, USA
\and\label{inst.NRAOSocorro}National Radio Astronomy Observatory, PO Box O, 1003 Lopezville Road, Socorro, NM 87801, USA
\and\label{inst.UVA}Department of Astronomy, University of Virginia, P.~O.~Box 400325, 530 McCormick Road, Charlottesville, VA 22904-4325
\and\label{inst.MPIfR}Max-Planck-Institut f\"ur Radioastronomie, Auf-dem-H\"ugel 69, 53121 Bonn, Germany    
\and\label{inst.Abdulaziz}Astron. Dept., Faculty of Science, King Abdulaziz University, P.O. Box 80203, Jeddah 21589, Saudi Arabia
\and\label{inst.Xinjiang}Xinjiang Astronomical Observatory, Chinese Academy of Sciences, 830011 Urumqi, China
\and\label{inst.OAN-IGN}Observatorio Astron{\'{o}}mico Nacional (OAN-IGN)-Observatorio de Madrid, Alfonso XII, 3, 28014-Madrid, Spain
\and\label{inst.CAB-INTA}Centro de Astrobiología (CSIC-INTA), Ctra. de Torrej\'on a Ajalvir km 4, 28850, Torrej\'on de Ardoz, Madrid, Spain
\and\label{inst.Arcetri}INAF Osservatorio Astrofisico di Arcetri, Largo Enrico Fermi 5, I-50125 Firenze, Italy
}

   \authorrunning{Huang et al.}
   \date{Submitted 09 December 2022; accepted 21 March 2023}

 
  \abstract
   {HNCO and SiO are well known shock tracers and have been observed in nearby galaxies, including the nearby (D=3.5 Mpc) starburst galaxy NGC 253. The simultaneous detection of these two species in regions where the star formation rate is high may be used to study the shock history of the gas. }
   {We perform a multi-line molecular study using these two shock tracers (SiO and HNCO) with the aim of characterizing the gas properties. We also explore the possibility of reconstructing the shock history in NGC 253's Central Molecular Zone (CMZ).  }
  {Six SiO transitions and eleven HNCO transitions were imaged at high resolution $1''.6$ (28 pc) with the Atacama Large Millimeter/submillimeter Array (ALMA) as part of the ALCHEMI Large Programme. Both non-LTE radiative transfer analysis and chemical modelling were performed in order to characterize the gas properties, and to investigate the chemical origin of the emission. }
  {The non-LTE radiative transfer analysis coupled with Bayesian inference shows clear evidence that the gas traced by SiO has different densities and temperatures than that traced by HNCO, with an indication that shocks are needed to produce both species. Chemical modelling further confirms such a scenario and suggests that fast and slow shocks are responsible for SiO and HNCO production, respectively, in most GMCs. We are also able to infer the physical characteristics of the shocks traced by SiO and HNCO for each GMC. }
 {Radiative transfer and chemical analysis of the SiO and HNCO in the CMZ of NGC 253 reveal a complex picture whereby most of the GMCs are subjected to shocks. We speculate on the possible shock scenarios responsible for the observed emission and provide potential history and timescales for each shock scenario. Higher spatial resolution observations of these two species are required in order to quantitatively differentiate between scenarios. }


   \maketitle
%
\section{Introduction}
Many key physical and chemical processes in the interstellar medium (ISM) influence the evolution of galaxies. 
These processes are often associated with star-formation, Active Galactic Nuclei (AGN), large-scale outflows, and shocks. 
In this context, starburst galaxies are prime laboratories for the investigation of these feedback mechanisms and their impact on the ISM. 
As for probing these physical and chemical processes in external galaxies, multi-line multi-species molecular observations are an ideal tool, given the wide range of critical densities associated with different molecular transitions, and the dependencies of chemical reactions on the energy budget of the ISM. 
Past observations have suggested several useful molecules in tracing specific regions within a galaxy, e.g. HCO and HOC\textsuperscript{+} being associated with the photon-dominated regions (PDRs) \citep[e.g.][]{Savage_Ziurys_2004,GB+2002,Gerin+2009_HCO_PDR,Martin+2009_PDR_ngc253}, and HCN and CS with dense gas \citep[e.g.][]{Gao_Solomon_2004,Bayet+2008,Aladro+2011_CS}. 
In reality, especially in the extragalactic context where the beam size often encompasses at least several parsecs, it is seldom the case that a single gas component can be identified by observations of just one or two molecular species \citep{Kauffmann+2017_hcn,Pety+2017_densegas,Viti_2017,Tafalla+2021}. This is because the same species can often be found in diverse environments, and multiple transitions of the same species do not necessarily come from the same gas component. 
As a result, molecular tracers that are uniquely sensitive to certain environments are considered extremely valuable in characterizing the physical and chemical conditions of the gas. 

As starburst activities inject a significant amount of energy into the ambient environment, starburst galaxies are particularly important targets to study the feedback mechanisms in the interstellar medium (ISM). 
Induced by high star-forming rates (SFR), strong stellar feedback can trigger outflows of ionized, neutral, and molecular gas. 
NGC 253 is a barred spiral galaxy that is almost edge-on with an inclination of $76^{\circ}$ \citep{McCormick+2013}. 
Being one of the nearest starburst systems \citep[D $\sim3.5 \pm 0.2$ Mpc,][]{Rekola+2005}, NGC 253 is also one of the most studied starburst galaxies. 
The central molecular zone (CMZ) of NGC 253 spans about $300\times100$ pc across \citep{Sakamoto+2011}, and contains more than 10 well studied Giant Molecular Clouds (GMCs), observed in the continuum as well as in molecular emissions \citep[e.g.][]{Sakamoto+2011,Leroy+2015,Leroy+2018,Levy+2022}. 
NGC 253 is a prototype of nuclear starburst with an SFR of $\sim2 $~M$_{\odot}$yr\textsuperscript{-1} coming from its central molecular zone \citep[CMZ,][]{Leroy+2015,Bendo+2015}, which is half of its global SF activity. 

A large-scale outflow in NGC 253 has been revealed by multi-wavelength observations: in X-rays \citep{Strickland+2000,Strickland+2002}, H$\alpha$ \citep{Westmoquette+2011}, molecular emission \citep{Turner1985,Bolatto+2013,Walter+2017,Krieger+2019}, and  dust \citep{Levy+2022}. 
This large-scale outflow is thought to be driven by the galaxy's starburst activity \citep{McCarthy+1987}, for there are no signs of AGN influence \citep{MS+2010,Lehmer+2013} despite there being a bright radio source associated with the nucleus of the galaxy \citep{TH1985}. 
Aside from coherent, large-scale outflows, the presence of shocks and turbulence are also complementary sources of mechanical energy in the ISM feedback processes. 
The signature of shocks in NGC 253 has been suggested by the detection of HNCO and SiO \citep{GB+2000_sio_253,Meier+2015_hncosio_253}, the detection of Class I methanol masers \citep{Humire+2022}, and the enhanced fractional abundances of CO\textsubscript{2} \citep{Harada+2022}. 

Both silicon monoxide, SiO, and isocyanic acid, HNCO, are well-known shock tracers \citep{sio_MP+1997,Huttemeister+1998,Zinchenko+2000,J-S+2008_shocktracers,Martin+2008_HNCO_galactic,hnco+RF+2010}, and have been observed in nearby galaxies \citep[e.g.,][]{GB+2000_sio_253,Meier_Turner_2005,Usero+2006,Martin+2009,GB+2010,Meier_Truner_2012_maffei2,Martin+2015_shocktracer_ngc1097,Meier+2015_hncosio_253,Kelly+2017,Huang+2022} {aside from Galactic sources such as star-forming regions} \citep[e.g.][]{Mendoza+2014,Podio+2017,HG+2019,Gorai+2020,Nazari+2021,Canelo+2021,Colzi+2021} {, evolved stars} \citep[e.g.][]{VP+2015,VP+2017,Rizzo+2021} {, and quiescent giant molecular clouds in the Galactic Centre} \citep{Zeng+2018}. %
The formation of HNCO has been suggested to be mainly on the icy mantles of dust grains \citep{Fedoseev+2015}, or possibly in the gas phase with subsequent freeze-out onto the dust grains when the temperatures are low \citep{LS+2015}.  
Regardless of how it forms, icy mantles sputtering associated with low-velocity ($\varv_{s}\leq20$ km s\textsuperscript{-1}) shocks can lead to an enhanced abundance of HNCO in the gas phase. 
The observations of HNCO towards a sample of Galactic Center sources \citep{Martin+2008_HNCO_galactic} has shown the high contrast observed in its abundance between regions under the influence of shocks and intense radiation fields, showing its potential as shock tracer. 
This high contrast was also shown in a sample of galaxies \citep{Martin+2009} despite the low resolution single dish observations used.
In fact, \citet{Kelly+2017} show that HNCO can also be thermally desorbed from the surface of dust grains when the gas and dust remain coupled at higher gas densities ($n_{\rm H2}\geq10^{4}$ cm\textsuperscript{-3}). 
Hence HNCO may not be a unique tracer of shock activity . 
On the other hand, a high abundance of silicon in the gas phase can only be explained by significant sputtering from the core of the dust grains by high-velocity ($\varv_{s}\geq50$ km s\textsuperscript{-1}) shocks \citep{Kelly+2017}. 
Once silicon is in the gas phase, it is expected to quickly react with molecular oxygen or a hydroxyl radical to form SiO \citep{Schilke+1997}. 
An enhanced gas-phase SiO abundance could thus be a sensitive indicator of the  heavily shocked regions. 

Potentially, the simultaneous detection of HNCO and SiO in a gas where shocks are believed to take place could provide us with a comprehensive picture of its shock history. 
Indeed, these two species have already been proposed for the characterization of different types of shocks (fast versus slow) in the AGN-hosting galaxies e.g - NGC 1068 \citep{Kelly+2017, Huang+2022} and NGC 1097 \citep{Martin+2015_shocktracer_ngc1097}, in the nearby weakly barred spiral galaxy IC 342 which hosts moderate starburst activities \citep{Meier_Turner_2005,Usero+2006}, and in the nearby starburst galaxy NGC 253 \citep{Meier+2015_hncosio_253}. 
For example in NGC 253, the subject of this work, HNCO 4\textsubscript{0,4}-3\textsubscript{0,3} was found distinctively prominent in the outer CMZ with the HNCO(4\textsubscript{0,4}-3\textsubscript{0,3})/SiO(2-1) intensity ratio dropping dramatically towards the inner disk, which suggested a decrease in shock strength as well as a dissipation of any shock signature by HNCO in the presence of strong radiation fields \citep{Meier+2015_hncosio_253}. 

In this work we present ALMA multi-transition observations of both HNCO and SiO towards NGC 253 observed as part of the ALMA large program, "ALMA Comprehensive High-resolution Extragalactic Molecular Inventory", \citep[ALCHEMI,][]{ALCHEMI_main_2021}. 
ALCHEMI covers wide and thorough spectral scans of the CMZ of NGC 253 in the frequency range of 84.2 to 373.2 GHz. 
ALCHEMI provides a comprehensive molecular view towards the CMZ of NGC 253, allowing for a systematic study of  both the physical and chemical properties of this nearby galaxy. 
In the broader sense, ALCHEMI provides a uniform molecular template for an extragalactic starburst environment where systematic uncertainties are minimized, as well as enabling a direct comparison of the ISM properties with the active star-forming environments within the Milky Way CMZ \citep{ALCHEMI_main_2021}. 
The great wealth of ALCHEMI data have so far unveiled many important properties of NGC 253, including the high cosmic-ray ionization rate (CRIR) nature of the galaxy \citep{Holdship+2021_SpectralRadex,Holdship+2022,Harada+2021,EB+2022}, the first detection of a phosphorus-bearing molecule in extragalactic sources \citep{Haasler+2022}, the identification of new methanol maser transitions \citep{Humire+2022}, and the use of HOCO\textsuperscript{+} as  a tracer of  the chemistry of CO\textsubscript{2} \citep{Harada+2022}.  

This work is structured as follows. 
In Sect. 2 we describe the detection of multiple transitions of HNCO and SiO in the ALCHEMI data set. 
In Sect. 3 we present the molecular line intensity maps, and the spectral line energy distribution (SLED) of HNCO and SiO, which are populated by the measured velocity-integrated line intensity from all the available excitation levels.
In Sect. 4 we describe the performed non-LTE (non Local Thermodynamic Equilibrium) radiative transfer analysis and chemical modelling in order to constrain the physical conditions of the gas and the chemical origin of the emission. 
In Sect. 5 we further explore the comparison and physical interpretation combining both radiative transfer modelling and chemical modelling results, and ponder upon the potential origins of the shocks. 
We summarize our findings in Sect. 6. 
\section{Observation and Data}
\subsection{ALCHEMI Data}
We briefly summarize the observational setup used to acquire the ALCHEMI survey data.
Full details regarding the data acquisition, calibration, and imaging are provided by \cite{ALCHEMI_main_2021}. 
The ALMA Large Program ALCHEMI (project code 2017.1.00161.L and 2018.1.00162.S) imaged the CMZ within NGC 253 in the ALMA frequency Bands 3, 4, 5, 6, and 7. 
The rest-frequency coverage of ALCHEMI ranged from 84.2 to 373.2\,GHz. 
The nominal phase center of the observations is $\alpha(I\rm CRS)$ = 00$^h$47$^m$33$^s$.26, $\delta( \rm ICRS)$ = $-25^\circ$17$^\prime$17$^{\prime\prime}.7$. 
A common rectangular area with size $50^{\prime\prime} \times 20^{\prime\prime}$ ($850\times340$\,pc) at a position angle of $65^\circ$ was imaged to cover the central nuclear region in NGC\,253. 
The final angular and spectral resolution of the image cubes generated from these measurements were $1.^{\prime\prime}6$ \citep[$\sim28$\,pc][]{ALCHEMI_main_2021} and $\sim10$ km\,s$^{-1}$ respectively. 
A common maximum recoverable angular scale of $15^{\prime\prime}$ was achieved after combining the 12\,m Array and Atacama Compact Array (ACA) measurements at all frequencies. 

From the ALCHEMI data we extracted the $\sim 1''.6$ (28 pc) resolution cubes of the CMZ of NGC\,253 for the 11 HNCO transitions and 6 SiO transitions listed in Table \ref{tab:Line_list}.  
In this work we only analyze the main isotopologue species for both molecules. 
Also note that we only studied the $K_{a}=0$ transitions of HNCO although in the ALCHEMI data sets some $K_{a}\ne 0$ transitions are also detected. 
This choice is made so that we focus only on the most robust detections with best signal-to-noise ratio (SNR) across the CMZ. 
{$K_{a}=0$ components are generally well above 10-$\sigma$ level in the detected emission, while for $K_{a}=1$ components, out of the 10 regions surveyed, only 2 GMCs are detected at $\lesssim 3\sim5-\sigma$.  The higher energy $K_{a} \geq 2$ components are significantly weaker. 
For the sake of consistency, we analyse only the $K_{a}=0$ components for all the GMC regions. }
For SiO only the vibrational ground states (v=0) were considered. 
Table~\ref{tab:Line_list} lists relevant spectroscopic information for all HNCO and SiO transitions studied in this article. 
The continuum subtraction and imaging processes performed for the data used in this paper are described in \cite{ALCHEMI_main_2021}. 
The representative spectra of all HNCO and SiO transitions used in the current work are shown in Appendix \ref{sec:spectra}. 
\begin{table*}[ht!]
  \centering
  \caption{HNCO and SiO transitions used in this work. }
  \label{tab:Line_list}
  \begin{tabular}{ccccccc}
  \hline
  \hline
    {Species} & {Transition $^{(a)}$} & {Rest Frequency} & $E_{u}$ & $A_{ul}$& $g_{u}$ & mJy beam\textsuperscript{-1} to K $^{(b)}$\\
    {}& {} & {[GHz]}& [K] & [s\textsuperscript{-1}] & & {}\\
    \hline
    \hline
    HNCO & 4\textsubscript{0,4}-3\textsubscript{0,3} & {87.9252} & {10.55} &     9.024e-06 &     9 & {0.062}\\
    {} & 5\textsubscript{0,5}-4\textsubscript{0,4} & {109.9057} & {15.82} &      1.803e-05 &     11& {0.040}\\
    {} & 6\textsubscript{0,6}-5\textsubscript{0,5} & {131.8857} & {22.15} &      3.163e-05 &     13& {0.027}\\
    {} & 7\textsubscript{0,7}-6\textsubscript{0,6} & {153.8651} & {29.54} &      5.078e-05 &     15& {0.020}\\
    {} & 8\textsubscript{0,8}-7\textsubscript{0,7} & {175.8437} & {37.98} &      7.643e-05 &     17& {0.015}\\
    {} & 9\textsubscript{0,9}-8\textsubscript{0,8} & {197.8215} & {47.47} &      1.095e-04 &     19& {0.012}\\
    {} & 10\textsubscript{0,10}-9\textsubscript{0,9} & {219.7983} & {58.02} &    1.51e-04  &     21& {0.010}\\
    {} & 12\textsubscript{0,12}-11\textsubscript{0,11} & {263.7486} & {82.28} &  2.631e-04 &     25& {0.007}\\
    {} & 13\textsubscript{0,13}-12\textsubscript{0,12} & {285.7220} & {95.99} &  3.355e-04 &     27& {0.006}\\
    {} & 14\textsubscript{0,14}-13\textsubscript{0,13} & {307.6939} & {110.76} & 4.201e-04 &     29& {0.005}\\
    {} & 15\textsubscript{0,15}-14\textsubscript{0,14} & {329.6644} & {126.58} & 5.178e-04 &     31& {0.004}\\
    \hline
    {SiO} & 2-1 & {86.847} & {6.25} & 2.927e-05 &       5  &  {0.063}\\
    {} & 3-2 & {130.269} & {12.50}  & 1.058e-04 &       7  &  {0.028}\\
    {} & 4-3 & {173.688} & {20.84}  & 2.602e-04 &       9  &  {0.015}\\
    {} & 5-4 & {217.105} & {31.26}  & 5.197e-04 &       11 &  {0.010}\\
    {} & 6-5 & {260.518} & {43.76}  & 9.118e-04 &       13 &  {0.007}\\
    {} & 7-6 & {303.927} & {58.35}  & 1.464e-03 &       15 &  {0.005}\\
    \hline
  \end{tabular}\\
  \footnotesize{We use HNCO and SiO molecular data from \citet{hnco_moldata_N+1995,hnco_moedata_S+2018,sio_moldata_B+2018} via the LAMDA database \citep{LAMDA_2005}. $^{(a)}$ The transition's quantum number labeling and the rest frequency data are from the Cologne Database for Molecular Spectroscopy (CDMS) catalogue \footnote{https://cdms.astro.uni-koeln.de/} \citep{CDMS_2001,CDMS_2005,CDMS_2016}. $^{(b)}$The conversion factor from [mJy/beam] to [K] described in Sect. \ref{sec:LTE} in each transition is provided. }
\end{table*}

\subsection{Extraction of spectral-line emission}
\label{sec:CubeLineMoment}

In order to extract integrated spectral line intensities from our data cubes we use \texttt{CubeLineMoment}\footnote{\url{https://github.com/keflavich/cube-line-extractor}} \citep{Mangum+2019}. 
\texttt{CubeLineMoment} employs a set of spectral and spatial masks to extract integrated intensities for a defined list of target spectral frequencies. 
As noted by \cite{Mangum+2019}, the \texttt{CubeLineMoment} masking process uses a brighter spectral line, whose velocity structure over the galaxy is most representative of our science target lines (HNCO and SiO) in the same spectral cube, as a velocity tracer of the gas component inspected. 
Final products from the \texttt{CubeLineMoment} analysis include moment 0 (integrated intensity; Jy km\,s\textsuperscript{
-1}), 1 (average velocity; km\,s\textsuperscript{
-1}) and 2 (velocity dispersion; km\,s\textsuperscript{
-1}) images masked below $3\sigma$ threshold (channel-based). 

We have selected 12 GMC regions with aperture size of the ALCHEMI beam size for further quantitative analysis that will be described from Sect. \ref{sec:LTE} onward: GMC 1a, 1b, 2a, 2b, 3, 4, 5, 6, 7, 8a, 9a, and 10. 
The choice of these regions is based on the regions identified by \citet{Leroy+2015} using dense gas tracers, and details are also discussed further by \citet{EB+2022}. 
In particular, the line intensity peaks in some of the GMCs on our line intensity maps are often shown to be offset from the nominal GMC positions identified by \citet{Leroy+2015} in the outskirts of the CMZ. 
From our HNCO and SiO maps, we adopt the closest peaks to the GMC positions referred by \citet{Leroy+2015} for the data with the optimal signal-to-noise ratio (SNR). 
These newly designated locations are GMC 1a, 1b, 2a, 2b, 8a, and 9a. 
These selected 12 GMC locations  are listed in Table \ref{tab:GMC_locations} and marked in white solid circles on the maps shown in Fig. \ref{fig:mom0_H_I}-\ref{fig:mom0_S}. 
\begin{table}[ht!]
  \centering
  \caption{All the selected NGC\,253 GMC positions described in Sect. \ref{sec:CubeLineMoment}. }
  \label{tab:GMC_locations}
  \begin{tabular}{ccc}
  \hline
    {GMC} & {R.A.(ICRS)} & {Dec.(ICRS)} \\
    {}& {(00$^h$ 47$^m$)} & {($-25^\circ$ 17$^\prime$)}\\
    \hline
    \hline
    GMC\,1a & 31$^s$.9344 & 28$^{\prime\prime}$.822 \\
    GMC\,1b & 32$^s$.0494 & 25$^{\prime\prime}$.827 \\
    GMC\,2a & 32$^s$.1985 & 21$^{\prime\prime}$.379 \\
    GMC\,2b & 32$^s$.3449 & 18$^{\prime\prime}$.886 \\
    GMC\,3 & 32$^s$.8056 & 21$^{\prime\prime}$.552 \\
    GMC\,4 & 32$^s$.9736 & 19$^{\prime\prime}$.968 \\
    GMC\,5 & 33$^s$.2112 & 17$^{\prime\prime}$.412 \\
    GMC\,6 & 33$^s$.3312 & 15$^{\prime\prime}$.756 \\
    GMC\,7 & 33$^s$.6432 & 13$^{\prime\prime}$.272 \\
    GMC\,8a & 33$^s$.9443 & 10$^{\prime\prime}$.888 \\
    GMC\,9a & 34$^s$.1287 & 12$^{\prime\prime}$.040 \\
    GMC\,10 & 34$^s$.2360 & 07$^{\prime\prime}$.836 \\
    \hline
  \end{tabular}
\end{table}

\subsubsection{Interloper assessment}
For the selected transitions used in this work, our line intensity extraction with \texttt{CubeLineMoment} also includes assessment for potential contamination from neighboring lines. 
Following the procedure described by \citet{Holdship+2021_SpectralRadex}, only two lines were found with line contamination beyond the flux uncertainties: HNCO 12\textsubscript{0,12}-11\textsubscript{0,11} with 34\% contamination from HC\textsubscript{3}N 29-28 v=0 (263.79230800 GHz), and SiO 7-6 with 35\% contamination from OCS 25-24 v=0 (303.99326170 GHz). 
The correction concerning these line contaminations was applied to the measured line intensities before further analysis was performed. 

\section{Velocity-integrated line intensities}
\label{sec:LTE}
\begin{figure*}
  \centering
  \begin{tabular}[b]{@{}p{0.8\textwidth}@{}}
    \centering\includegraphics[width=1.0\linewidth]{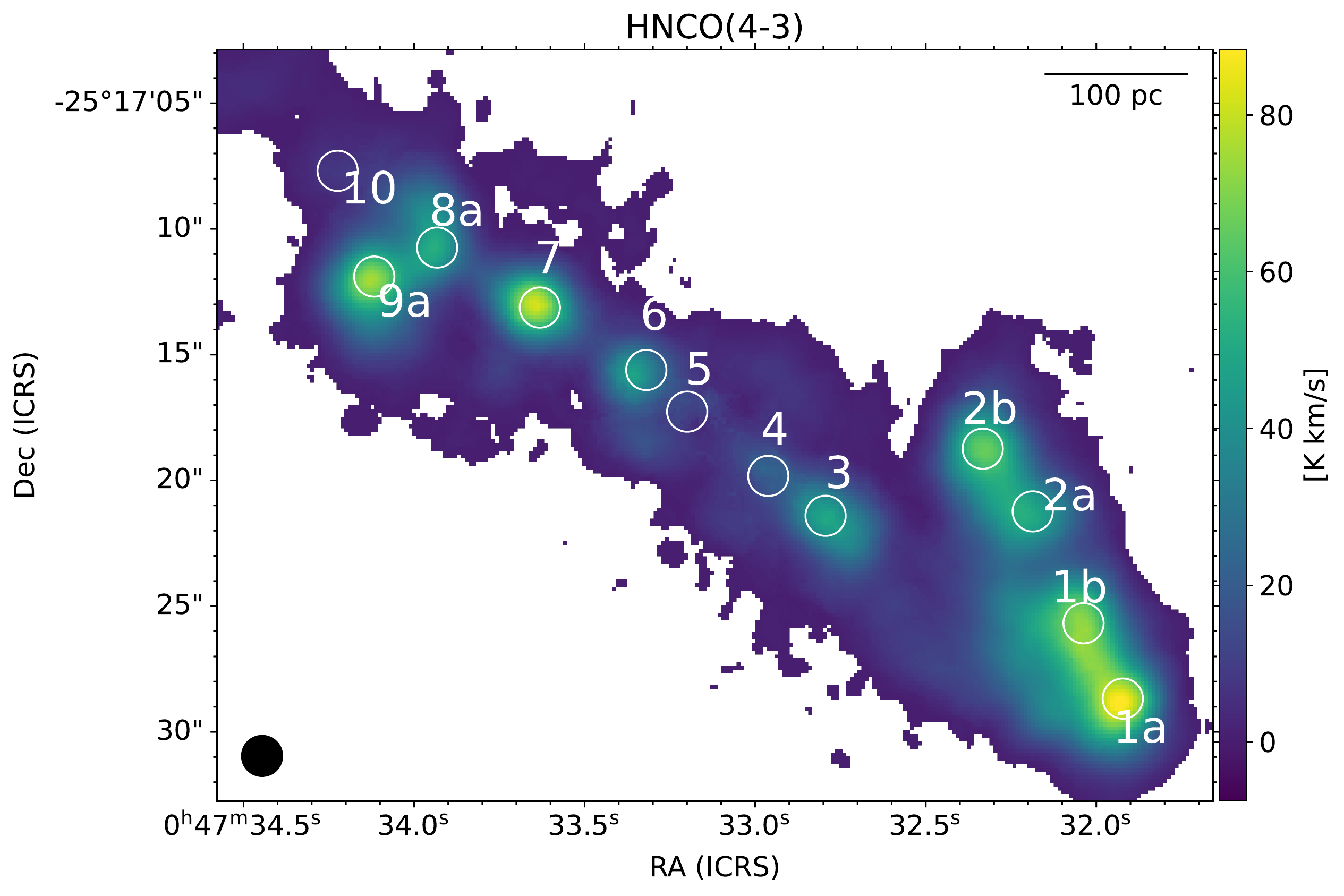} \\
    \centering\small (a) 
  \end{tabular}%
  \quad
  \begin{tabular}[b]{@{}p{0.8\textwidth}@{}}
    \centering\includegraphics[width=1.0\linewidth]{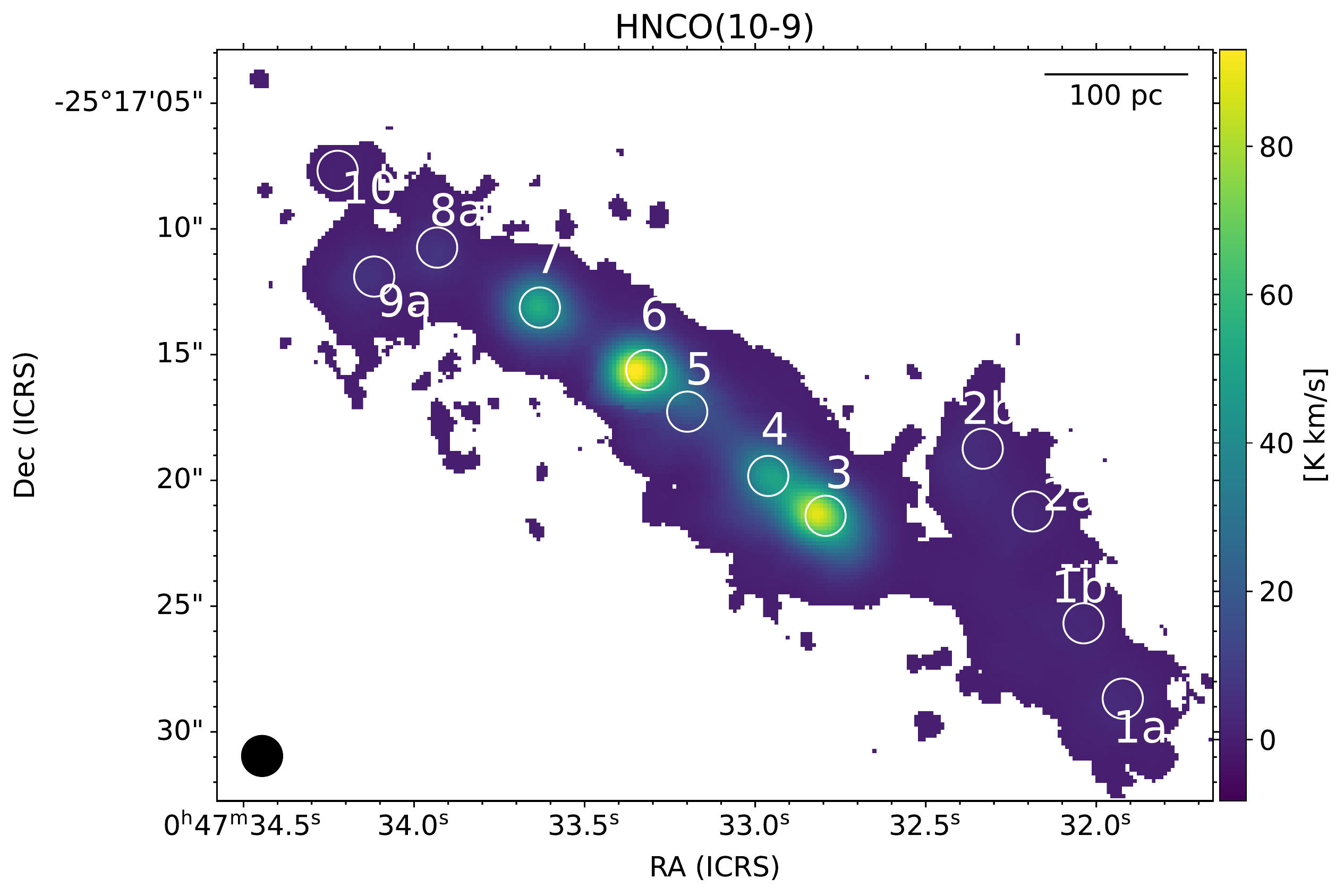} \\
    \centering\small (b) 
  \end{tabular}
  \caption{Velocity-integrated line intensities in [K km s\textsuperscript{-1}] of the HNCO transitions: (4\textsubscript{0,4}-3\textsubscript{0,3}) and (10\textsubscript{0,10}-9\textsubscript{0,9}), in (a) and (b) respectively. These two transitions are representative of the drastic variation in the trend of brightness from outer to inner GMCs. The rest of the HNCO line intensity maps are provided in Appendix \ref{sec:appen_mom0}. The studied GMC regions as listed in Tab. \ref{tab:GMC_locations} are labeled in white texts on the map. The ALCHEMI $1''.6 \times 1''.6$ beam is displayed in the lower-left corner of the map. 
  }
  \label{fig:mom0_H_I}
\end{figure*}

\begin{figure*}
  \centering
  \begin{tabular}[b]{@{}p{0.8\textwidth}@{}}
    \centering\includegraphics[width=1.0\linewidth]{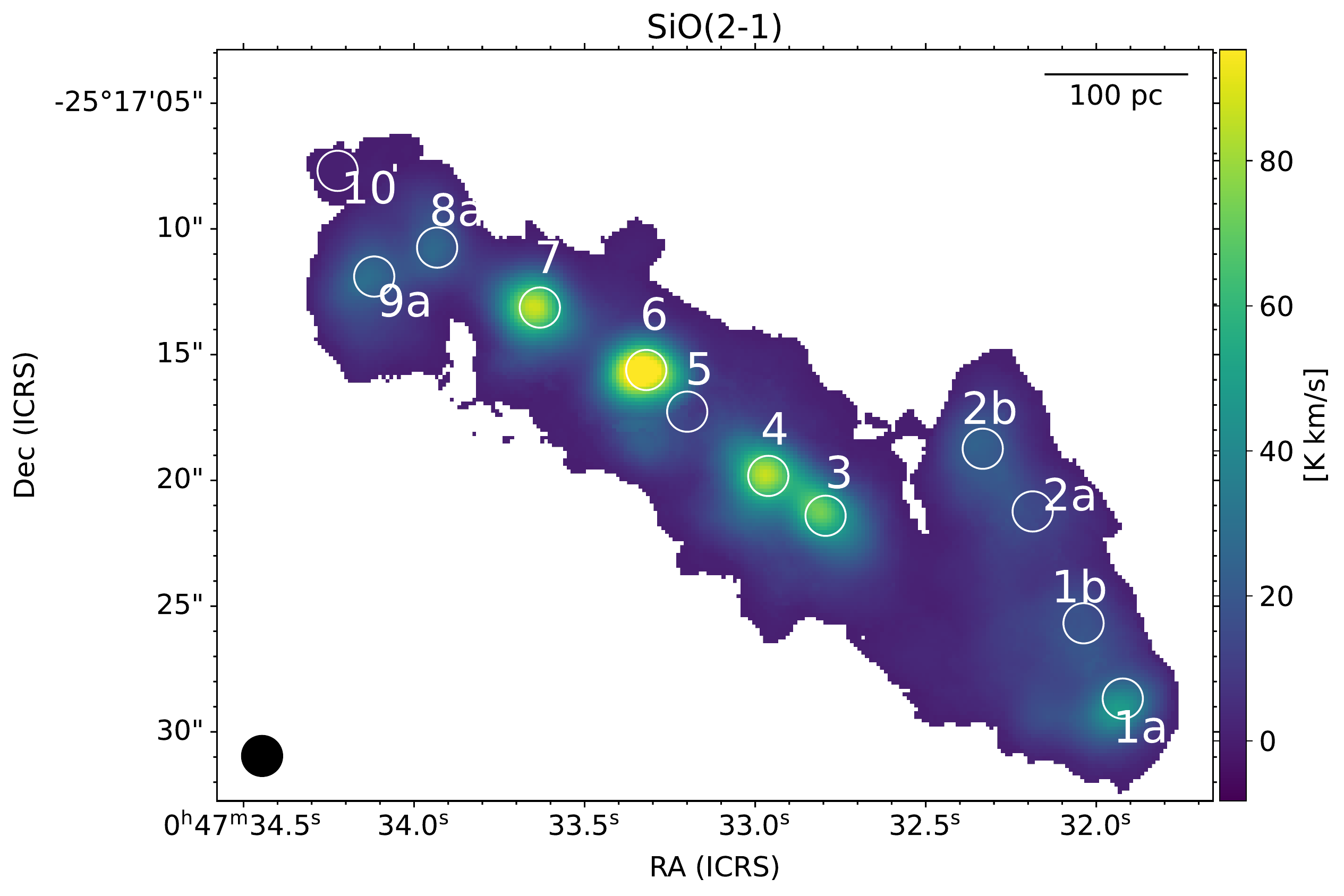} \\
  \end{tabular}%
  \quad
  \caption{Velocity-integrated line intensities of the SiO (2-1) transition in [K km s\textsuperscript{-1}]. The remaining SiO line intensity maps are provided in Appendix \ref{sec:appen_mom0}. The studied GMC regions as listed in Tab. \ref{tab:GMC_locations} are labeled in white texts on the map. The ALCHEMI $1''.6 \times 1''.6$ beam is displayed in the lower-left corner of the map. }
  \label{fig:mom0_S}
\end{figure*}

In Figs.~\ref{fig:mom0_H_I} and \ref{fig:mom0_S} we present the velocity-integrated line intensity maps from HNCO (4\textsubscript{0,4}-3\textsubscript{0,3}), HNCO (10\textsubscript{0,10}-9\textsubscript{0,9}), and SiO 2-1 transitions. 
The remaining intensity maps from other observed transitions of these two species studied in the current work are presented in Appendix \ref{sec:appen_mom0} (Figs. \ref{fig:mom0_add_I}-\ref{fig:mom0_add_III}). 
For each transition map, we also overlaid all the 12 GMC regions listed in Table \ref{tab:GMC_locations}, and the ALCHEMI beam size in the lower left corner in each map. 
The line intensities have been converted from [Jy/beam km/s] to [K km/s]; the conversion factors are listed in Table \ref{tab:Line_list}. 

Looking at the spatial distribution of the line emission, the brightest emission often occurs in the inner CMZ (e.g., GMC 3/6/7) for most transitions of SiO and HNCO, except for HNCO 4\textsubscript{0,4}-3\textsubscript{0,3} where GMC 1a in the outermost CMZ is the brightest region as shown in Fig. \ref{fig:mom0_H_I}(a). 
\citet{Meier+2015_hncosio_253} found HNCO 4\textsubscript{0,4}-3\textsubscript{0,3} distinctively prominent in the outer CMZ, which is consistent with what we find here. 
On the other hand, they also found that 
the HNCO(4\textsubscript{0,4}-3\textsubscript{0,3})/SiO(2-1) ratio drops towards the inner CMZ. 
They suggested that this could be due to the decreasing shock strength and the erased shock chemistry of HNCO in the presence of a dominating central radiation field, or due to the different dependencies on temperature in the partition function of each species for SiO is a linear molecule and HNCO is an asymmetric top \citep{Meier+2015_hncosio_253}. 
We want to highlight that such trends in the HNCO/SiO intensity ratio do not apply to the higher-J transition pairs, as we have already seen that the higher-J HNCO transitions are brighter in the inner CMZ (see also the comparison in Fig. \ref{fig:mom0_H_I}), as are all the SiO transitions. 
Such complexity cannot be captured by single-transition observations, thus justifies the importance of the multi-line observations and analysis we perform here. 
The trend of the intensity ratio of these two species from our data will be briefly discussed in Sec. \ref{sec:intensity_ratio}. 


\begin{figure*}
        \centering
    \includegraphics[scale=0.46]{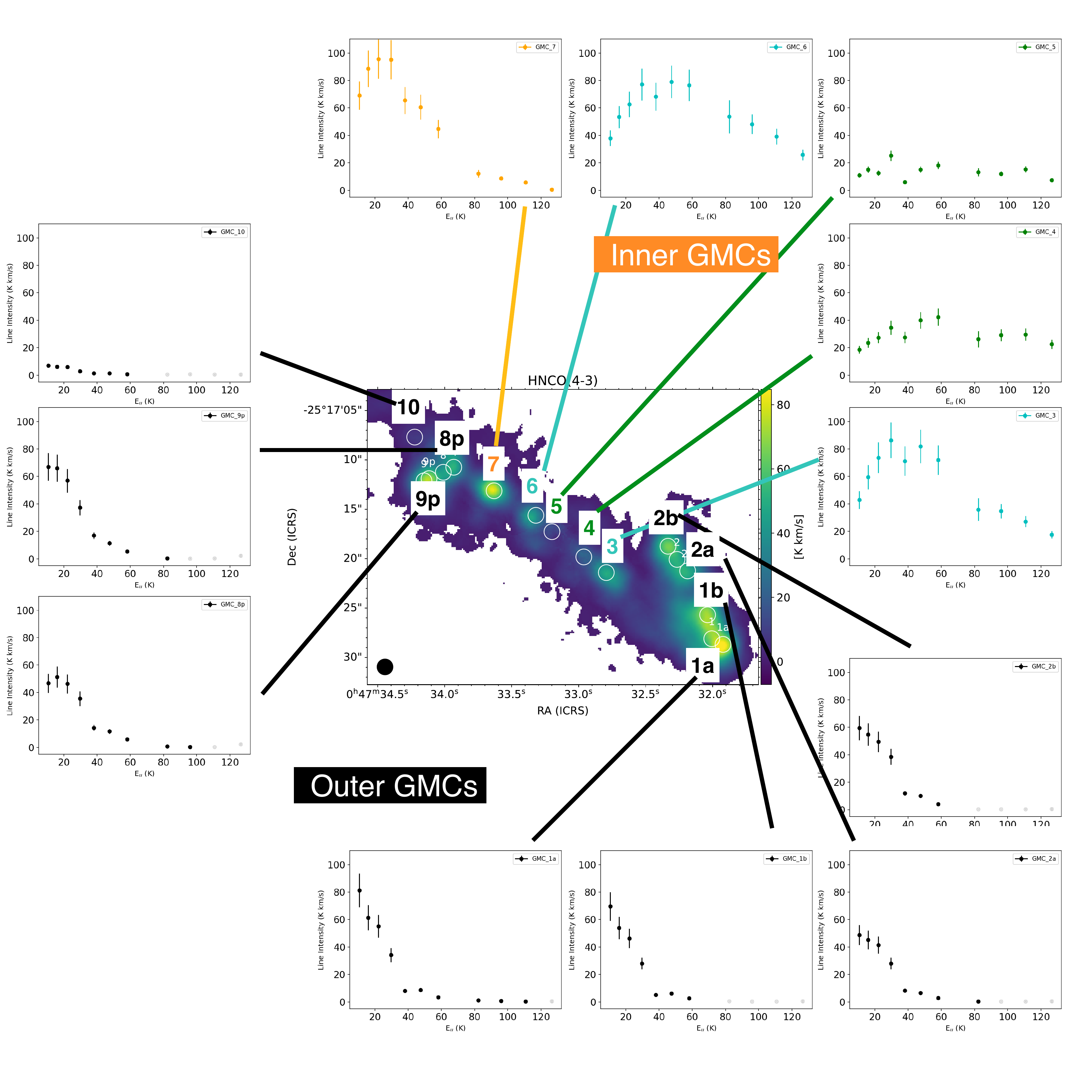}
    \caption{The SLED over all available energy levels of HNCO. The units are in [K km s\textsuperscript{-1}]. The GMCs are categorized by the "shape" of the ladders and labeled accordingly in different colors as described in Sect. \ref{sec:LTE}: GMC3/6 (in cyan), GMC4/5 (in green), and GMC7 (in orange), and the outer GMCs - GMC 1a/1b/2a/2b/8a/9a/10 (in black). The "shadowed" markers in each diagram represent non-detected points, displayed with upper limit only. }
    \label{fig:allGMCs_ladder_H}
\end{figure*}

\begin{figure*}
        \centering
    \includegraphics[scale=0.46]{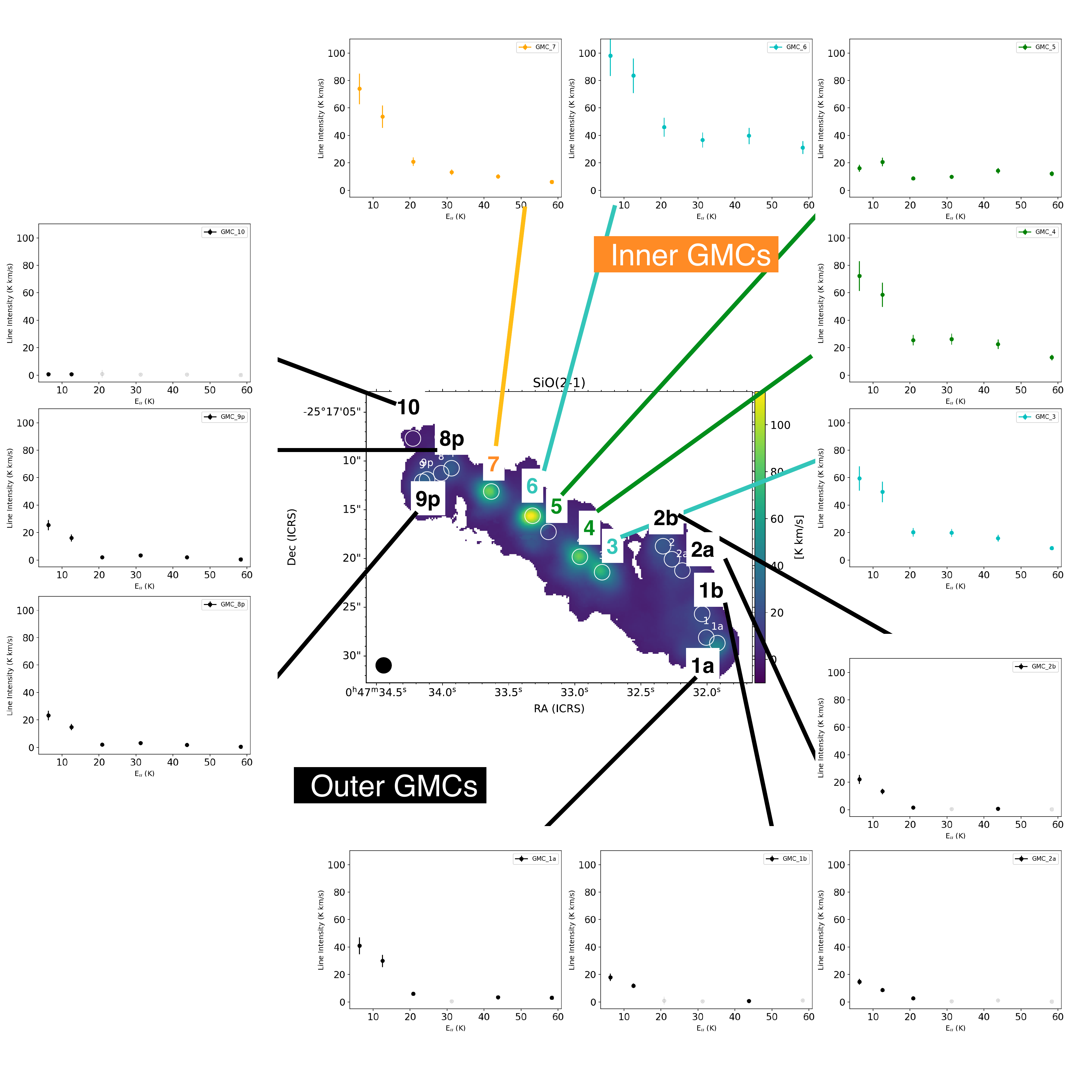}
    \caption{As in Fig. 3 but for SiO. 
    }
    \label{fig:allGMCs_ladder_S}
\end{figure*}

Within each GMC region, we further extract the beam-averaged line intensities across all available J transitions, and build up the so-called spectral line energy distribution (SLED) by populating the line intensities onto the same diagram, ordered by the transition upper-level energy, $E_{u}$
[K]. 
Fig. \ref{fig:allGMCs_ladder_H} and Fig. \ref{fig:allGMCs_ladder_S} show the HNCO and SiO SLEDs, respectively.  
In both sets of SLEDs, we group all the collected results by color to highlight the similarities and differences in shapes (the "magnitude" of the ladder, and where the ladder peaks in terms of $E_{u}$) among the GMCs.
In particular, it is clear that the excitation conditions for HNCO vary substantially from one GMC to another. 
Globally there is a distinction between the inner (GMC 3, 4, 5, 6, 7, represented with colored SLEDs in Fig. \ref{fig:allGMCs_ladder_H}) and the outer (SLEDs in black in Fig. \ref{fig:allGMCs_ladder_H}) regions of the CMZ. 
In the inner CMZ, we find also variation in the shapes of the molecular ladders, hinting that interesting physics and chemistry are taking place. 
Such variance among the subset of the GMCs in the inner CMZ can be grouped as: GMC3 and GMC6 (in cyan), GMC4 and GMC5 (in green), and GMC7 (in orange). 
GMC3/6 and GMC4/5 both have the brightest emission at mid-J excitation, between J=7-6 and J=12-11, across all HNCO transitions. 
Yet the GMC3/6 group has overall larger line intensities than GMC4/5. 
GMC7 tends to peak at lower $E_{u}$, suggesting that HNCO may be tracing colder gas in the inner CMZ. 
The SiO SLEDs (Fig. \ref{fig:allGMCs_ladder_S}) have very similar shapes across all the GMCs, as seen, also, in the intensity maps over different J levels. 

We note that both HNCO and SiO do not trace well GMC5 in most transitions. 
Also absorption features due to the fact that line emission is observed against the strong continuum towards the center of the galaxy as well as self-absorption in the potentially optically thick regime in GMC5 have been reported from multiple ALCHEMI studies with different species \citep[][]{Meier+2015_hncosio_253,Humire+2022}. 
Such features are also seen in our data from GMC5. 
To interpret the values extracted from this region requires extra caution and a proper radiative transfer modelling involving strong continuum emission in the background, which is beyond the scope of the current study.  
Additionally in GMC10, there is a lack of detection in most transitions as shown in grey points in the SLEDs in Fig. \ref{fig:allGMCs_ladder_H}-\ref{fig:allGMCs_ladder_S}. 
As a result, we will not discuss these two regions (GMC 5, 10) any further.

\subsection{Line intensity ratios}
\label{sec:intensity_ratio}
Variations in specific molecular line ratios are often used as probes of the physical characteristics and energetic processes in galaxies \citep[see detailed discussion by][]{Krips+2008}. This method relies on the assumption that the ratio of spectral line intensities may be proportional to the ratio of column densities, under the assumption of optically thin LTE. 
%

The SiO/HNCO molecular line intensity ratios have often been used as an indicator of shock strength in the past \citep[e.g.][]{Meier+2015_hncosio_253,Kelly+2017}, with SiO often referred to as a strong-shock tracer and HNCO a weak-shock tracer. 
However, \citet{Meier+2015_hncosio_253} find the fact that the ratio SiO(2-1) / HNCO (4\textsubscript{0,4}-3\textsubscript{0,3}) is higher in the inner CMZ of NC253 than in the outer regions may not be explainable by the shock strength alone. 
In the discussion by  \citet{Meier+2015_hncosio_253}, two additional arguments were brought up to explain the trend of [SiO(2-1) / HNCO (4\textsubscript{0,4}-3\textsubscript{0,3})] aside from the different shock strengths: the erased shock signatures due to PDR/UV fields and the different dependence of the partition function ($Z$) over temperatures (at LTE). 


The partition function of either species, {$Z_{\rm HNCO}$} and {$Z_{\rm SiO}$} can be approximated to be proportional to $\propto T^{3/2}$ and $\propto T$ for asymmetric tops and linear rotors respectively.
To test whether the difference in partition function could explain the observed trends in NGC 253 apart from the differentiation in their chemical abundances, we choose two pairs of SiO and HNCO transitions, SiO (2-1)  and HNCO (4\textsubscript{0,4}-3\textsubscript{0,3}) and SiO (7-6) and HNCO (10\textsubscript{0,10}-9\textsubscript{0,9}), based on close upper-level excitation energy state within each pair ($\sim 10$ K and $\sim 58$ K) (see Figure \ref{fig:ratio_maps}). 
In fact, the line intensity ratios from these two pairs cannot be well explained  by optically thin LTE: the latter would imply 
[SiO(7-6) / HNCO (10\textsubscript{0,10}-9\textsubscript{0,9})] > [SiO(2-1) / HNCO (4\textsubscript{0,4}-3\textsubscript{0,3})] in all GMCs. In Appendix \ref{sec:ratio_maps} (Fig. \ref{fig:ratio_maps}) we see the opposite trend (see also Tab.~\ref{tab:ratio}) between the two pairs of ratios. 

{Finally we remark that our conclusion that the gas is not in optically thin LTE remains valid regardless of the dependence of the partition functions on the temperature. The assumption of optically thin LTE implies a single excitation temperature. In our case when we compare different pairs of SiO/HNCO ratios from different J levels, the associated partition function should remain the same for each molecule. In this regard we conclude that the regions we investigated are generally not exhibiting optically thin LTE emission. }
{One final caveat concerning the line intensity ratio and the partition function is that, at higher gas temperatures, the vibrational contribution of the partition function may be relevant and is transition-level dependent. This 
may affect our derivation of  physical quantities such as the column density \citep[e.g.][]{CDMS_2016,PartitionFn_HNCO_2019} but as this may be relevant only for this initial LTE analysis, we do not discuss this further. }

In the following modelling section, we therefore remove the optically thin LTE assumption from our analysis.
\renewcommand{\arraystretch}{1.5}
\begin{table*}[ht!]
  \centering
  \caption{SiO/HNCO line intensity ratios}
  \label{tab:ratio}
  \begin{tabular}{c|cccccccc}
  \hline
    Ratio pair & 1a & 2b & 3 & 4 & 6 & 7 & 8a & 9a \\
    \hline
    ${I_{\rm SiO(2-1)}}/{I_{\rm HNCO(4_{0,4}-3_{0,3})}}$ & 0.50 &  0.38  &  1.39 &  4.01 &  2.64 & 1.07 &  0.50 &  0.38\\
    $\sigma_{\rm ratio}$ & (0.107) & (0.079)  & (0.294) & (0.826)  & (0.548) & (0.227) & (0.106) & (0.081)\\
    \hline
    ${I_{\rm SiO(7-6)}}/{I_{\rm HNCO(10_{0,10}-9_{0,9})}}$ & 0.82 &  - - &  0.12 &  0.31 &  0.41 & 0.14 &  0.11 &  - -\\
    $\sigma_{\rm ratio}$& (0.179)  &  (- -) &  (0.026) & (0.065) & (0.086) & (0.029) & (0.025) & (- -)\\
    \hline
  \end{tabular}
\end{table*}
\renewcommand{\arraystretch}{1.0}
\section{Modelling analysis}
The relative distribution of the observed line intensities of SiO and HNCO across the CMZ and over the excitation levels is likely a consequence of chemical as well as physical differentiation across the CMZ. In this section we perform non-LTE radiative transfer and chemical modelling in parallel, in order to disentangle the chemical from the physical effects and ultimately attempt at characterizing the shock history of the gas. 

Unlike the analysis performed by \citet{Holdship+2021_SpectralRadex,Holdship+2022,EB+2022}, where they coupled the radiative transfer code \texttt{RADEX} with the chemical code \texttt{UCLCHEM} and performed a Bayesian inference, we intentionally separate the two modelling processes and perform them in parallel. One of the advantages of this approach is that we obtain independently the best fit for the abundances, which we would require to be consistent for either model to be validated.

We describe our non-LTE radiative transfer modelling analysis with \texttt{RADEX} in Sect. \ref{sec:radex_tech}, and the chemical modelling with \texttt{UCLCHEM} in Sect. \ref{sec:chem_model}. 
\label{sec:modellings}
\subsection{Non-LTE radiative transfer analysis}
\label{sec:radex_tech}
For the non-LTE analysis, we use the radiative transfer code \texttt{RADEX} \citep{radex_vandertak_2007} via the Python package \texttt{SpectralRadex}\footnote{https://spectralradex.readthedocs.io} \citep{Holdship+2021_SpectralRadex} using HNCO and SiO molecular data \citep{hnco_moldata_N+1995,hnco_moedata_S+2018,sio_moldata_B+2018} from the LAMDA database \citep{LAMDA_2005}. This allows us to account for how the gas density and temperature affect the excitation of the transitions in the non-LTE regime and to constrain three physical parameters of interest: gas density ($n_{\rm H2}$), gas temperature ($T_{\rm kin}$), and the modeled species total column density ($N_{\rm species}$) as well as the beam filling factor ($\eta_{\rm ff}$). 

We coupled the \texttt{RADEX} modelling with a Bayesian inference process in inferring gas properties for properly sampling the parameter space and obtaining reliable uncertainties. 
The posterior probability distributions and the Bayesian evidence are derived with the nested sampling Monte Carlo algorithm MLFriends \citep{ultranest16,ultranest19} using the
\texttt{UltraNest} package\footnote{\url{https://johannesbuchner.github.io/UltraNest/}} \citep{ultranest21}.
Similarly to the approach adopted by \citet{Huang+2022}, we assume priors of uniform or log-uniform distribution within the determined ranges (given in Table \ref{tab:table_prior}) and assume that the uncertainty on our measured intensities is Gaussian so that our likelihood is given by $P(\theta | d) \sim \exp(-\frac{1}{2}\chi^2),$ where $\chi^2$ is the chi-squared statistic between our measured intensities and the \texttt{RADEX} output for a set of parameters $\theta$. 
\begin{table}[ht!]
  \centering
  \caption{Prior range adopted for our parameter space explored in the \texttt{RADEX}-Bayesian inference process described in Sect. \ref{sec:radex_tech}. The beam filling factor is defined as:  $\eta_{ff}=\frac{\theta^{2}_{S}}{\theta^{2}_{MB}+\theta^{2}_{S}}$}
  \label{tab:table_prior}
  \begin{tabular}{c|cc}
  \hline
    Variable  & Range & Distribution type\\
    \hline
    Gas density, $n_{\rm H2}$ [cm\textsuperscript{-3}] & $10^{2}-10^{8}$ & Log-uniform\\
    Gas temperature, $T_{\rm kin}$ [K] & $10-800$ & Uniform\\
    $N$(SiO) [cm\textsuperscript{-2}] & $10^{12}-10^{18}$ & Log-uniform\\
    $N$(HNCO) [cm\textsuperscript{-2}] & $10^{12}-10^{18}$ & Log-uniform\\
    Beam filling factor, $\eta_{ff}$ & $0.0-1.0$ & Uniform\\
    \hline
  \end{tabular}
\end{table}

In general this analysis is confined by the assumption that all of the molecular transitions arise from a single and homogeneous gas component. 
This assumption can only offer an averaged view for the gas properties given the limited resolution of the observations, and the different critical densities of the available transitions. 
Despite that, if variations in gas temperature and density within each GMC region are not too steep, the inference should still be able to give us an indication of the average gas properties in the non-LTE study. 

\begin{figure*}
  \centering
  \includegraphics[width=16cm]{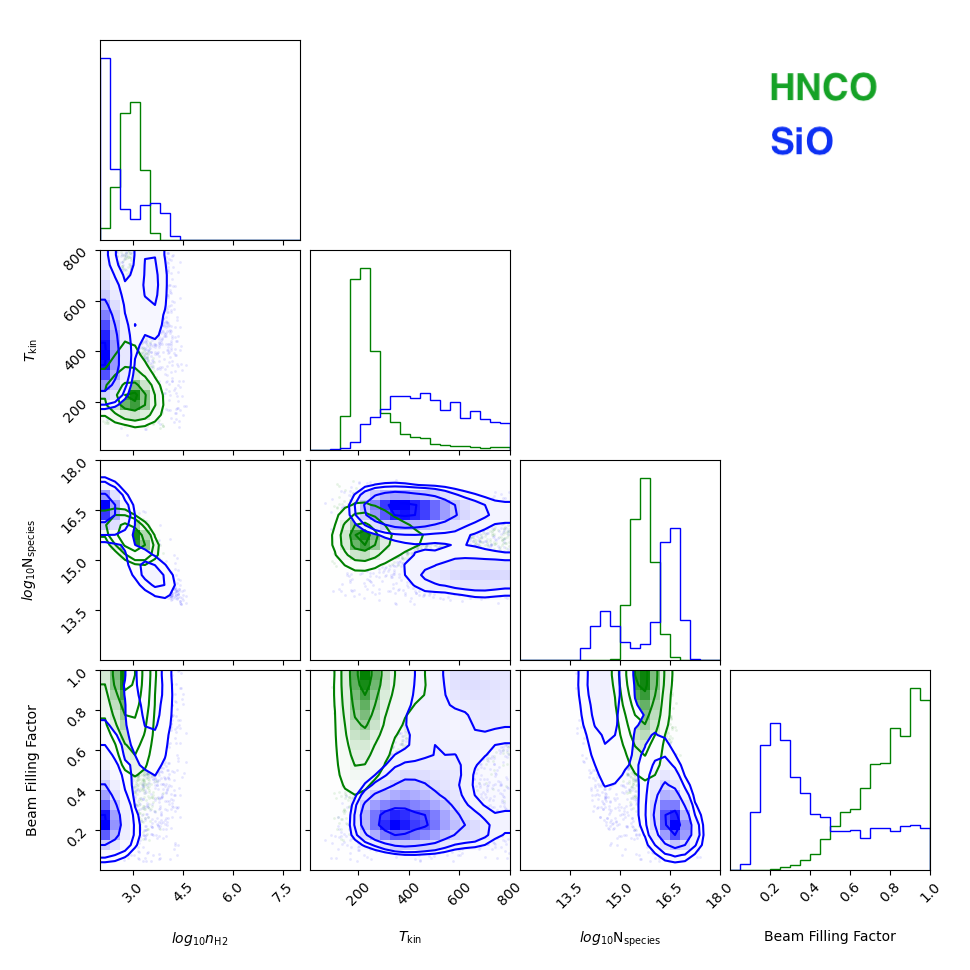}
  \caption{Bayesian inference results for the gas properties traced by HNCO (green) and SiO (blue) of the GMC 1a region. The corner plots show the sampled distributions for each parameter, as displayed on the x-axis. The 1-D distributions on the diagonal are the posterior distributions for each explored parameter; the rest 2-D distributions are the joint posterior for corresponding parameter pairs on the x- and y- axes. }
  \label{fig:RADEX_corner_GMC1a}
\end{figure*}

\begin{figure*}
  \centering
  \includegraphics[width=16cm]{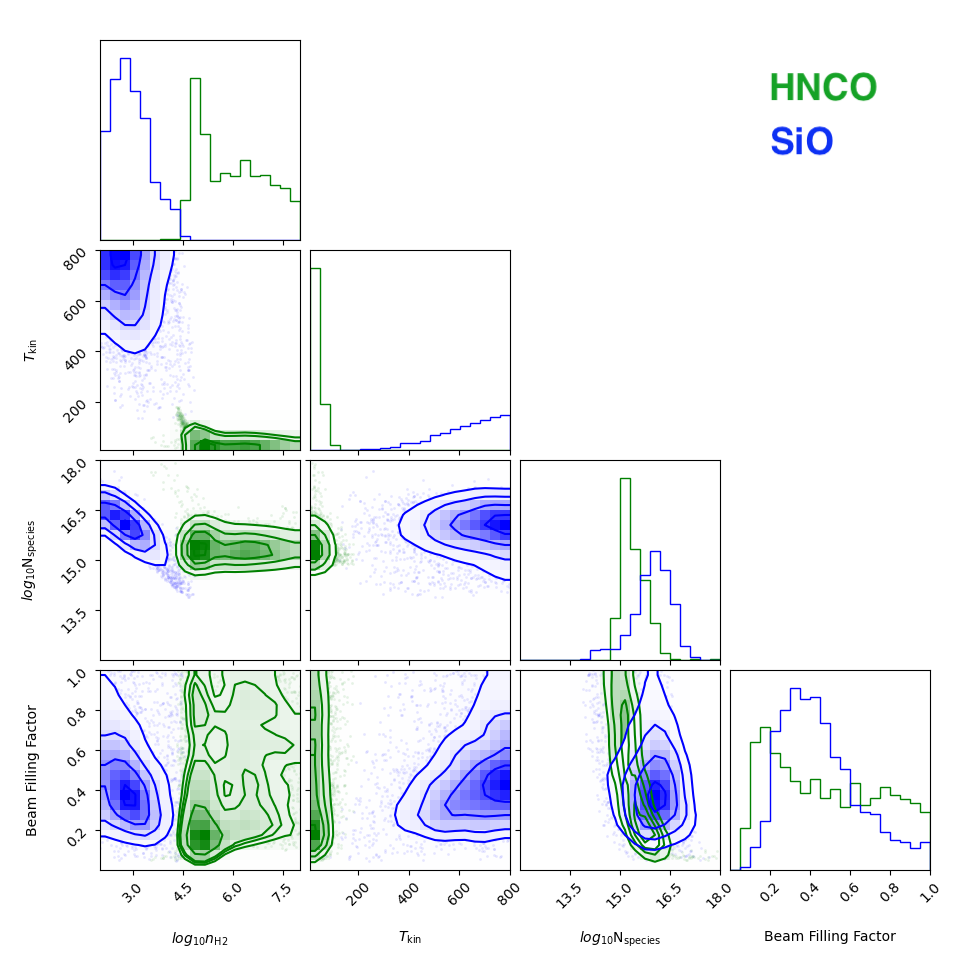}
  \caption{As Figure~\ref{fig:RADEX_corner_GMC1a} but for GMC7, as a representative region for the inner GMCs. 
  }
  \label{fig:RADEX_corner_GMC7}
\end{figure*}

\begin{figure*}
  \centering
  \includegraphics[width=16cm]{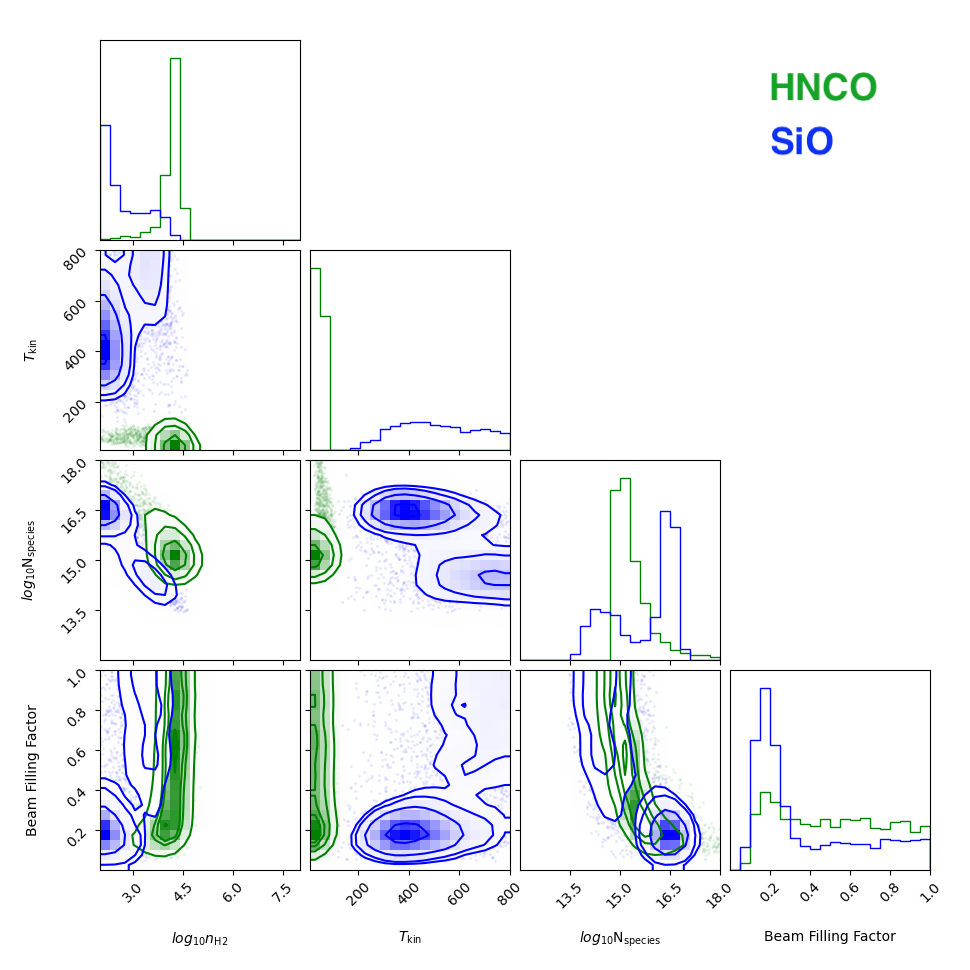}
  \caption{As Figure~\ref{fig:RADEX_corner_GMC1a} but for GMC9a, as a representative region for the outer GMCs.
  }
  \label{fig:RADEX_corner_GMC9p}
\end{figure*}

In Figs. \ref{fig:RADEX_corner_GMC1a}-\ref{fig:RADEX_corner_GMC9p} we show the most representative cases of the posterior distribution among the sampled GMC regions. 
The remaining GMC cases are shown in Appendix \ref{sec:rest_corners}. 
The inferred gas properties from our GMCs are also listed in Table \ref{tab:table_inferred_property}. 
Owing to the high quality of the ALCHEMI data and the ample number of transitions per species, most of the inferred gas properties are  well constrained, which is a real benefit for further physical interpretation. 
In some circumstances when RADEX fitting struggles to find a good solution, it may produce a solution where both $n_{H_{2}}$ and $T_{kin}$ are at the edges of the parameter space, with either (low-$n_{H_{2}}$, high-$T_{kin}$) or (high-$n_{H_{2}}$, low-$T_{kin}$); this is a well known degeneracy in the $n_{H_{2}}$-$T_{kin}$ space. 
In our results, however, this does not seem to happen often. 
{It is also worth noting that for some GMCs (e.g. GMC 6 and 1a) we did find cases with an optical depth, as predicted by RADEX-Bayesian analysis,  greater than unity, for  some of the SiO and/or HNCO transitions; this is compatible with our conclusion in Sect. \ref{sec:intensity_ratio} concerning non-optically thin LTE conditions. }

Overall HNCO is tracing denser and cooler gas components compared to SiO across all the GMCs. 
On the other hand, HNCO total column density is systematically smaller than SiO. 
This hints at the two species tracing distinctively different gas components. 
The species total column density is a function of gas density, emission region size, and the species' fractional (relative to hydrogen) abundance. 
If one assumes the beam filling factors from HNCO and SiO are comparable (e.g. GMC 9a - Fig. \ref{fig:RADEX_corner_GMC9p}), suggesting the emission region traced by both species is also comparable, this suggests that the HNCO fractional abundance may be  systematically smaller than that of SiO. 

We also want to highlight that the gas temperature probed by SiO in most GMCs is hot ($T > 400$ K), hinting at the presence of shock heating. 
The two exceptions are GMC 2b and 8a, but both are still pretty warm ($T>200$ K). 
If both requires shocks, these suggest that the two groups of GMCs are simply at different post-shock cooling stages or that they are affected by different types of shocks. 

There are a few GMCs that are particularly interesting. 
The gas traced by HNCO in GMC 1a points to a quite low gas density ($n_{\rm H2} \sim 10^{3}$ cm\textsuperscript{-3}) and high gas temperature ($T_{\rm kin}\sim 230$ K) compared to the rest of the GMCs traced by HNCO. 
The chemical modelling performed by \citet{Kelly+2017} with a standard cosmic ray ionization rate (CRIR, =$1.3 \times 10^{-17}$ s\textsuperscript{-1}) \citep{Holdship+2017} showed that thermal sublimation cannot account for the HNCO enhancement with this low density for the gas and dust are not well coupled. 
In the case of NGC 253, however, multiple works have revealed high CRIR ($\sim 10^{-14}-10^{-11}$ s\textsuperscript{-1}) for all GMCs \citep{Holdship+2021_SpectralRadex,Harada+2021,Holdship+2022,EB+2022}. We shall discuss this further in Sec. \ref{sec:Cshock_chem}.
On the other hand, the gas traced by HNCO in GMC 7 points to a very low gas temperature ($T\sim 24$ K) compared to all the other GMCs.  
In Sec. \ref{sec:Hotcore_chem},
we will explore the possibility of thermal sublimation of HNCO  at low temperatures apart from the shock scenarios. 
It is also interesting to compare GMCs of comparable temperatures ($T_{\rm kin}>50$), as traced by HNCO in GMCs 4 and 6: here we find that GMC 6 has a higher abundance of HNCO (by a factor of two) compared to GMC 4.
This was in fact already hinted in the SLEDs shown in Sec. \ref{sec:LTE} where the HNCO SLED of GMC 6 showed greater brightness overall than GMC 4 despite their peaks all leaning towards a similar $E_{u}$.
Finally, caution needs to be taken in interpreting our HNCO observations in GMC3, as although it appears that both gas density and gas temperature are constrained, in reality the $n_{\rm H2}$ peak and $T_{\rm kin}$ peak point to two degenerate sets of solutions. We list the best fit with best likelihood values in Table \ref{tab:table_inferred_property}.

\renewcommand{\arraystretch}{1.5}
\begin{table*}[ht!]
  \centering
  \caption{The inferred gas properties traced by HNCO and SiO from the Bayesian inference processes  over four selected regions across GMC1a-9a (columns 2-5). For poorly constrained cases we identify the upper or lower limit of the distribution and for such cases we place the 95 (for upper limit) or 5 (for lower limit) percentile values in  parenthesis. We also list the relevant shock timescales derived in Sect. \ref{sec:timelines_shocks} (columns 6-8). }
  \label{tab:table_inferred_property}
  \begin{tabular}{c|cccccccc}
  \hline
    GMC & Species & {$log_{10}(n_{H_{2}})$} & {$T_{kin}$} & {$log_{10}(N_{\rm species})$} & $\eta_{ff}$ & $\tau_{\rm dissipation}$ & $\tau_{shock}$ & $\tau_{shock,joint}$ \\
    {} & {} & {[cm\textsuperscript{-3}]} & {[K]} & {[cm\textsuperscript{-2}]} & {} & [yr] & {} & [yr] \\
    \hline
    \hline
    1a & HNCO &{$2.92^{+0.32}_{-0.32}$} & {$230.1^{+95.25}_{-43.67}$} &{$15.68^{+0.32}_{-0.30}$} & {$0.82^{+0.13}_{-0.21}$} & {$\sim10^4$} & {$\geq\tau_{\rm dissipation}$} & {$\sim10^5$}\\
    {} & SiO & {$2.37^{+1.15}_{-0.30}$} & {$472.10^{+194.15}_{-155.44}$} &{$16.33^{+0.36}_{-1.71}$} & {$0.36^{+0.43}_{-0.16}$} & {$\sim10^5$} & {$\sim\tau_{\rm dissipation}$} & {- -}\\
    \hline
    2b & HNCO & {$4.33^{+0.15}_{-0.19}$} & {$39.82^{+8.07}_{-6.01}$} &{$15.02^{+0.50}_{-0.22}$} & {$0.54^{+0.31}_{-0.33}$} & {$\sim10^3$} & {$\gg\tau_{\rm dissipation}$} & {$\geq10^5$}\\
    {} & SiO & {$2.21^{+1.20}_{-0.16}$} & {$201.39^{+187.57}_{-86.56}$} &{$16.58^{+0.18}_{-2.21}$} & {$0.17^{+0.61}_{-0.05}$} & {$\sim10^5$} & {$\geq\tau_{\rm dissipation}$} & {- -}\\
    \hline
    3$^{(a)}$ & HNCO &{$\geq {6}$} & {$<100$} &{$15.44^{+0.59}_{-0.43}$} & {$0.28^{+0.45}_{-0.20}$} & {$10^1$} & {$\gg\tau_{\rm dissipation}$} & {$\sim10^3$}\\
    {} & SiO &{$3.93^{+0.42}_{-0.75}$} & {$649.40^{+108.83}_{-195.23}$} &{$15.73^{+0.88}_{-0.35}$} & {$0.09^{+0.08}_{-0.04}$} & {$\sim10^3$} & {$\sim\tau_{\rm dissipation}$} & {}\\
    \hline
    4 & HNCO &{$6.73^{+0.84}_{-0.92}$} & {$64.90^{+38.56}_{-7.69}$} &{$15.09^{+1.14}_{-0.39}$} & {$0.31^{+0.46}_{-0.29}$} & {$\sim10^1$} & {$\gg\tau_{\rm dissipation}$} & {$\sim10^3$}\\
    {} & SiO &{$4.07^{+0.41}_{-0.93}$} & {$614.08^{+135.24}_{-248.67}$} &{$15.93^{+1.31}_{-0.46}$} & {$0.06^{+0.06}_{-0.02}$} & {$\sim10^3$} & {$\sim\tau_{\rm dissipation}$} & {- -}\\
    \hline
    6 & HNCO &{$6.01^{+1.19}_{-0.93}$} & {$57.09^{+152.19}_{-13.03}$} &{$15.36^{+0.74}_{-0.39}$} & {$0.33^{+0.45}_{-0.27}$} & {$\sim10^1$} & {$\gg\tau_{\rm dissipation}$} & {$\sim10^3$}\\
    {} & SiO &{$4.20^{+0.50}_{-0.99}$} & {$515.90^{+199.81}_{-263.75}$} &{$16.53^{+0.96}_{-0.64}$} & {$0.05^{+0.02}_{-0.01}$} & {$\sim10^3$} & {$\sim\tau_{\rm dissipation}$} & {- -}\\
    \hline
    7 & HNCO &{$5.86^{+1.28}_{-0.98}$} & {$24.28^{+33.24}_{-4.73}$} &{$15.32^{+0.47}_{-0.27}$} & {$0.45^{+0.36}_{-0.27}$} & {$\sim10^1$} & {$\gg\tau_{\rm dissipation}$} & {$\sim10^4$}\\
    {} & SiO &{$2.85^{+0.64}_{-0.51}$} & {$656.24^{+101.93}_{-152.47}$} &{$16.02^{+0.48}_{-0.59}$} & {$0.42^{+0.24}_{-16}$} & {$\sim10^4$} & {$\sim\tau_{\rm dissipation}$} & {- -}\\
    \hline
    8a & HNCO &{$4.13^{+0.14}_{-0.15}$} & {$85.41^{+19.33}_{-15.61}$} &{$14.95^{+0.43}_{-0.19}$} & {$0.57^{+0.29}_{-0.32}$} & {$\sim10^3$} & {$\gg\tau_{\rm dissipation}$} & {$\geq10^5$}\\
    {} & SiO &{$2.16^{+1.10}_{-0.12}$} & {$250.24^{+196.22}_{-101.71}$} &{$16.58^{+0.15}_{-2.05}$} & {$0.16^{+0.54}_{-0.05}$} & {$\sim10^5$} & {$\geq\tau_{\rm dissipation}$} & {- -}\\
    \hline
    9a & HNCO &{$4.20^{+0.16}_{-0.25}$} & {$49.71^{+9.62}_{-7.73}$} &{$15.17^{+0.62}_{-0.25}$} & {$0.51^{+0.33}_{-0.30}$} & {$\sim10^3$} & {$\gg\tau_{\rm dissipation}$} & {$\sim10^5$}\\
    {} & SiO &{$2.51^{+1.08}_{-0.41}$} & {$501.03^{+193.35}_{-160.78}$} &{$16.21^{+0.37}_{-1.79}$} & {$0.25^{+0.51}_{-0.10}$} & {$\sim10^5$} & {$\sim\tau_{\rm dissipation}$} & {- -}\\
    \hline
  \end{tabular}\\
  \footnotesize{$^{(a)}$ The bi-modality of inferred gas properties in GMC 3 traced by HNCO is explained in Sec. \ref{sec:radex_tech}. }
\end{table*}

\subsection{Chemical modelling}
\label{sec:chem_model}
The RADEX and Bayesian inference process is "blind" to chemistry, in so far as the chemistry behind each species is not taken into consideration: this may result in chemically unfeasible "best fit" parameters. In this section we therefore perform chemical modelling with the open source time dependent gas-grain \texttt{UCLCHEM}\footnote{https://uclchem.github.io} code \citep{Holdship+2017} in order to further disentangle and constrain the chemical origin of the observed HNCO and SiO emissions. 
\texttt{UCLCHEM} is a gas-grain chemical modelling code that incorporates user-defined chemical networks to produce chemical abundances along user-defined physics modules that can simulate a variety of physical conditions. 

Compared to the older version of \texttt{UCLCHEM} used by \citet{Kelly+2017}, the latest \texttt{UCLCHEM  v3.1} includes updated chemistry and physics modules.
Of importance to this study, \texttt{UCLCHEM v3.1} includes an improved sputtering module in the parameterized C-shock model following \citet{J-S+2008_shocktracers} as well as 
a 3-phase chemistry where chemistry is computed for the gas phase, the grain surfaces, and the bulk ice. These improvements ensure a better treatment of  the sputtering  of refractory species such as Si-bearing molecules during the shock process. 
Aside from these technical differences, we also tailored our modelling for NGC 253 by using a much higher CRIR, $\zeta=10^{3-5}\zeta_{0}$ \citep[][$\zeta_{0}=1.3\times 10^{-17}$ s\textsuperscript{-1} is the standard galactic CRIR]{Holdship+2021_SpectralRadex,Harada+2021,Holdship+2022,EB+2022}, 
than the standard galactic CRIR. 

Chemical modelling with \texttt{UCLCHEM} typically involves 2 evolutionary stages. 
In our models, Stage 1 starts with the gas in diffuse atomic/ionic 
form and follows the chemical evolution of the gas and ices  undergoing free-fall collapse up to a final density at the end of Stage 1. Based on the recipe described by \citet{Holdship+2017}, the initial elemental abundances are assumed to be solar. The temperature is set at 10 K. The output of this stage is a model of a typical quiescent molecular cloud. 
In our case we assume that the gas-phase elemental abundance of Si has been depleted from solar level \citep[Si/H$\sim 4.07\times 10^{-5}$, ][]{Lodders+2003_solarabund} so that 99\% of the Si is incorporated into the grain cores.

In Stage 2, we explore both shock (Sect. \ref{sec:Cshock_chem}) and non-shock (Sect. \ref{sec:Hotcore_chem}) scenarios. For the shock models we run a grid where we vary the following parameters: pre-shock gas density, post-shock gas temperature, shock velocity, and cosmic ray ionization rate (CRIR or $\zeta$). 
In Table \ref{tab:table_chemgrids} we list the parameter space explored in our stage-2 chemical modelling analysis. 

For the post-shock gas we assume that the post-shock gas temperature $T_{post-shock}$ is $\sim$ $50$ K in order to cover most low temperatures measured in Sect. \ref{sec:radex_tech} as well as the dust temperatures ($T_{d}=35$ K) measured by \citet{Leroy+2015} and the kinetic temperature ($T_{k}\sim50$ K) measured in less-dense gas ($n_{\rm H2}\sim10^{4}$ cm\textsuperscript{-3}) by \citet{Mangum+2019}. 

\renewcommand{\arraystretch}{1.0}
\begin{table*}[ht!]
  \centering
  \caption{The parameter space  explored in our chemical modelling. Note that  $\zeta_{0}=1.3\times10^{-17}$ s\textsuperscript{-1} and $X_{Si, gas,\odot}=4.07\times10^{-5}$.}
  \label{tab:table_chemgrids}
  \begin{tabular}{c|c}
  \hline
    Variable  & Grid \\
    \hline
    Pre-shock/Initial gas density, $n_{\rm H2}$ [cm\textsuperscript{-3}] & [$10^{3}$, $10^{4}$, $10^{5}$, $10^{6}$] \\
    C-shock velocity, $\varv_{shock}$ [km s\textsuperscript{-1}] & [$5.0$, $10.0$, $20.0$, $30.0$, $40.0$, $50.0$, $60.0$] \\
    CRIR $\zeta$ [$\zeta_{0}$] & [$10^{3}$, $10^{5}$] \\
    Physical model & Shock or non-shocked scenario \\
    Post-shock gas temperature, $T$ [K] & $50$ \\
    \hline
  \end{tabular}
\end{table*}
\subsubsection{The effects of the passage of C-shock(s) on the GMCs of NGC 253}
\label{sec:Cshock_chem}
\begin{figure*}
  \centering
  \begin{tabular}[b]{@{}p{0.45\textwidth}@{}}
    \centering\includegraphics[width=1.0\linewidth]{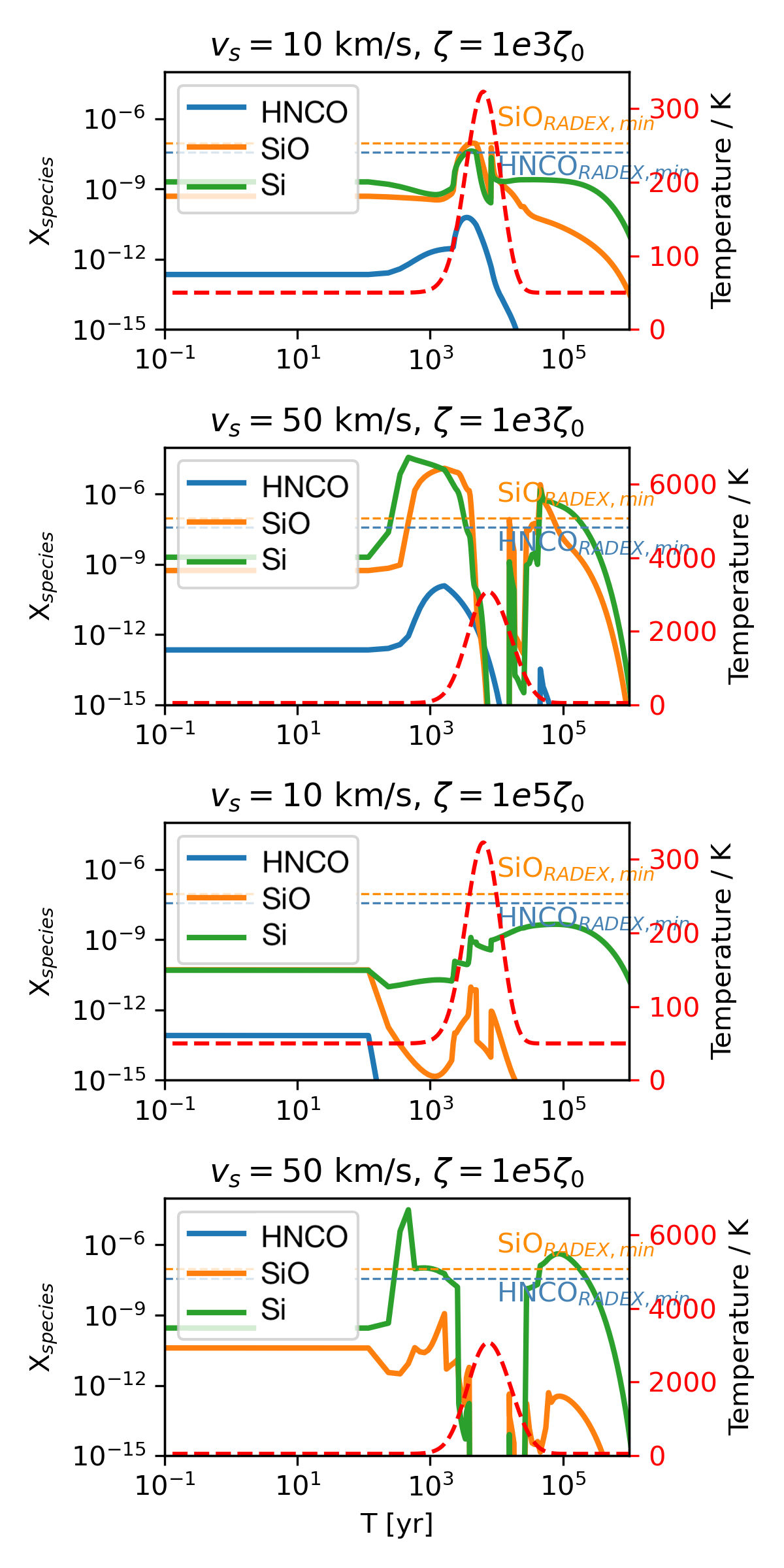} \\
    \centering\small (a)
  \end{tabular}
  \begin{tabular}[b]{@{}p{0.45\textwidth}@{}}
    \centering\includegraphics[width=1.0\linewidth]{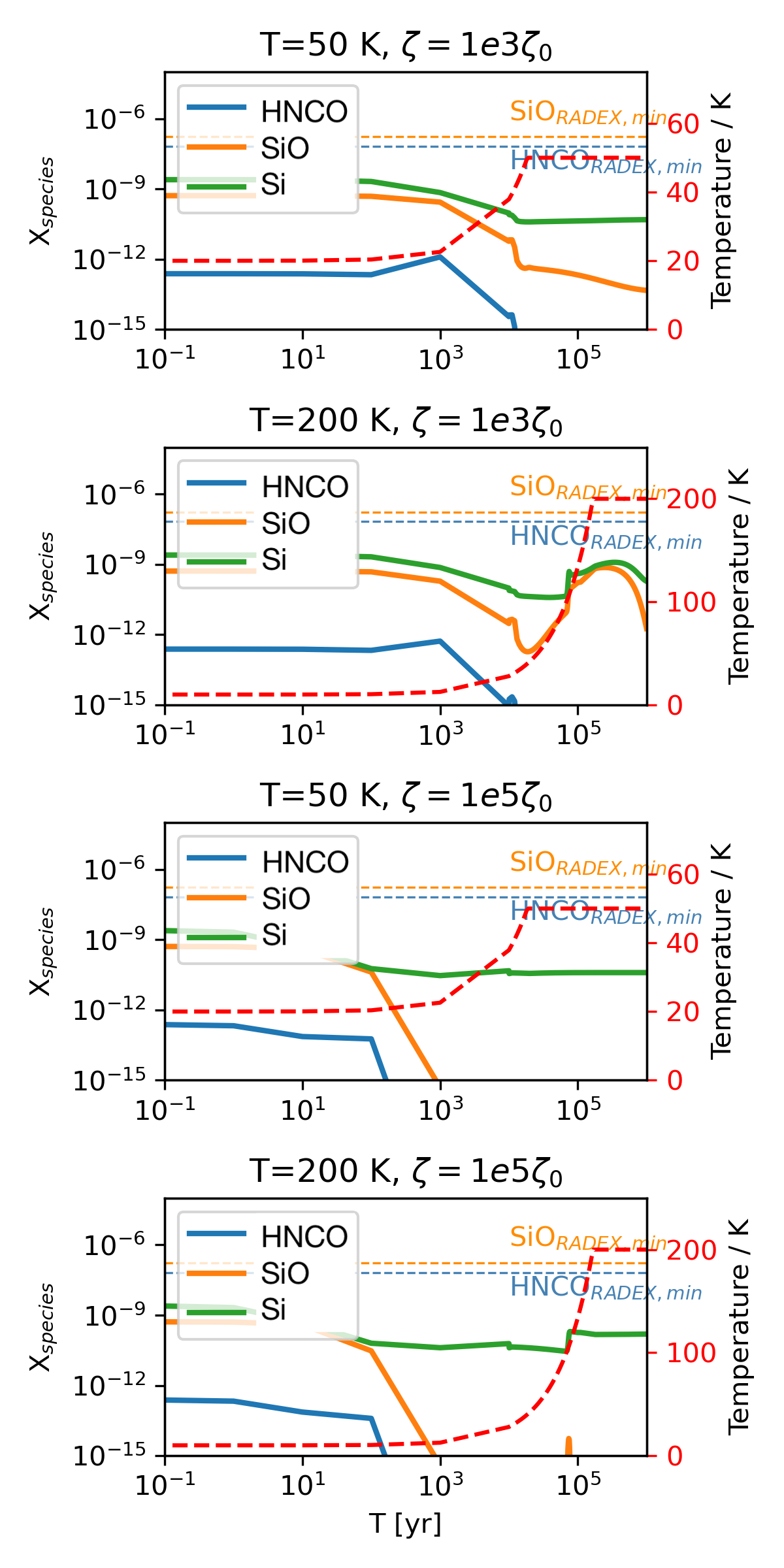} \\
    \centering\small (b)
  \end{tabular}
  \caption{Chemical abundances as a function of time for a shock  (left panel)  and a non-shocked gas  (right panel). The pre-shock gas density in the shock models and the gas density in non-shock models is $10^{3}$ cm\textsuperscript{-3}.
  The red dashed curve represents the temperature profile, with the temperature scale on the vertical axis on the right, also in red. For the shock models, within each panel we present from top to bottom: [shock velocity ($v_{s}$)=10 km/s, CRIR ($\zeta$)=$10^3\zeta_{0}$], [shock velocity ($v_{s}$)=50 km/s, CRIR ($\zeta$)=$10^3\zeta_{0}$], [shock velocity ($v_{s}$)=10 km/s, CRIR ($\zeta$)=$10^5\zeta_{0}$], and [shock velocity ($v_{s}$)=50 km/s, CRIR =$10^5\zeta_{0}$]. The two selected temperatures for the non-shock models are 50 K and 200 K. The dashed, colored horizontal lines indicate the lower limit of the species fractional abundances "measured" from our RADEX-Bayesian inference based on observational data and with an assumed hydrogen column density - see Sec. \ref{sec:RADEX-CHEM_relation}, for HNCO (blue) and SiO (orange) respectively. The fractional abundance values used are the minimum derived values  among GMCs. 
  For example, the SiO reference line (orange) is  from values "measured" at GMC 7 as listed in Tab. \ref{tab:table_inferred_property} which provides the lowest "measured" fractional SiO abundance among all cases with gas density $n_{\rm H2}$ lower than $10^{3}$ cm\textsuperscript{-3}. }
  \label{fig:UCLCHEM_Cshock_den03}
\end{figure*}

\begin{figure*}
  \centering
  \begin{tabular}[b]{@{}p{0.43\textwidth}@{}}
    \centering\includegraphics[width=1.0\linewidth]{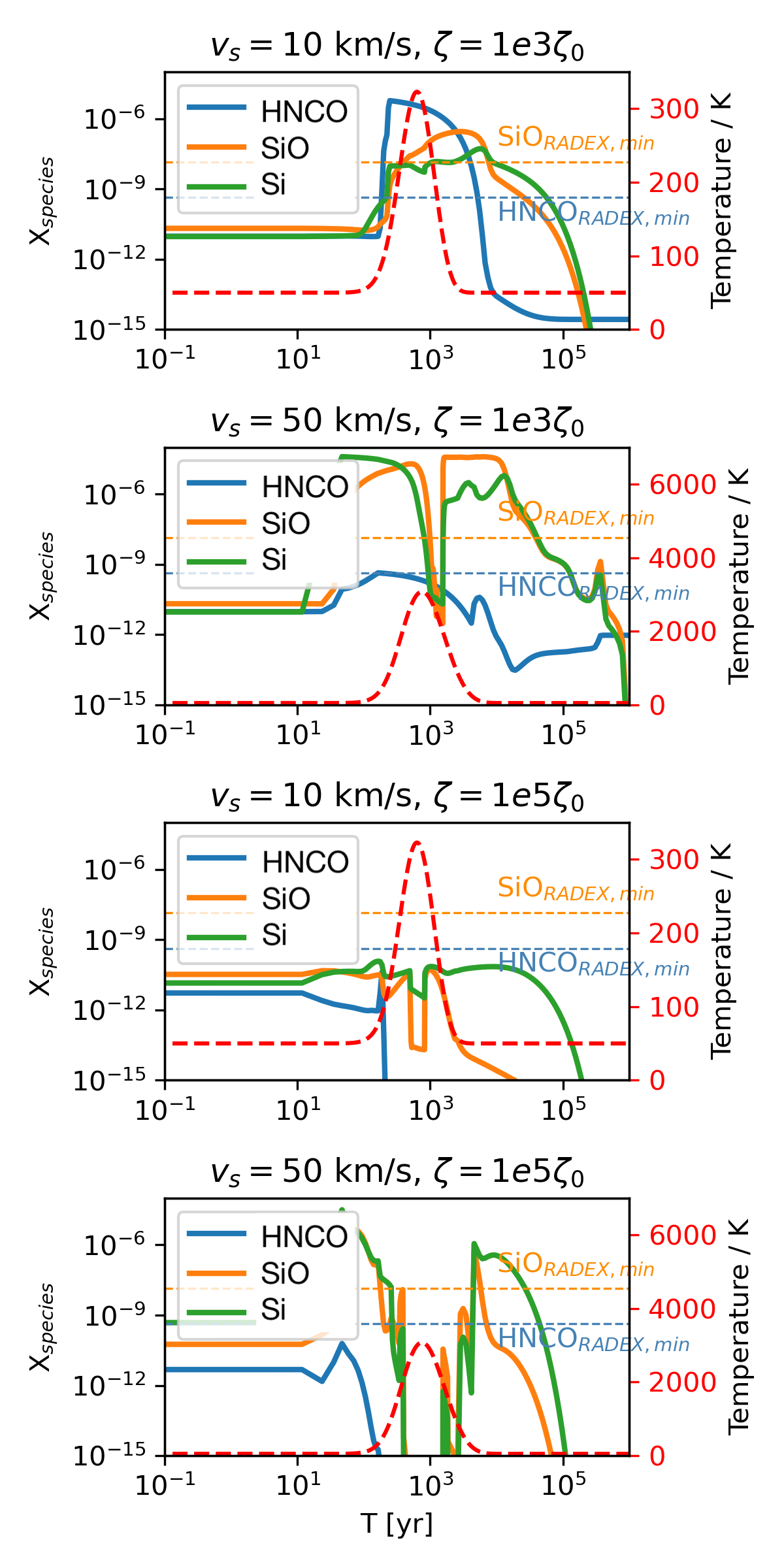} \\
    \centering\small (a)
  \end{tabular}
  \begin{tabular}[b]{@{}p{0.43\textwidth}@{}}
    \centering\includegraphics[width=1.0\linewidth]{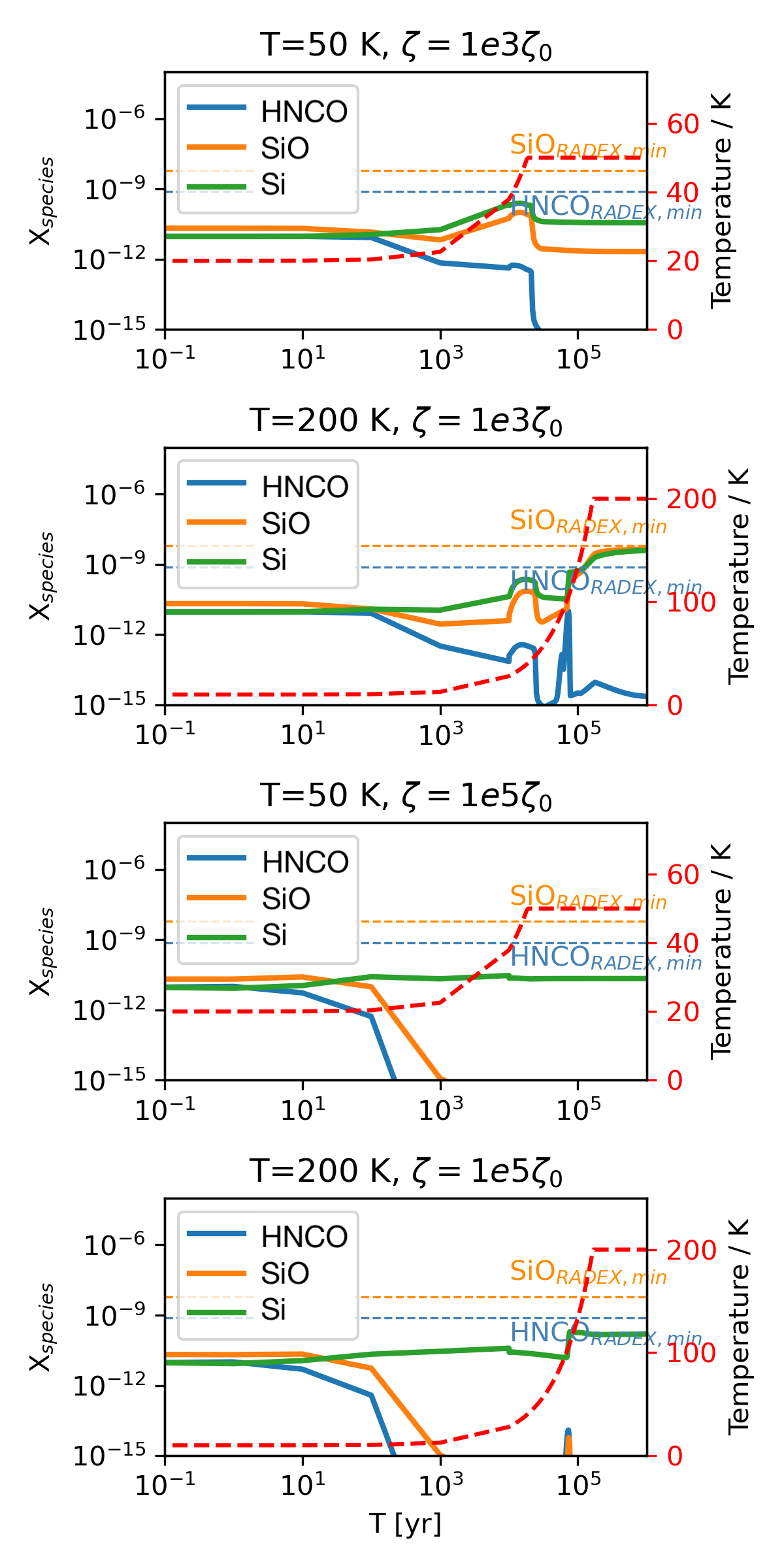} \\
    \centering\small (b)
  \end{tabular}
  \caption{As in  Figure~\ref{fig:UCLCHEM_Cshock_den03} but for a  pre-shock gas density in (a) and a gas density in (b) of  $10^{4}$ cm\textsuperscript{-3}. }
  \label{fig:UCLCHEM_Cshock_den04}
\end{figure*}

\begin{figure*}
  \centering
  \begin{tabular}[b]{@{}p{0.43\textwidth}@{}}
    \centering\includegraphics[width=1.0\linewidth]{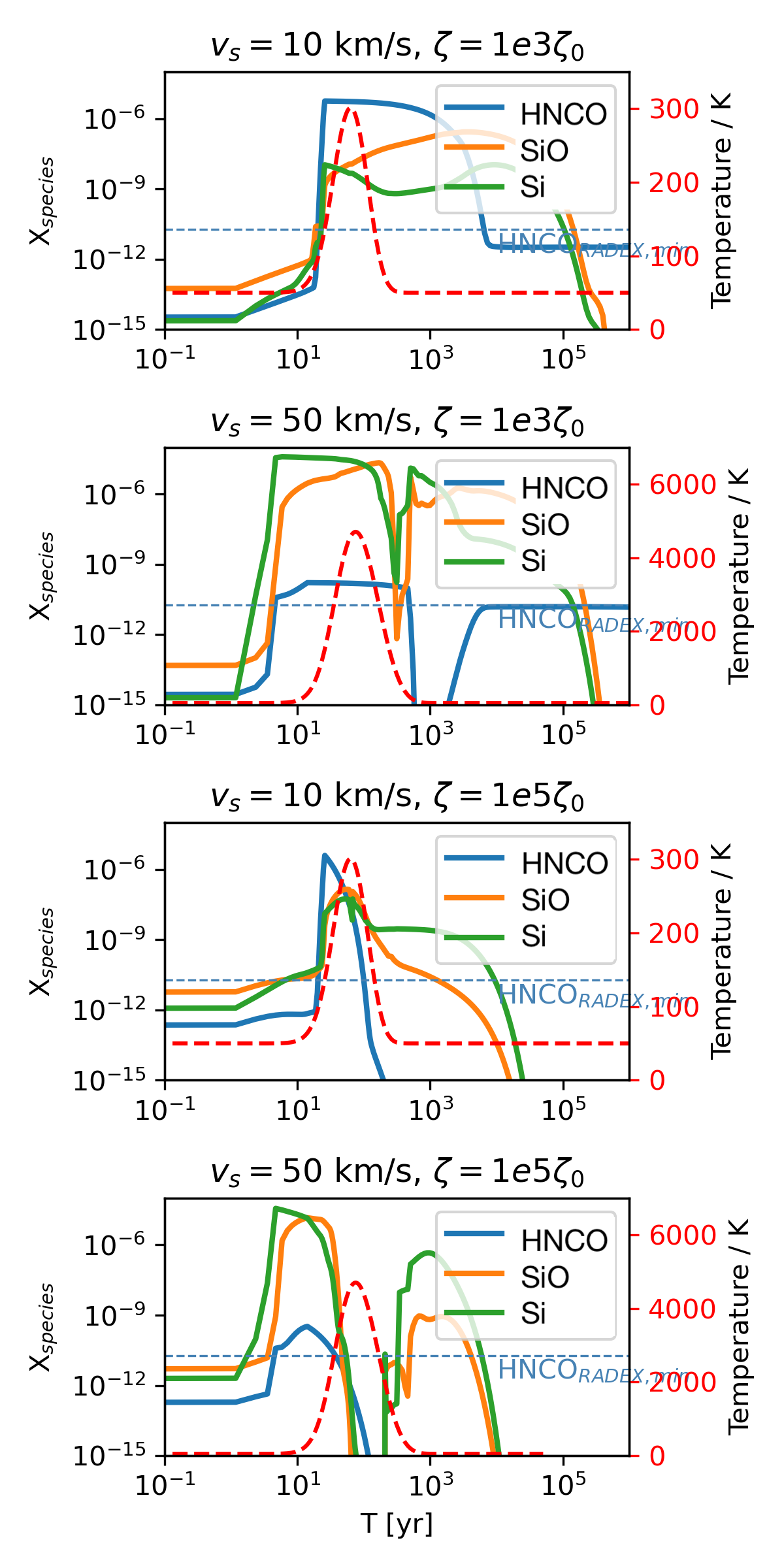} \\
    \centering\small (a)
  \end{tabular}
  \begin{tabular}[b]{@{}p{0.43\textwidth}@{}}
    \centering\includegraphics[width=1.0\linewidth]{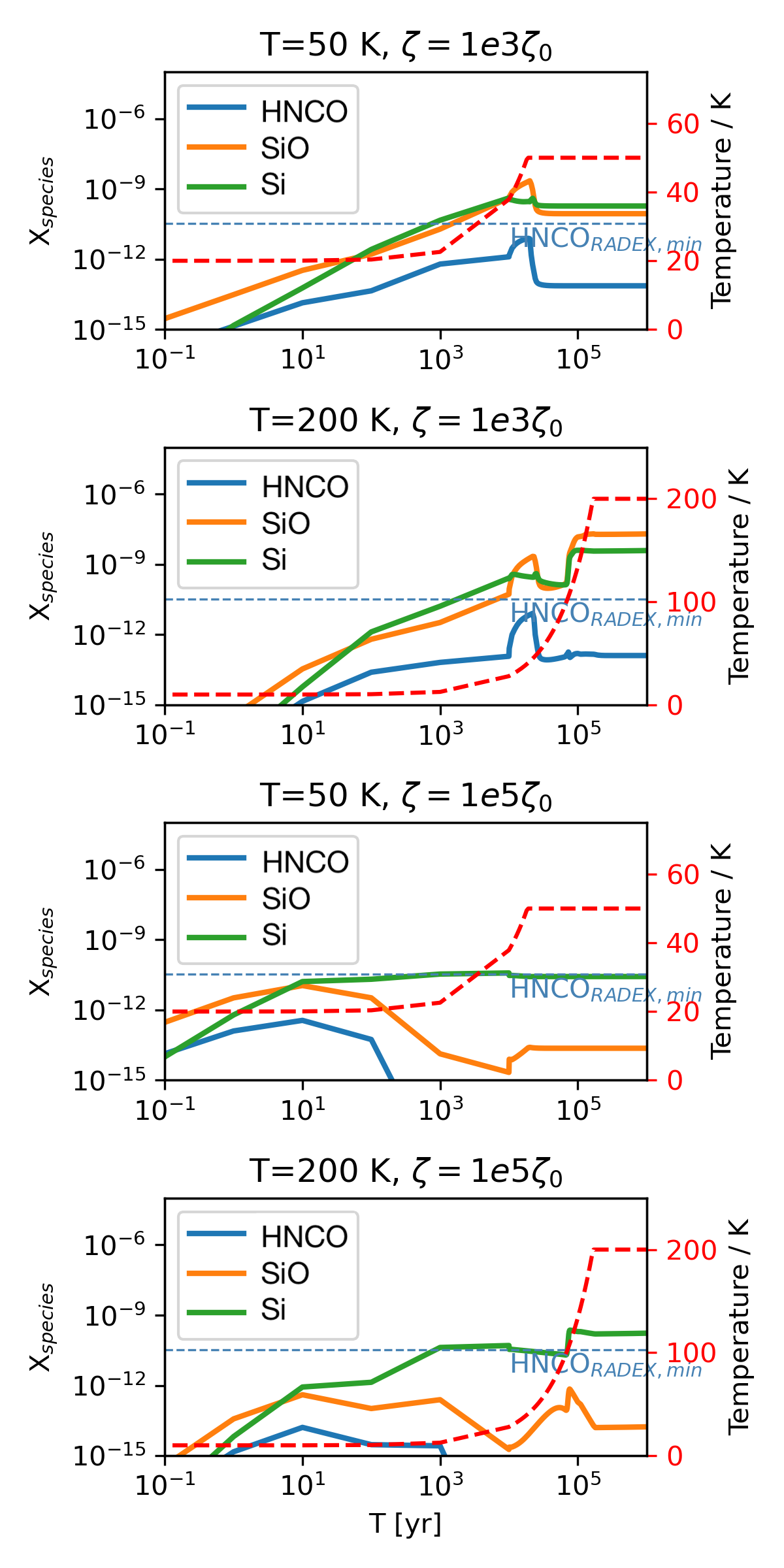} \\
    \centering\small (b)
  \end{tabular}
  \caption{As in Figure~\ref{fig:UCLCHEM_Cshock_den03} but for a  pre-shock gas density in (a) and a gas density in (b) of  $10^{5}$ cm\textsuperscript{-3}. The auxiliary line that indicates the lower limit of SiO (orange) fractional abundances "measured" from our RADEX-Bayesian inference is missing compared to Fig. \ref{fig:UCLCHEM_Cshock_den03}-\ref{fig:UCLCHEM_Cshock_den04} in the current case because observationally we do not find any GMC that is associated with such a high gas density. }
  \label{fig:UCLCHEM_Cshock_den05}
\end{figure*}

\begin{figure*}
  \centering
  \begin{tabular}[b]{@{}p{0.43\textwidth}@{}}
    \centering\includegraphics[width=1.0\linewidth]{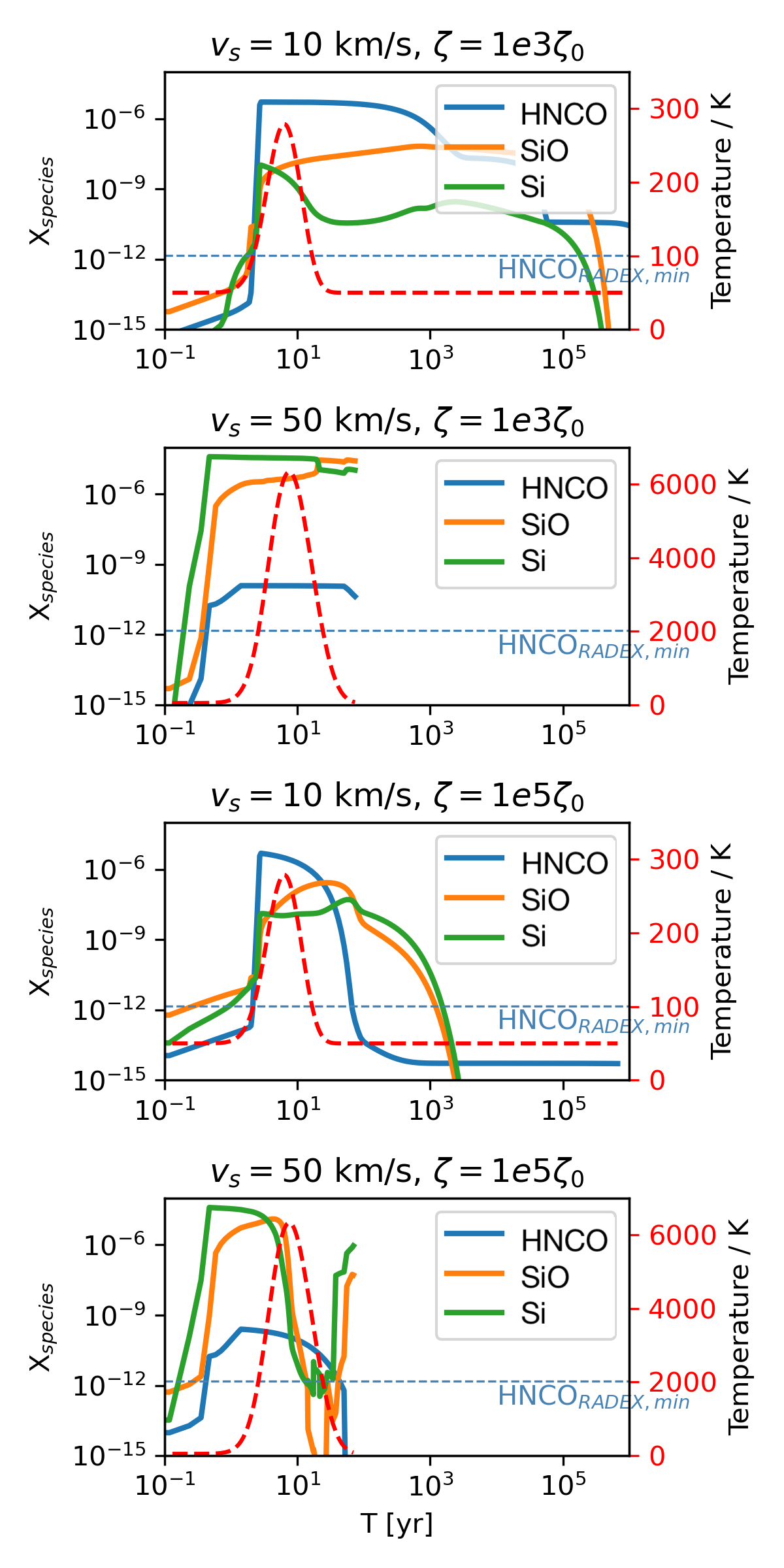} \\
  \end{tabular}
  \begin{tabular}[b]{@{}p{0.43\textwidth}@{}}
    \centering\includegraphics[width=1.0\linewidth]{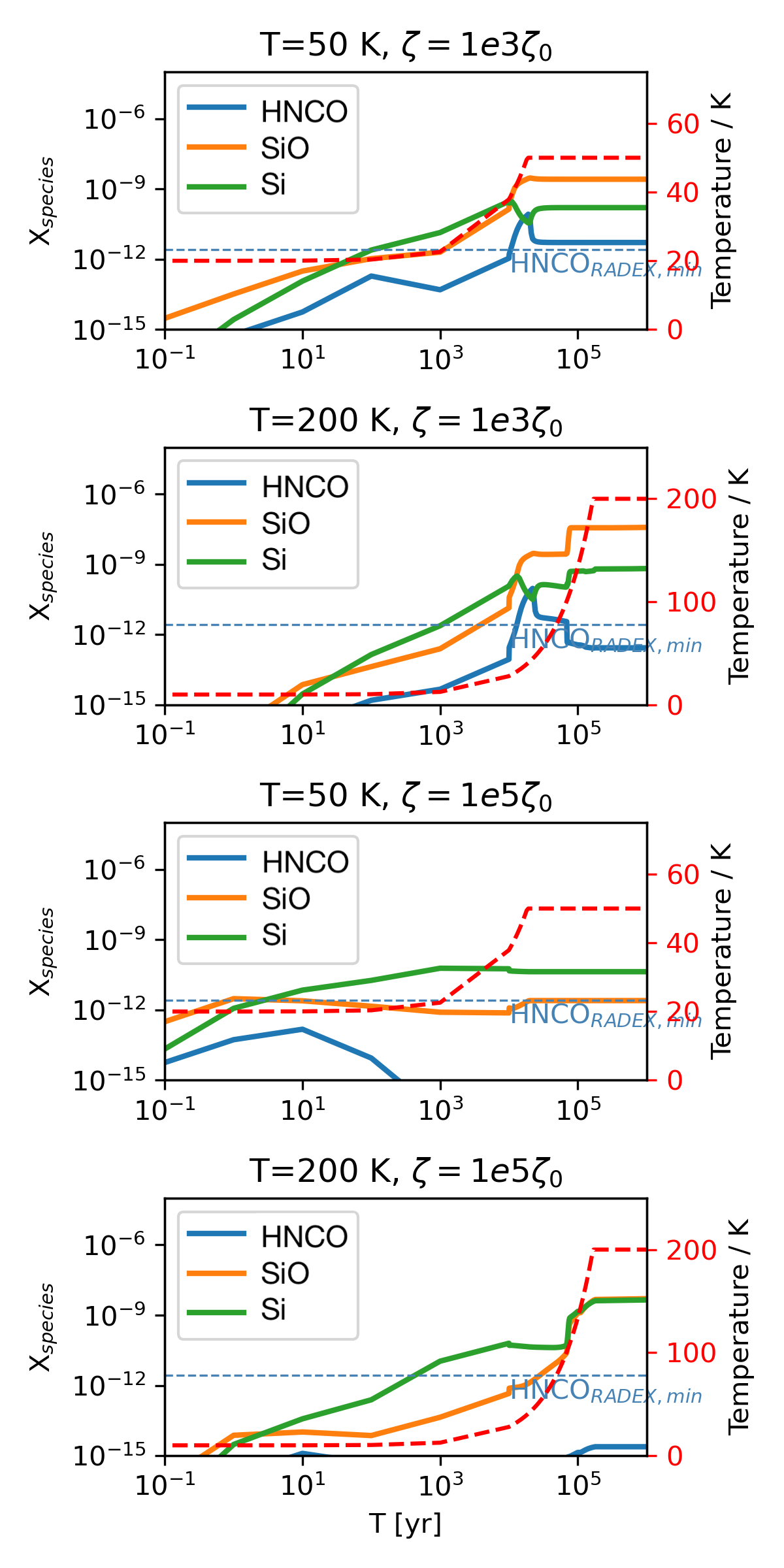} \\
    \centering\small (b)
  \end{tabular}
  \caption{As in Figure~\ref{fig:UCLCHEM_Cshock_den03} but for a  pre-shock gas density in (a) and a gas density in (b) of  $10^{6}$ cm\textsuperscript{-3}. The auxiliary line that indicates the lower limit of SiO (orange) fractional abundances "measured" from our RADEX-Bayesian inference is missing compared to Fig. \ref{fig:UCLCHEM_Cshock_den03}-\ref{fig:UCLCHEM_Cshock_den04} in the current case because observationally we do not find any GMC that is associated with such a high gas density. }
  \label{fig:UCLCHEM_Cshock_den06}

\end{figure*}

Figures \ref{fig:UCLCHEM_Cshock_den03}-\ref{fig:UCLCHEM_Cshock_den06} - left panels - show examples of our C-shock modellings. 
From our grid of models, we selected $v_{s}=10$ km s\textsuperscript{-1} and $v_{s}=50$ km s\textsuperscript{-1} as the most representative cases from our velocity grid for the slow and fast shock scenarios, respectively. 
In each figure we present cases associated with a specified pre-shock density, e.g. $n_{\rm H2}=10^3$ cm\textsuperscript{-3}, and each plot within the figure shows a case study with a specific combination of [shock velocity ($v_{s}$), CRIR ($\zeta$)]. 

In each case we plot the time evolution of the gas-phase molecular abundances of: HNCO (blue), SiO (orange), and Si (green). 
We also plot the gas temperature over time (red dashed) following the heating due to the shock as well as the minimum abundance imposed by the best fit from RADEX analysis (dashed horizontal lines) which will be discussed in Sec. \ref{sec:RADEX-CHEM_relation}.

Overall we see an enhancement of the SiO abundance during both slow ($v_{s}=10$ km/s) and fast ($v_{s}=50$ km/s) shocks, with the fast shocks leading to a much higher SiO abundance than achieved during the slow shocks. 
{In the fast-shock condition, our main formation route of SiO  depends on the gas density. In the lowest gas density case ($n_{\rm H2}= 10^{3}$ cm\textsuperscript{-3}) it is through the gas-phase reaction:
\begin{equation}
\label{eq:F_sio_den03}
    \text{SiOH}^{+} + \text{e}^{-} \longrightarrow \text{SiO} + \text{H},
\end{equation} while for higher density gas  ($n_{\rm H2}\geq 10^{4}$ cm\textsuperscript{-3}) the formation of SiO is mainly through the following gas-phase reaction:
\begin{equation}
\label{eq:F_sio_den04}
    \text{Si} + \text{OH} \longrightarrow \text{SiO} + \text{H}.
\end{equation} The main destruction route is via cosmic-ray induced photoreactions (expressed as "CRPHOT" in Eq. \ref{eq:D_sio_den03}) and/or other ionic particles (e.g. \ce{H3O+}, \ce{H3+}, \ce{H+}) such as:
\begin{equation}
\label{eq:D_sio_den03}
    \text{SiO} + \text{CRPHOT} \longrightarrow \text{Si} + \text{O}, 
\end{equation}
\begin{equation}
\label{eq:D_sio_den04}
    \text{SiO} + \text{\ce{H3O+}} \longrightarrow \text{Si} + \text{\ce{H2O}}, 
\end{equation}
}

{We also see that a high CRIR ($\zeta=10^{5}\zeta_{0}$) appears to further suppress the chemical abundances of SiO. 
In the higher density cases ($n_{\rm H2}\geq 10^{4}$ cm\textsuperscript{-3}) this is due to the fact that the most efficient formation route for SiO is via neutral-neutral reactions of Si with OH (Eq. \ref{eq:F_sio_den04}).  In a high CRIR environment both OH and Si are dissociated and ionized, respectively, more efficiently. 
In the lowest density case ($n_{\rm H2}= 10^{3}$ cm\textsuperscript{-3}) the formation route described in Eq. \ref{eq:F_sio_den03} is also much less efficient at higher CRIR. }

It is clear that SiO is heavily enhanced by fast-shock sputtering across the densities studied. 
The inferred gas densities traced by SiO in Sect. \ref{sec:radex_tech} are $n_{\rm H2}\leq10^4$ cm\textsuperscript{-3}, and in these regimes the SiO enhancements are all dominated by fast-shock chemistry. 


In contrast, in most cases, HNCO is  enhanced in slow ($v_{s}=10$ km s\textsuperscript{-1}) shocks rather than in fast shocks. 
{Our main formation route of HNCO  is on the dust grains followed by desorption:
\begin{equation}
\label{eq:F_hnco}
    \text{NH} + \text{CO} \longrightarrow \text{HNCO},
\end{equation} which is not viable when the ices are fully sputtered as in the fast-shock scenario, causing the associated low abundance of HNCO. Meanwhile the destruction of HNCO is mainly through two gas-phase routes: 
\begin{equation}
    \text{HNCO} + \text{CRPHOT} \longrightarrow \text{NH} + \text{CO},
\end{equation}
\begin{equation}
    \text{HNCO} + \text{H}^{+} \longrightarrow \text{NH}_{2}^{+} + \text{CO},
\end{equation}}
For cases with pre-shock gas density $n_{\rm H2}=10^3$ cm\textsuperscript{-3}, however, none of the shock scenarios leads to enough HNCO to be detectable. {The formation route described in Eq. \ref{eq:F_hnco} is never efficient enough at such low gas density. }
Yet, our earlier analysis points to some GMCs having a gas density of $\sim$ $n_{\rm H2}=10^3$ cm\textsuperscript{-3} e.g., GMC1a. Hence the HNCO in this cloud can not be matched by any shock model. We speculate that this failure to reproduce a high enough abundance of HNCO may be due to one of the following reasons:  
1) an incomplete gas or surface HNCO network in our chemical modelling; 
2) a best likelihood case in the RADEX-inferred gas properties not necessarily being chemically feasible e.g. a fit in the [$n_{\rm H2}$,$T_{\rm kin}$,column density] space can be chemically unfeasible if the best fit n and T can not lead to abundances predicted by chemical modelling that leads back to the best fit of column density; we investigate this option in Section 5.1.  
3) observationally we may be also averaging over a multi-component gas, which are denser and more compact within our beam; 
4) HNCO may not be tracing the shocked gas in these low-density GMCs but may be enhanced by a different physical/chemical process. 
We explore this latter point in Section \ref{sec:Hotcore_chem} with non-shock chemistry. 

Finally we note that \citet{EB+2022} found that the inferred CRIR ($\zeta$) is  bi-modal for GMC 1, with a main peak at $10^{3}\zeta_{0}$ and a second peak at $10\zeta_{0}$, although the model with low CRIR was considered likely to be unphysical \citep{EB+2022}. 
Hence in Fig. \ref{fig:append_den03Cshock_varZeta} we also show the C-shock chemical modelling with a CRIR of $\zeta=\zeta_{0}$ and $\zeta=10\zeta_{0}$. 
We indeed also find that with a lower CRIR, the HNCO abundance increases. However it is still about ten times lower than the "observed" value, as further discussed in Sect. \ref{sec:RADEX-CHEM_relation}. 
The chemical model with low CRIR is still not able to explain our measurements toward GMC 1a.  

\subsubsection{Chemical evolution of non-shocked gas}
\label{sec:Hotcore_chem}
As mentioned earlier, HNCO may be the product of thermal sublimation, rather than shock sputtering, as long as the gas and dust are  well coupled. 
In order to test whether such mechanism can efficiently  boost the HNCO abundance in the gas where densities are as low as 10$^3$ cm$^{-3}$ we use  \texttt{UCLCHEM} to model the chemical evolution of a gas that is warmed up  without the presence of shocks. This scenario could be representative of gas being warmed by the presence of star forming processes (including outflows), and/or X-ray or cosmic rays. The physical module within \texttt{UCLCHEM} allows for the temperature to increase over time.

We adopt two typical maximum temperatures inferred from HNCO observations: 50 K and 200 K. These two selected temperatures capture most of the temperature ranges traced by HNCO as presented in Sect. \ref{sec:radex_tech}, and the low temperatures measured in the literature as discussed earlier \citep{Leroy+2015,Mangum+2019}. 

We show the results of this modelling in the panels on the right  in Figures \ref{fig:UCLCHEM_Cshock_den03}-\ref{fig:UCLCHEM_Cshock_den06}. 
From these figures, it is clear that neither HNCO nor SiO are enhanced enough to match any of the observations. 
We test this scenario also with a lower CRIR  ($\zeta=1\zeta_{0}$, Fig. \ref{fig:append_hotcore_Zeta1_50K200K}) and find that only cases with high temperatures (200 K) and dense gas ($n_{\rm H2}\geq10^4$ cm\textsuperscript{-3}) show a noticeable enhancement of HNCO abundance. However, none of the best-fit physical properties of any of our GMCs are consistent with such combinations of density and temperature. 
We also note that for GMCs (GMC 2b, 7) that probed gas temperature traced by HNCO even lower than the lowest-T case tested here, 50 K, given none of the 50 K cases could reproduce sufficient HNCO, we expect no thermal sublimation being feasible at these GMCs. 

In summary, we do not find a reasonable non-shock model that can produce the high HNCO abundance we observe from the CMZ of NGC 253, particularly for gas densities as low as 10$^3$ cm$^{-3}$.
\section{Discussion}

\subsection{Are the RADEX-inferred gas properties chemically feasible?}
\label{sec:RADEX-CHEM_relation}

The RADEX-inferred gas properties are the gas density ($n_{\rm H2}$), gas temperature ($T_{\rm kin}$), and the column density of the species ($N_{\rm HNCO}$ and $N_{\rm SiO}$). 
Chemical models, on the other hand, are provided with initial densities and temperatures and compute the chemical abundances of HNCO and SiO as a function of time. In this section, we attempt to determine whether the physical conditions inferred by our radiative transfer inference analysis are compatible with the chemistry. The column density of a species can also be related to its chemical abundances predicted from chemical modelling, via the following "on the spot" approximation \citep{Dyson1997}:

\begin{equation} \label{eq1}
\begin{split}
N_{\rm species} & = \eta_{ff} \times N_{\rm H2} \times X_{\rm species} \\
 & \sim \eta_{ff}\times n_{\rm H2} \times 2R_{\rm GMC,i} \times X_{\rm species} (n_{\rm H2},v_{s},...) 
 \end{split}
\end{equation}
where $\eta_{ff}$ is the beam filling factor, $n_{\rm H2}$ is the hydrogen column density, $X_{\rm species}$ is the fractional (with respect to the total number of atomic hydrogen nuclei) abundance of our species and $R_{\rm GMC,i}$ is the radius of the GMC in consideration. 
We assume spherical clouds and approximate the line-of-sight depth with the plane-of-sky "diameter" of a GMC, $2\times R_{\rm GMC,i}$, to be multiplied with gas volume density for the estimate of column density. 
Since the chemically predicted abundance is dependent on various factors, we express it as $X_{\rm species} (n_{\rm H2},v_{s},...)$ in the last equality. 
$N_{\rm species}$ as defined by the above expression \it{must} \rm then be less or equal to:
\begin{equation} \label{eq2}
N_{\rm species}\leq 1.0\times n_{H2,Bayesian} \times 28 \rm{ pc} \times X_{\rm species}(n_{H2,Bayesian},v_{s},...)_{max}
\end{equation}
where the beam filling factor, which may range between 0.0 and 1.0, was here assumed to be equal to 1 in order to obtain an  upper limit in this estimate.  The cloud diameter was taken to be the size of the beam, and we consider only the maximum abundance derived from the chemical models. 

In other words, the above relationship determines whether the column density (estimated by the radiative transfer analysis) can be reproduced by the chemical models computed with the parameters derived from the radiative transfer modelling. Such verification is reported in 
Tables \ref{tab:table_ineqH}-\ref{tab:table_ineqS} for all our GMCs. 

Indeed, this verification confirms what we qualitatively found in Sect. \ref{sec:Cshock_chem}, namely that in cases where the best fit for the pre-shock gas density is $n_{\rm H2}=10^{3}$ cm\textsuperscript{-3} (case solely for GMC 1a) we cannot reconcile the HNCO column density inferred by the radiative transfer analysis with what is chemically feasible. 
For all the other cases the relationship in Eq. 2 is satisfied. 

In addition, we mentioned in Sect. \ref{sec:Cshock_chem} that we saw an enhancement of SiO abundance during both slow ($v_{s}=10$ km s\textsuperscript{-1}) and fast ($v_{s}=50$ km/s) shocks, but the fast shocks lead to a much higher SiO abundance than the slow shocks.
In fact, we find that the "enhanced" SiO abundance via slow shocks is insufficient compared to the RADEX-inferred results for the low density case, $n_{\rm H2}=10^{3}$ cm\textsuperscript{-3} (top figure in the left panel of Figure \ref{fig:UCLCHEM_Cshock_den03}) using the same relation in Eq. \ref{eq2}. 
For $n_{\rm H2}=10^{4}$ cm\textsuperscript{-3} case, although the predicted SiO yielded from slow shocks is sufficient, the high temperature measured from RADEX in these cases (GMC 4 and 6), $T>500$ K, are not expected in the slow-shock models. 
This again reinforces the conclusion we drew in Sect. \ref{sec:Cshock_chem}, that the abundance enhancement of SiO is dominated by fast-shock sputtering across the densities studied. 

\begin{table}[ht!]
  \centering
  \caption{Comparison between the observations-inferred species column density and the chemical modellings for HNCO. LHS and RHS refers to the left-hand-side and right-hand-side of the inequality described in Eq. \ref{eq2}. }
  \label{tab:table_ineqH}
  \begin{tabular}{c|cc|c}
  \hline
    GMC  & LHS & RHS & LHS $\leq$ RHS? \\
    \hline
    1a & $10^{15.68}$ & $10^{12.64}$ & NO\\
    2b & $10^{15.02}$ & $10^{19.04}$ & YES\\
    3 & $10^{15.44}$ & $10^{19.82}$ & YES\\
    4 & $10^{15.09}$ & $10^{21.38}$ & YES\\
    6 & $10^{15.36}$ & $10^{20.66}$ & YES\\
    7 & $10^{15.32}$ & $10^{20.55}$ & YES\\
    8a & $10^{14.95}$ & $10^{18.84}$ & YES\\
    9a & $10^{15.17}$ & $10^{18.91}$ & YES\\
    \hline
  \end{tabular}
\end{table}
\begin{table}[ht!]
  \centering
  \caption{Comparison between the observations-inferred species column density and the chemical modellings for SiO. LHS and RHS refers to the left-hand-side and right-hand-side of the inequality described in Eq. \ref{eq2}. }
  \label{tab:table_ineqS}
  \begin{tabular}{c|cc|c}
  \hline
    GMC  & LHS & RHS & LHS $\leq$ RHS? \\
    \hline
    1a & $10^{16.33}$ & $10^{17.41}$ & YES\\
    2b & $10^{16.58}$ & $10^{17.25}$ & YES\\
    3 & $10^{15.73}$ & $10^{18.97}$ & YES\\
    4 & $10^{15.93}$ & $10^{19.60}$ & YES\\
    6 & $10^{16.53}$ & $10^{19.73}$ & YES\\
    7 & $10^{16.02}$ & $10^{17.89}$ & YES\\
    8a & $10^{16.58}$ & $10^{17.20}$ & YES\\
    9a & $10^{16.21}$ & $10^{17.55}$ & YES\\
    \hline
  \end{tabular}
\end{table}

\subsection{Physical interpretation of the gas properties}
\label{sec:timelines_shocks}
The current understanding of the shock chemistry as traced by HNCO and SiO is built upon many previous studies. 
The good correlation between HNCO and SiO revealed in Galactic dense molecular cores by \citet{Zinchenko+2000} hinted that both species trace shocks, although  the absence of HNCO in the higher velocity wings observed in the SiO spectral profile also hinted to the fact that high-velocity shock conditions may suppress the HNCO abundance. 
A follow-up survey over sources towards the Galactic Center performed by \citet{Martin+2008_HNCO_galactic} reveals that  HNCO can  however also be heavily destroyed by UV radiation in PDR regions (later also found in NGC 253 in \citet{Martin+2009}). 
On the other hand, chemical modelling of HNCO and SiO in NGC 1068 performed by \citet{Kelly+2017} confirmed that the HNCO abundance can be suppressed in high-velocity shocks due to the destruction of its precursor, the molecule NO. 
The analysis of HNCO and SiO in this work seem to lead to the conclusion that in at least most of the GMCs HNCO and SiO emission can only be explained by the presence of shocks. 

With this clear association between shock chemistry and our observed HNCO and SiO emission, the highly varying inferred gas properties across the GMCs may be the result of independent shock episodes.  
In this section, we use the physical properties inferred by the radiative transfer analysis coupled with the assumption that each GMC is subjected to the passage of a shock to estimate a rough timescale of the history of the shocks. Specifically, the "dissipation time" ($\tau_{\rm dissipation}$) is defined as the timescale when the velocity of ions and neutrals is equal, and can be viewed as the time scale over which the shock dissipates, or alternatively the shock-influence timescale. 
$\tau_{\rm dissipation}$ is estimated by dividing the shock dissipation distance described by \citet{Holdship+2017} by the shock velocity, and depends solely on the pre-shock gas density. 
The larger the gas density is, the shorter the timescale. 
From the post-shock temperature of the gas (derived from the \texttt{UCLCHEM} modelling) we can also roughly estimate the post-shock cooling timescale ($\tau_{shock}$) relative to $\tau_{\rm dissipation}$. The two timescales can be qualitatively related in the following way:

\begin{equation}
    \label{eq:3}
    \begin{cases}
    \text{Hot (T>400 K): }\tau_{shock}\sim\tau_{\rm dissipation}\quad; \\
    \text{Half-way cooling (T}\sim200\text{ K): }\tau_{shock}\geq\tau_{\rm dissipation}\quad; \\
    \text{Cold (T<100 K): }\tau_{shock}\gg\tau_{\rm dissipation}
    \end{cases}
\end{equation}

If the gas component remains hot, with the inferred gas temperature ($T_{\rm kin}$) from the RADEX-Bayesian inference described in Sect. \ref{sec:radex_tech} being higher than 400 K, the region is possibly still under the influence of a shock episode, therefore the age of the shock ($\tau_{\rm shock}$) is comparable to the dissipation timescale ($\tau_{\rm dissipation}$). 
The same logic is applied to the remaining two cases - the half-way cooling and cold-gas conditions. 
Using Eq. \ref{eq:3} the inferred age of the shock ($\tau_{\rm shock}$) from each species for each GMC is listed in column 7 of Table\ref{tab:table_inferred_property}. 

Making one further assumption that the HNCO and SiO observed arise from the same shock episode, we can also take the intersection of both shock timescales and derive the joint shock timescale, $\tau_{\rm shock,joint}$. 
If there is no intersection between the SiO-shock timescale and HNCO-shock timescale, it may indeed be that the two molecules arise from different shock episodes, noting however that an "intersection" between the two timescales, does not necessarily prove the opposite. 
The resulting "joint" shock timescale ($\tau_{\rm shock,joint}$) from this qualitative comparison are listed in the final column of Table \ref{tab:table_inferred_property} and displayed qualitatively in the bottom panel of Fig. \ref{fig:shematic_3shock_scenarios}. 
The $\tau_{\rm shock,joint}$ of GMCs (GMC 4, 6, 7) from the inner CMZ tends to be smaller than the outer GMCs. 
In other words, this suggests the shocks in the inner GMCs tend to be younger ($\tau_{\rm shock,joint}\sim 10^{3}$ yr) than in the outer GMCs ($\tau_{\rm shock,joint}\geq 10^{5}$ yr). 

\subsection{Origin(s) of the shocks}
\label{sec:shock_origin}
\begin{figure*}
        \centering
    \includegraphics[scale=0.47]{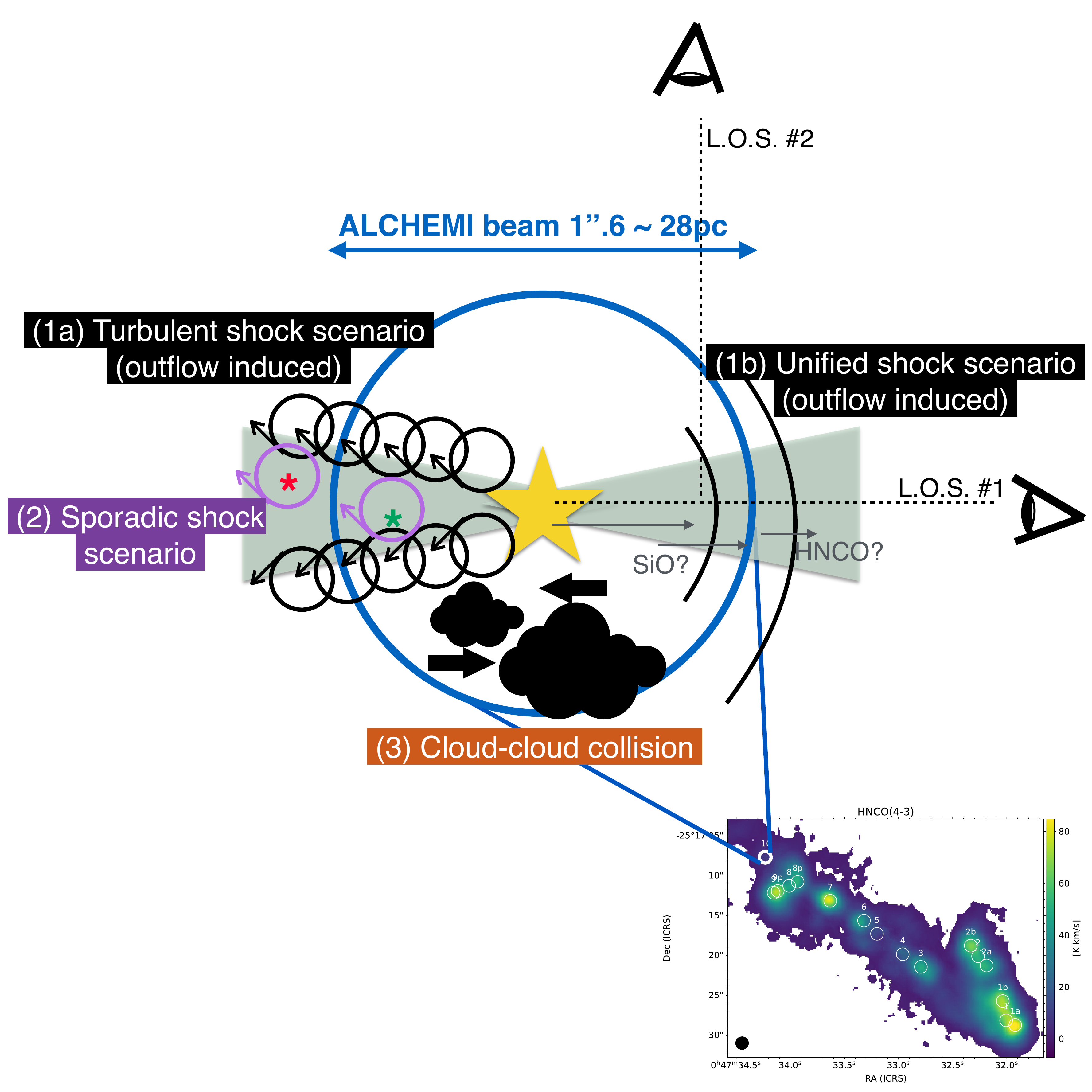}\\
    \includegraphics[scale=0.8]{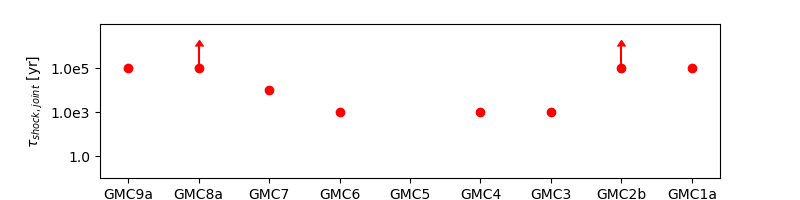}
    \caption{A sketch (not to scale) of the three possible sources for star-formation induced shocks in the GMCs within the NGC\,253 CMZ. We illustrate these three sources of shocks described in Sect. \ref{sec:shock_origin}, and the relative layout of two possible line-of-sight (L.O.S.) orientations relative to the layout of shocked gas in each GMC. We assume the size of individual GMCs is comparable or larger than the ALCHEMI beam $1''.6\sim28$ pc. The shocks probed by HNCO and SiO 
    can be either approximated as (a) centralized mini-starburst episode(s) (labels in black, with 
    the yellow star at center and 
    the associated outflow in pale green), (b) sporadic/fragmented turbulent shock events that are scattered both temporally and spatially throughout the GMC (label in purple, with green and red asterisks), and (c) the cloud-cloud collision leading to shock episodes (label in orange). }
    \label{fig:shematic_3shock_scenarios}
\end{figure*}

In this section, we qualitatively explore the possible origin(s) of the shocked environments probed via our HNCO and SiO observations.
We speculate three main possible scenarios that can lead to shocks within individual GMCs as illustrated in Fig. \ref{fig:shematic_3shock_scenarios}: 1) shocks induced by outflowing material from "burst(s)" of star formation in the central region of each GMC (marked in black labels); 2) turbulent shocks induced by star-formation episodes in scattered locations within each GMC (marked with purple label); and 3) shocks induced by cloud-cloud collisions (marked with orange label). 
In Fig. \ref{fig:shematic_3shock_scenarios} we also place the relative layout of two possible line-of-sight (L.O.S.) orientations relative to the physical setting in individual GMCs, which is not necessarily aligned with the galactic plane of NGC 253. 
We use the ALCHEMI beam ($1''.6 \sim 28$ pc) as the GMC size - plotted in solid blue circle - noting that of course GMCs may be differ from that. 

As the first potential source of shocks, the outflow can either induce: (1a) turbulent shocks on the working surface between the outflow and the ambient material (setup 1a in Fig. \ref{fig:shematic_3shock_scenarios}), or (1b) could also directly push the ambient material along the normal direction, and create "bow" shocks along the outflow (setup 1b in Fig. \ref{fig:shematic_3shock_scenarios}).  
Both setups are marked in black labels in Fig. \ref{fig:shematic_3shock_scenarios}. 

For the second potential shock source, the HNCO and SiO emission could also be an ensemble from random locations and determining a timeline such as $\tau_{shock, joint}$ in Sec. \ref{sec:timelines_shocks} for the shock episodes would not be possible. This scenario is labelled in purple circles and text in Fig. \ref{fig:shematic_3shock_scenarios}. 

Finally, we show the scenario where the shocks are caused by cloud-cloud collisions, which may occur especially near the outer GMCs (GMC 1a, 2b, 8a, 9a) because these regions are believed to be located at the intersections of a few dynamical orbits of the galaxy, including bar orbits and nuclear ring \citep[see][and reference therein]{Harada+2022,Humire+2022}. 
We note that cloud-cloud collisions in our own Galaxy and Large Magellanic Cloud (LMC) seem to occur at a moderate velocity $\sim 10-20$ km s\textsuperscript{-1} \citep{Li+2018,Fukui+2015} hinting only at episodes of weak shocks. 
Cloud-cloud collisions at higher velocities, however, can still occur at the intersections of dynamical orbits as proposed by \citet{Harada+2019} for M83, and we expect similar case applies to NGC 253. 

We briefly discuss these scenarios below.
\citet{GB+2017} speculate on the presence of non-dissociative shocks (traced by C\textsubscript{2}H) generated by the highly turbulent interfaces between the outflow and neighbouring molecular gas in the circumnuclear disk (at few 100-pc scale) as well as in the starburst regions of the AGN-host galaxy NGC 1068.
In our individual GMCs, although at the much smaller physical scale of few pc, we may be witnessing a similar scenario, but on smaller scales.  
From our intensity maps, however, we do not see the extended morphology that suggests such structure (e.g. see Figure 2 by \citet{GB+2017}); there is also the possibility that our spatial resolution with HNCO and SiO observations is just not sufficient to resolve such morphology within the beam-sized clump. 
Also, \citet{Holdship+2021_SpectralRadex} showed that the enhanced C\textsubscript{2}H abundance could arise from either a high CRIR or shocks that occur within a timescale of $10^{5}$ years, with the latter being less likely due to the timescale being very short. 
From our inferred shock timescales the shock scenario does not appear to be entirely impossible. 
The other possible setup is that the star-formation induced outflow can directly push against the ambient material along the normal/perpendicular direction, and create shocks along the outflow propagation. 
This creates a "single" shock episode that sweeps across the gas traced by both HNCO and SiO, possibly in different layers since they seem to trace quite different gas densities in most of our GMCs. 
In this case, if the shocked gas components probed by HNCO and SiO can trace back to the same shock episode, we can further pin down the "age" of such shocks with $\tau_{\rm shock, joint}$ as discussed in Sect. \ref{sec:timelines_shocks}. 
Such inferred age of this hypothesized "single" shock in each GMC is listed in Tab. \ref{tab:table_inferred_property}, also shown qualitatively in the bottom panel of Fig. \ref{fig:shematic_3shock_scenarios}. 
Of course episodic shocks may be happening in random locations within the GMCs if star formation is ongoing.
To determine the location of such sporadic shock episodes, we would need higher spatial resolution observations.  

As a final note, globally the shock episodes throughout the CMZ may link to large-scale dynamical structures, such as an interface with the large-scale outflow. 
However the spatial extent of our SiO and HNCO observations do not seem to be strongly tied with this possibility.

\label{sec:indep_shock}
\section{Conclusions}
We analyzed six SiO transitions and eleven HNCO transitions imaged at GMC-scales in the CMZ of NGC 253 with ALMA, as part of the ALMA Large program ALCHEMI. 
We briefly summarize below our main conclusions: 
  \begin{enumerate}
      \item {Unlike the SiO SLEDs, the HNCO SLEDs differ in shape across the GMCs, hinting at substantial variations in at least the excitation conditions in the gas traced by HNCO across the GMCs. }
   \item {Through radiative transfer modelling using RADEX coupled with a Bayesian inference process, we have successfully characterized the gas properties traced by these two molecular species and found them to be distinctively different. }
      \item {Through radiative transfer and chemical modelling, we find that the most likely physical scenario has the SiO emission arising from fast-shocks while the HNCO emission arises from slow-shocks. }
     \item {We are able to infer the physical characteristics of the shocks traced by SiO and HNCO for each GMC, in particular the age of shocks traced by each species in each GMC. }
     \item {We propose three possible shock scenarios that could explain the observed SiO and HNCO emission (see Figure \ref{fig:shematic_3shock_scenarios}). Higher spatial resolution observations are needed in order to discern among these shock scenarios. }
\end{enumerate}

\begin{acknowledgements}
      KYH, SV, JH, and MB are funded by the European Research Council (ERC) Advanced Grant MOPPEX 833460.vii. 
      SGB acknowledges support from the research project PID2019-106027GA-C44 of the Spanish Ministerio de Ciencia e Innovaci{\'o}n. 
      L.C. acknowledges financial support through the Spanish grant PID2019-105552RB-C41 funded by MCIN/AEI/10.13039/501100011033. 
      KYH and SV acknowledge the help from Marcus Keil and Ross O'Donoghue in working with UCLCHEM. 
      KYH acknowledges assistance from Allegro, the European ALMA Regional Center node in the Netherlands. 
      This paper makes use of the following ALMA data: ADS/JAO.ALMA\#2017.1.00161.L and ADS/JAO.ALMA\#2018.1.00162.S. ALMA is a partnership of ESO (representing its member states), NSF (USA) and NINS (Japan), together with NRC (Canada), MOST and ASIAA (Taiwan), and KASI (Republic of Korea), in cooperation with the Republic of Chile. The Joint ALMA Observatory is operated by ESO, AUI/NRAO and NAOJ.  
\end{acknowledgements}
\bibliographystyle{aa}
\bibliography{ngc1068,fundamentals,ngc253}

\newcommand{\noop}[1]{}
\begin{thebibliography}{88}
\expandafter\ifx\csname natexlab\endcsname\relax\def\natexlab#1{#1}\fi

\bibitem[{{Aladro} {et~al.}(2011){Aladro}, {Mart{\'\i}n-Pintado},
  {Mart{\'\i}n}, {Mauersberger}, \& {Bayet}}]{Aladro+2011_CS}
{Aladro}, R., {Mart{\'\i}n-Pintado}, J., {Mart{\'\i}n}, S., {Mauersberger}, R.,
  \& {Bayet}, E. 2011, \aap, 525, A89

\bibitem[{{Balan{\c{c}}a} {et~al.}(2018){Balan{\c{c}}a}, {Dayou}, {Faure},
  {Wiesenfeld}, \& {Feautrier}}]{sio_moldata_B+2018}
{Balan{\c{c}}a}, C., {Dayou}, F., {Faure}, A., {Wiesenfeld}, L., \&
  {Feautrier}, N. 2018, \mnras, 479, 2692

\bibitem[{{Bayet} {et~al.}(2008){Bayet}, {Lintott}, {Viti},
  {Mart{\'\i}n-Pintado}, {Mart{\'\i}n}, {Williams}, \& {Rawlings}}]{Bayet+2008}
{Bayet}, E., {Lintott}, C., {Viti}, S., {et~al.} 2008, \apjl, 685, L35

\bibitem[{{Behrens} {et~al.}(2022){Behrens}, {Mangum}, {Holdship}, {Viti},
  {Harada}, {Martin}, {Sakamoto}, {Muller}, {Tanaka}, {Nakanishi},
  {Herrero-Illana}, {Yoshimura}, {Aladro}, {Colzi}, {Emig}, {Henkel}, {Huang},
  {Humire}, {Meier}, \& {Rivilla}}]{EB+2022}
{Behrens}, E., {Mangum}, J.~G., {Holdship}, J., {et~al.} 2022, arXiv e-prints,
  arXiv:2209.06244

\bibitem[{{Bendo} {et~al.}(2015){Bendo}, {Beswick}, {D'Cruze}, {Dickinson},
  {Fuller}, \& {Muxlow}}]{Bendo+2015}
{Bendo}, G.~J., {Beswick}, R.~J., {D'Cruze}, M.~J., {et~al.} 2015, \mnras, 450,
  L80

\bibitem[{{Bolatto} {et~al.}(2013){Bolatto}, {Warren}, {Leroy}, {Walter},
  {Veilleux}, {Ostriker}, {Ott}, {Zwaan}, {Fisher}, {Weiss}, {Rosolowsky}, \&
  {Hodge}}]{Bolatto+2013}
{Bolatto}, A.~D., {Warren}, S.~R., {Leroy}, A.~K., {et~al.} 2013, \nat, 499,
  450

\bibitem[{{Buchner}(2016)}]{ultranest16}
{Buchner}, J. 2016, Statistics and Computing, 26, 383

\bibitem[{{Buchner}(2019)}]{ultranest19}
{Buchner}, J. 2019, \pasp, 131, 108005

\bibitem[{{Buchner}(2021)}]{ultranest21}
{Buchner}, J. 2021, The Journal of Open Source Software, 6, 3001

\bibitem[{{Canelo} {et~al.}(2021){Canelo}, {Bronfman}, {Mendoza}, {Duronea},
  {Merello}, {Carvajal}, {Fria{\c{c}}a}, \& {Lepine}}]{Canelo+2021}
{Canelo}, C.~M., {Bronfman}, L., {Mendoza}, E., {et~al.} 2021, \mnras, 504,
  4428

\bibitem[{{Carvajal} {et~al.}(2019){Carvajal}, {Favre}, {Kleiner},
  {Ceccarelli}, {Bergin}, \& {Fedele}}]{PartitionFn_HNCO_2019}
{Carvajal}, M., {Favre}, C., {Kleiner}, I., {et~al.} 2019, \aap, 627, A65

\bibitem[{{Colzi} {et~al.}(2021){Colzi}, {Rivilla}, {Beltr{\'a}n},
  {Jim{\'e}nez-Serra}, {Mininni}, {Melosso}, {Cesaroni}, {Fontani},
  {Lorenzani}, {S{\'a}nchez-Monge}, {Viti}, {Schilke}, {Testi}, {Alonso}, \&
  {Kolesnikov{\'a}}}]{Colzi+2021}
{Colzi}, L., {Rivilla}, V.~M., {Beltr{\'a}n}, M.~T., {et~al.} 2021, \aap, 653,
  A129

\bibitem[{{Dyson} \& {Williams}(1997)}]{Dyson1997}
{Dyson}, J.~E. \& {Williams}, D.~A. 1997, {The physics of the interstellar
  medium}

\bibitem[{{Endres} {et~al.}(2016){Endres}, {Schlemmer}, {Schilke}, {Stutzki},
  \& {M{\"u}ller}}]{CDMS_2016}
{Endres}, C.~P., {Schlemmer}, S., {Schilke}, P., {Stutzki}, J., \&
  {M{\"u}ller}, H. S.~P. 2016, Journal of Molecular Spectroscopy, 327, 95

\bibitem[{{Fedoseev} {et~al.}(2015){Fedoseev}, {Ioppolo}, {Zhao}, {Lamberts},
  \& {Linnartz}}]{Fedoseev+2015}
{Fedoseev}, G., {Ioppolo}, S., {Zhao}, D., {Lamberts}, T., \& {Linnartz}, H.
  2015, \mnras, 446, 439

\bibitem[{{Fukui} {et~al.}(2015){Fukui}, {Harada}, {Tokuda}, {Morioka},
  {Onishi}, {Torii}, {Ohama}, {Hattori}, {Nayak}, {Meixner}, {Sewi{\l}o},
  {Indebetouw}, {Kawamura}, {Saigo}, {Yamamoto}, {Tachihara}, {Minamidani},
  {Inoue}, {Madden}, {Galametz}, {Lebouteiller}, {Mizuno}, \&
  {Chen}}]{Fukui+2015}
{Fukui}, Y., {Harada}, R., {Tokuda}, K., {et~al.} 2015, \apjl, 807, L4

\bibitem[{{Gao} \& {Solomon}(2004)}]{Gao_Solomon_2004}
{Gao}, Y. \& {Solomon}, P.~M. 2004, \apj, 606, 271

\bibitem[{{Garc{\'\i}a-Burillo} {et~al.}(2000){Garc{\'\i}a-Burillo},
  {Mart{\'\i}n-Pintado}, {Fuente}, \& {Neri}}]{GB+2000_sio_253}
{Garc{\'\i}a-Burillo}, S., {Mart{\'\i}n-Pintado}, J., {Fuente}, A., \& {Neri},
  R. 2000, \aap, 355, 499

\bibitem[{{Garc{\'\i}a-Burillo} {et~al.}(2002){Garc{\'\i}a-Burillo},
  {Mart{\'\i}n-Pintado}, {Fuente}, {Usero}, \& {Neri}}]{GB+2002}
{Garc{\'\i}a-Burillo}, S., {Mart{\'\i}n-Pintado}, J., {Fuente}, A., {Usero},
  A., \& {Neri}, R. 2002, \apjl, 575, L55

\bibitem[{{Garc{\'\i}a-Burillo} {et~al.}(2010){Garc{\'\i}a-Burillo}, {Usero},
  {Fuente}, {Mart{\'\i}n-Pintado}, {Boone}, {Aalto}, {Krips}, {Neri},
  {Schinnerer}, \& {Tacconi}}]{GB+2010}
{Garc{\'\i}a-Burillo}, S., {Usero}, A., {Fuente}, A., {et~al.} 2010, \aap, 519,
  A2

\bibitem[{{Garc{\'\i}a-Burillo} {et~al.}(2017){Garc{\'\i}a-Burillo}, {Viti},
  {Combes}, {Fuente}, {Usero}, {Hunt}, {Mart{\'\i}n}, {Krips}, {Aalto},
  {Aladro}, {Ramos Almeida}, {Alonso-Herrero}, {Casasola}, {Henkel},
  {Querejeta}, {Neri}, {Costagliola}, {Tacconi}, \& {van der Werf}}]{GB+2017}
{Garc{\'\i}a-Burillo}, S., {Viti}, S., {Combes}, F., {et~al.} 2017, \aap, 608,
  A56

\bibitem[{{Gerin} {et~al.}(2009){Gerin}, {Goicoechea}, {Pety}, \&
  {Hily-Blant}}]{Gerin+2009_HCO_PDR}
{Gerin}, M., {Goicoechea}, J.~R., {Pety}, J., \& {Hily-Blant}, P. 2009, \aap,
  494, 977

\bibitem[{{Gorai} {et~al.}(2020){Gorai}, {Bhat}, {Sil}, {Mondal}, {Ghosh},
  {Chakrabarti}, \& {Das}}]{Gorai+2020}
{Gorai}, P., {Bhat}, B., {Sil}, M., {et~al.} 2020, \apj, 895, 86

\bibitem[{{Haasler} {et~al.}(2022){Haasler}, {Rivilla}, {Mart{\'\i}n},
  {Holdship}, {Viti}, {Harada}, {Mangum}, {Sakamoto}, {Muller}, {Tanaka},
  {Yoshimura}, {Nakanishi}, {Colzi}, {Hunt}, {Emig}, {Aladro}, {Humire},
  {Henkel}, \& {van der Werf}}]{Haasler+2022}
{Haasler}, D., {Rivilla}, V.~M., {Mart{\'\i}n}, S., {et~al.} 2022, \aap, 659,
  A158

\bibitem[{{Harada} {et~al.}(2022){Harada}, {Mart{\'\i}n}, {Mangum}, {Sakamoto},
  {Muller}, {Rivilla}, {Henkel}, {Meier}, {Colzi}, {Yamagishi}, {Tanaka},
  {Nakanishi}, {Herrero-Illana}, {Yoshimura}, {Humire}, {Aladro}, {van der
  Werf}, \& {Emig}}]{Harada+2022}
{Harada}, N., {Mart{\'\i}n}, S., {Mangum}, J.~G., {et~al.} 2022, \apj, 938, 80

\bibitem[{{Harada} {et~al.}(2021){Harada}, {Mart{\'\i}n}, {Mangum}, {Sakamoto},
  {Muller}, {Tanaka}, {Nakanishi}, {Herrero-Illana}, {Yoshimura}, {M{\"u}hle},
  {Aladro}, {Colzi}, {Rivilla}, {Aalto}, {Behrens}, {Henkel}, {Holdship},
  {Humire}, {Meier}, {Nishimura}, {van der Werf}, \& {Viti}}]{Harada+2021}
{Harada}, N., {Mart{\'\i}n}, S., {Mangum}, J.~G., {et~al.} 2021, \apj, 923, 24

\bibitem[{{Harada} {et~al.}(2019){Harada}, {Sakamoto}, {Mart{\'\i}n},
  {Watanabe}, {Aladro}, {Riquelme}, \& {Hirota}}]{Harada+2019}
{Harada}, N., {Sakamoto}, K., {Mart{\'\i}n}, S., {et~al.} 2019, \apj, 884, 100

\bibitem[{{Hern{\'a}ndez-G{\'o}mez} {et~al.}(2019){Hern{\'a}ndez-G{\'o}mez},
  {Sahnoun}, {Caux}, {Wiesenfeld}, {Loinard}, {Bottinelli}, {Hammami}, \&
  {Menten}}]{HG+2019}
{Hern{\'a}ndez-G{\'o}mez}, A., {Sahnoun}, E., {Caux}, E., {et~al.} 2019,
  \mnras, 483, 2014

\bibitem[{{Holdship} {et~al.}(2022){Holdship}, {Mangum}, {Viti}, {Behrens},
  {Harada}, {Mart{\'\i}n}, {Sakamoto}, {Muller}, {Tanaka}, {Nakanishi},
  {Herrero-Illana}, {Yoshimura}, {Aladro}, {Colzi}, {Emig}, {Henkel},
  {Nishimura}, {Rivilla}, \& {van der Werf}}]{Holdship+2022}
{Holdship}, J., {Mangum}, J.~G., {Viti}, S., {et~al.} 2022, arXiv e-prints,
  arXiv:2204.03668

\bibitem[{{Holdship} {et~al.}(2017){Holdship}, {Viti}, {Jim{\'e}nez-Serra},
  {Makrymallis}, \& {Priestley}}]{Holdship+2017}
{Holdship}, J., {Viti}, S., {Jim{\'e}nez-Serra}, I., {Makrymallis}, A., \&
  {Priestley}, F. 2017, \aj, 154, 38

\bibitem[{{Holdship} {et~al.}(2021){Holdship}, {Viti}, {Mart{\'\i}n}, {Harada},
  {Mangum}, {Sakamoto}, {Muller}, {Tanaka}, {Yoshimura}, {Nakanishi},
  {Herrero-Illana}, {M{\"u}hle}, {Aladro}, {Colzi}, {Emig},
  {Garc{\'\i}a-Burillo}, {Henkel}, {Humire}, {Meier}, {Rivilla}, \& {van der
  Werf}}]{Holdship+2021_SpectralRadex}
{Holdship}, J., {Viti}, S., {Mart{\'\i}n}, S., {et~al.} 2021, \aap, 654, A55

\bibitem[{{Huang} {et~al.}(2022){Huang}, {Viti}, {Holdship},
  {Garc{\'\i}a-Burillo}, {Kohno}, {Taniguchi}, {Mart{\'\i}n}, {Aladro},
  {Fuente}, \& {S{\'a}nchez-Garc{\'\i}a}}]{Huang+2022}
{Huang}, K.~Y., {Viti}, S., {Holdship}, J., {et~al.} 2022, arXiv e-prints,
  arXiv:2202.05005

\bibitem[{{Humire} {et~al.}(2022){Humire}, {Henkel}, {Hern{\'a}ndez-G{\'o}mez},
  {Mart{\'\i}n}, {Mangum}, {Harada}, {Muller}, {Sakamoto}, {Tanaka},
  {Yoshimura}, {Nakanishi}, {M{\"u}hle}, {Herrero-Illana}, {Meier}, {Caux},
  {Aladro}, {Mauersberger}, {Viti}, {Colzi}, {Rivilla}, {Gorski}, {Menten},
  {Huang}, {Aalto}, {van der Werf}, \& {Emig}}]{Humire+2022}
{Humire}, P.~K., {Henkel}, C., {Hern{\'a}ndez-G{\'o}mez}, A., {et~al.} 2022,
  \aap, 663, A33

\bibitem[{{H\"uttemeister} {et~al.}(1998){H\"uttemeister}, {Dahmen},
  {Mauersberger}, {Henkel}, {Wilson}, \& {Martin-Pintado}}]{Huttemeister+1998}
{H\"uttemeister}, S., {Dahmen}, G., {Mauersberger}, R., {et~al.} 1998, \aap,
  334, 646

\bibitem[{{Jim{\'e}nez-Serra} {et~al.}(2008){Jim{\'e}nez-Serra}, {Caselli},
  {Mart{\'\i}n-Pintado}, \& {Hartquist}}]{J-S+2008_shocktracers}
{Jim{\'e}nez-Serra}, I., {Caselli}, P., {Mart{\'\i}n-Pintado}, J., \&
  {Hartquist}, T.~W. 2008, \aap, 482, 549

\bibitem[{{Kauffmann} {et~al.}(2017){Kauffmann}, {Goldsmith}, {Melnick},
  {Tolls}, {Guzman}, \& {Menten}}]{Kauffmann+2017_hcn}
{Kauffmann}, J., {Goldsmith}, P.~F., {Melnick}, G., {et~al.} 2017, \aap, 605,
  L5

\bibitem[{{Kelly} {et~al.}(2017){Kelly}, {Viti}, {Garc{\'\i}a-Burillo},
  {Fuente}, {Usero}, {Krips}, \& {Neri}}]{Kelly+2017}
{Kelly}, G., {Viti}, S., {Garc{\'\i}a-Burillo}, S., {et~al.} 2017, \aap, 597,
  A11

\bibitem[{{Krieger} {et~al.}(2019){Krieger}, {Bolatto}, {Walter}, {Leroy},
  {Zschaechner}, {Meier}, {Ott}, {Weiss}, {Mills}, {Levy}, {Veilleux}, \&
  {Gorski}}]{Krieger+2019}
{Krieger}, N., {Bolatto}, A.~D., {Walter}, F., {et~al.} 2019, \apj, 881, 43

\bibitem[{{Krips} {et~al.}(2008){Krips}, {Neri}, {Garc{\'\i}a-Burillo},
  {Mart{\'\i}n}, {Combes}, {Graci{\'a}-Carpio}, \& {Eckart}}]{Krips+2008}
{Krips}, M., {Neri}, R., {Garc{\'\i}a-Burillo}, S., {et~al.} 2008, \apj, 677,
  262

\bibitem[{{Lehmer} {et~al.}(2013){Lehmer}, {Wik}, {Hornschemeier}, {Ptak},
  {Antoniou}, {Argo}, {Bechtol}, {Boggs}, {Christensen}, {Craig}, {Hailey},
  {Harrison}, {Krivonos}, {Leyder}, {Maccarone}, {Stern}, {Venters}, {Zezas},
  \& {Zhang}}]{Lehmer+2013}
{Lehmer}, B.~D., {Wik}, D.~R., {Hornschemeier}, A.~E., {et~al.} 2013, \apj,
  771, 134

\bibitem[{{Leroy} {et~al.}(2015){Leroy}, {Bolatto}, {Ostriker}, {Rosolowsky},
  {Walter}, {Warren}, {Donovan Meyer}, {Hodge}, {Meier}, {Ott}, {Sandstrom},
  {Schruba}, {Veilleux}, \& {Zwaan}}]{Leroy+2015}
{Leroy}, A.~K., {Bolatto}, A.~D., {Ostriker}, E.~C., {et~al.} 2015, \apj, 801,
  25

\bibitem[{{Leroy} {et~al.}(2018){Leroy}, {Bolatto}, {Ostriker}, {Walter},
  {Gorski}, {Ginsburg}, {Krieger}, {Levy}, {Meier}, {Mills}, {Ott},
  {Rosolowsky}, {Thompson}, {Veilleux}, \& {Zschaechner}}]{Leroy+2018}
{Leroy}, A.~K., {Bolatto}, A.~D., {Ostriker}, E.~C., {et~al.} 2018, \apj, 869,
  126

\bibitem[{{Levy} {et~al.}(2022){Levy}, {Bolatto}, {Leroy}, {Sormani}, {Emig},
  {Gorski}, {Lenki{\'c}}, {Mills}, {Tarantino}, {Teuben}, {Veilleux}, \&
  {Walter}}]{Levy+2022}
{Levy}, R.~C., {Bolatto}, A.~D., {Leroy}, A.~K., {et~al.} 2022, \apj, 935, 19

\bibitem[{{Li} {et~al.}(2018){Li}, {Tan}, {Christie}, {Bisbas}, \&
  {Wu}}]{Li+2018}
{Li}, Q., {Tan}, J.~C., {Christie}, D., {Bisbas}, T.~G., \& {Wu}, B. 2018,
  \pasj, 70, S56

\bibitem[{{Lodders}(2003)}]{Lodders+2003_solarabund}
{Lodders}, K. 2003, \apj, 591, 1220

\bibitem[{{L{\'o}pez-Sepulcre} {et~al.}(2015){L{\'o}pez-Sepulcre}, {Jaber},
  {Mendoza}, {Lefloch}, {Ceccarelli}, {Vastel}, {Bachiller}, {Cernicharo},
  {Codella}, {Kahane}, {Kama}, \& {Tafalla}}]{LS+2015}
{L{\'o}pez-Sepulcre}, A., {Jaber}, A.~A., {Mendoza}, E., {et~al.} 2015, \mnras,
  449, 2438

\bibitem[{{Mangum} {et~al.}(2019){Mangum}, {Ginsburg}, {Henkel}, {Menten},
  {Aalto}, \& {van der Werf}}]{Mangum+2019}
{Mangum}, J.~G., {Ginsburg}, A.~G., {Henkel}, C., {et~al.} 2019, \apj, 871, 170

\bibitem[{{Mart{\'\i}n} {et~al.}(2015){Mart{\'\i}n}, {Kohno}, {Izumi}, {Krips},
  {Meier}, {Aladro}, {Matsushita}, {Takano}, {Turner}, {Espada}, {Nakajima},
  {Terashima}, {Fathi}, {Hsieh}, {Imanishi}, {Lundgren}, {Nakai}, {Schinnerer},
  {Sheth}, \& {Wiklind}}]{Martin+2015_shocktracer_ngc1097}
{Mart{\'\i}n}, S., {Kohno}, K., {Izumi}, T., {et~al.} 2015, \aap, 573, A116

\bibitem[{{Mart{\'\i}n} {et~al.}(2021){Mart{\'\i}n}, {Mangum}, {Harada},
  {Costagliola}, {Sakamoto}, {Muller}, {Aladro}, {Tanaka}, {Yoshimura},
  {Nakanishi}, {Herrero-Illana}, {M{\"u}hle}, {Aalto}, {Behrens}, {Colzi},
  {Emig}, {Fuller}, {Garc{\'\i}a-Burillo}, {Greve}, {Henkel}, {Holdship},
  {Humire}, {Hunt}, {Izumi}, {Kohno}, {K{\"o}nig}, {Meier}, {Nakajima},
  {Nishimura}, {Padovani}, {Rivilla}, {Takano}, {van der Werf}, {Viti}, \&
  {Yan}}]{ALCHEMI_main_2021}
{Mart{\'\i}n}, S., {Mangum}, J.~G., {Harada}, N., {et~al.} 2021, \aap, 656, A46

\bibitem[{{Mart{\'\i}n} {et~al.}(2009{\natexlab{a}}){Mart{\'\i}n},
  {Mart{\'\i}n-Pintado}, \& {Mauersberger}}]{Martin+2009}
{Mart{\'\i}n}, S., {Mart{\'\i}n-Pintado}, J., \& {Mauersberger}, R.
  2009{\natexlab{a}}, \apj, 694, 610

\bibitem[{{Mart{\'\i}n} {et~al.}(2009{\natexlab{b}}){Mart{\'\i}n},
  {Mart{\'\i}n-Pintado}, \& {Viti}}]{Martin+2009_PDR_ngc253}
{Mart{\'\i}n}, S., {Mart{\'\i}n-Pintado}, J., \& {Viti}, S. 2009{\natexlab{b}},
  \apj, 706, 1323

\bibitem[{{Mart{\'\i}n} {et~al.}(2008){Mart{\'\i}n}, {Requena-Torres},
  {Mart{\'\i}n-Pintado}, \& {Mauersberger}}]{Martin+2008_HNCO_galactic}
{Mart{\'\i}n}, S., {Requena-Torres}, M.~A., {Mart{\'\i}n-Pintado}, J., \&
  {Mauersberger}, R. 2008, \apj, 678, 245

\bibitem[{{Mart{\'\i}n-Pintado} {et~al.}(1997){Mart{\'\i}n-Pintado}, {de
  Vicente}, {Fuente}, \& {Planesas}}]{sio_MP+1997}
{Mart{\'\i}n-Pintado}, J., {de Vicente}, P., {Fuente}, A., \& {Planesas}, P.
  1997, \apjl, 482, L45

\bibitem[{{McCarthy} {et~al.}(1987){McCarthy}, {van Breugel}, \&
  {Heckman}}]{McCarthy+1987}
{McCarthy}, P.~J., {van Breugel}, W., \& {Heckman}, T. 1987, \aj, 93, 264

\bibitem[{{McCormick} {et~al.}(2013){McCormick}, {Veilleux}, \&
  {Rupke}}]{McCormick+2013}
{McCormick}, A., {Veilleux}, S., \& {Rupke}, D. S.~N. 2013, \apj, 774, 126

\bibitem[{{Meier} \& {Turner}(2005)}]{Meier_Turner_2005}
{Meier}, D.~S. \& {Turner}, J.~L. 2005, \apj, 618, 259

\bibitem[{{Meier} \& {Turner}(2012)}]{Meier_Truner_2012_maffei2}
{Meier}, D.~S. \& {Turner}, J.~L. 2012, \apj, 755, 104

\bibitem[{{Meier} {et~al.}(2015){Meier}, {Walter}, {Bolatto}, {Leroy}, {Ott},
  {Rosolowsky}, {Veilleux}, {Warren}, {Wei{\ss}}, {Zwaan}, \&
  {Zschaechner}}]{Meier+2015_hncosio_253}
{Meier}, D.~S., {Walter}, F., {Bolatto}, A.~D., {et~al.} 2015, \apj, 801, 63

\bibitem[{{Mendoza} {et~al.}(2014){Mendoza}, {Lefloch}, {L{\'o}pez-Sepulcre},
  {Ceccarelli}, {Codella}, {Boechat-Roberty}, \& {Bachiller}}]{Mendoza+2014}
{Mendoza}, E., {Lefloch}, B., {L{\'o}pez-Sepulcre}, A., {et~al.} 2014, \mnras,
  445, 151

\bibitem[{{M{\"u}ller} {et~al.}(2005){M{\"u}ller}, {Schl{\"o}der}, {Stutzki},
  \& {Winnewisser}}]{CDMS_2005}
{M{\"u}ller}, H. S.~P., {Schl{\"o}der}, F., {Stutzki}, J., \& {Winnewisser}, G.
  2005, Journal of Molecular Structure, 742, 215

\bibitem[{{M{\"u}ller} {et~al.}(2001){M{\"u}ller}, {Thorwirth}, {Roth}, \&
  {Winnewisser}}]{CDMS_2001}
{M{\"u}ller}, H.~S.~P., {Thorwirth}, S., {Roth}, D.~A., \& {Winnewisser}, G.
  2001, \aap, 370, L49

\bibitem[{{M{\"u}ller-S{\'a}nchez} {et~al.}(2010){M{\"u}ller-S{\'a}nchez},
  {Gonz{\'a}lez-Mart{\'\i}n}, {Fern{\'a}ndez-Ontiveros}, {Acosta-Pulido}, \&
  {Prieto}}]{MS+2010}
{M{\"u}ller-S{\'a}nchez}, F., {Gonz{\'a}lez-Mart{\'\i}n}, O.,
  {Fern{\'a}ndez-Ontiveros}, J.~A., {Acosta-Pulido}, J.~A., \& {Prieto}, M.~A.
  2010, \apj, 716, 1166

\bibitem[{{Nazari} {et~al.}(2021){Nazari}, {van Gelder}, {van Dishoeck},
  {Tabone}, {van't Hoff}, {Ligterink}, {Beuther}, {Boogert}, {Caratti o
  Garatti}, {Klaassen}, {Linnartz}, {Taquet}, \& {Tychoniec}}]{Nazari+2021}
{Nazari}, P., {van Gelder}, M.~L., {van Dishoeck}, E.~F., {et~al.} 2021, \aap,
  650, A150

\bibitem[{{Niedenhoff} {et~al.}(1995){Niedenhoff}, {Yamada}, {Belov}, \&
  {Winnewisser}}]{hnco_moldata_N+1995}
{Niedenhoff}, M., {Yamada}, K.~M.~T., {Belov}, S.~P., \& {Winnewisser}, G.
  1995, Journal of Molecular Spectroscopy, 174, 151

\bibitem[{{Pety} {et~al.}(2017){Pety}, {Guzm{\'a}n}, {Orkisz}, {Liszt},
  {Gerin}, {Bron}, {Bardeau}, {Goicoechea}, {Gratier}, {Le Petit}, {Levrier},
  {{\"O}berg}, {Roueff}, \& {Sievers}}]{Pety+2017_densegas}
{Pety}, J., {Guzm{\'a}n}, V.~V., {Orkisz}, J.~H., {et~al.} 2017, \aap, 599, A98

\bibitem[{{Podio} {et~al.}(2017){Podio}, {Codella}, {Lefloch}, {Balucani},
  {Ceccarelli}, {Bachiller}, {Benedettini}, {Cernicharo}, {Faginas-Lago},
  {Fontani}, {Gusdorf}, \& {Rosi}}]{Podio+2017}
{Podio}, L., {Codella}, C., {Lefloch}, B., {et~al.} 2017, \mnras, 470, L16

\bibitem[{{Rekola} {et~al.}(2005){Rekola}, {Richer}, {McCall}, {Valtonen},
  {Kotilainen}, \& {Flynn}}]{Rekola+2005}
{Rekola}, R., {Richer}, M.~G., {McCall}, M.~L., {et~al.} 2005, \mnras, 361, 330

\bibitem[{{Rizzo} {et~al.}(2021){Rizzo}, {Cernicharo}, \&
  {Garc{\'\i}a-Mir{\'o}}}]{Rizzo+2021}
{Rizzo}, J.~R., {Cernicharo}, J., \& {Garc{\'\i}a-Mir{\'o}}, C. 2021, \apjs,
  253, 44

\bibitem[{{Rodr{\'\i}guez-Fern{\'a}ndez}
  {et~al.}(2010){Rodr{\'\i}guez-Fern{\'a}ndez}, {Tafalla}, {Gueth}, \&
  {Bachiller}}]{hnco+RF+2010}
{Rodr{\'\i}guez-Fern{\'a}ndez}, N.~J., {Tafalla}, M., {Gueth}, F., \&
  {Bachiller}, R. 2010, \aap, 516, A98

\bibitem[{Sahnoun {et~al.}(2018)Sahnoun, Wiesenfeld, Hammami, \&
  Jaidane}]{hnco_moedata_S+2018}
Sahnoun, E., Wiesenfeld, L., Hammami, K., \& Jaidane, N. 2018, The Journal of
  Physical Chemistry A, 122, 3004, pMID: 29480723

\bibitem[{{Sakamoto} {et~al.}(2011){Sakamoto}, {Mao}, {Matsushita}, {Peck},
  {Sawada}, \& {Wiedner}}]{Sakamoto+2011}
{Sakamoto}, K., {Mao}, R.-Q., {Matsushita}, S., {et~al.} 2011, \apj, 735, 19

\bibitem[{{Savage} \& {Ziurys}(2004)}]{Savage_Ziurys_2004}
{Savage}, C. \& {Ziurys}, L.~M. 2004, \apj, 616, 966

\bibitem[{{Schilke} {et~al.}(1997){Schilke}, {Walmsley}, {Pineau des Forets},
  \& {Flower}}]{Schilke+1997}
{Schilke}, P., {Walmsley}, C.~M., {Pineau des Forets}, G., \& {Flower}, D.~R.
  1997, \aap, 321, 293

\bibitem[{{Sch{\"o}ier} {et~al.}(2005){Sch{\"o}ier}, {van der Tak}, {van
  Dishoeck}, \& {Black}}]{LAMDA_2005}
{Sch{\"o}ier}, F.~L., {van der Tak}, F.~F.~S., {van Dishoeck}, E.~F., \&
  {Black}, J.~H. 2005, \aap, 432, 369

\bibitem[{{Strickland} {et~al.}(2000){Strickland}, {Heckman}, {Weaver}, \&
  {Dahlem}}]{Strickland+2000}
{Strickland}, D.~K., {Heckman}, T.~M., {Weaver}, K.~A., \& {Dahlem}, M. 2000,
  \aj, 120, 2965

\bibitem[{{Strickland} {et~al.}(2002){Strickland}, {Heckman}, {Weaver},
  {Hoopes}, \& {Dahlem}}]{Strickland+2002}
{Strickland}, D.~K., {Heckman}, T.~M., {Weaver}, K.~A., {Hoopes}, C.~G., \&
  {Dahlem}, M. 2002, \apj, 568, 689

\bibitem[{{Tafalla} {et~al.}(2021){Tafalla}, {Usero}, \&
  {Hacar}}]{Tafalla+2021}
{Tafalla}, M., {Usero}, A., \& {Hacar}, A. 2021, \aap, 646, A97

\bibitem[{{Turner}(1985)}]{Turner1985}
{Turner}, B.~E. 1985, \apj, 299, 312

\bibitem[{{Turner} \& {Ho}(1985)}]{TH1985}
{Turner}, J.~L. \& {Ho}, P.~T.~P. 1985, \apjl, 299, L77

\bibitem[{{Usero} {et~al.}(2006){Usero}, {Garc{\'\i}a-Burillo},
  {Mart{\'\i}n-Pintado}, {Fuente}, \& {Neri}}]{Usero+2006}
{Usero}, A., {Garc{\'\i}a-Burillo}, S., {Mart{\'\i}n-Pintado}, J., {Fuente},
  A., \& {Neri}, R. 2006, \aap, 448, 457

\bibitem[{{van der Tak} {et~al.}(2007){van der Tak}, {Black}, {Sch{\"o}ier},
  {Jansen}, \& {van Dishoeck}}]{radex_vandertak_2007}
{van der Tak}, F.~F.~S., {Black}, J.~H., {Sch{\"o}ier}, F.~L., {Jansen}, D.~J.,
  \& {van Dishoeck}, E.~F. 2007, \aap, 468, 627

\bibitem[{{Velilla Prieto} {et~al.}(2015){Velilla Prieto}, {S{\'a}nchez
  Contreras}, {Cernicharo}, {Ag{\'u}ndez}, {Quintana-Lacaci}, {Alcolea},
  {Bujarrabal}, {Herpin}, {Menten}, \& {Wyrowski}}]{VP+2015}
{Velilla Prieto}, L., {S{\'a}nchez Contreras}, C., {Cernicharo}, J., {et~al.}
  2015, \aap, 575, A84

\bibitem[{{Velilla Prieto} {et~al.}(2017){Velilla Prieto}, {S{\'a}nchez
  Contreras}, {Cernicharo}, {Ag{\'u}ndez}, {Quintana-Lacaci}, {Bujarrabal},
  {Alcolea}, {Balan{\c{c}}a}, {Herpin}, {Menten}, \& {Wyrowski}}]{VP+2017}
{Velilla Prieto}, L., {S{\'a}nchez Contreras}, C., {Cernicharo}, J., {et~al.}
  2017, \aap, 597, A25

\bibitem[{{Viti}(2017)}]{Viti_2017}
{Viti}, S. 2017, \aap, 607, A118

\bibitem[{{Walter} {et~al.}(2017){Walter}, {Bolatto}, {Leroy}, {Veilleux},
  {Warren}, {Hodge}, {Levy}, {Meier}, {Ostriker}, {Ott}, {Rosolowsky},
  {Scoville}, {Weiss}, {Zschaechner}, \& {Zwaan}}]{Walter+2017}
{Walter}, F., {Bolatto}, A.~D., {Leroy}, A.~K., {et~al.} 2017, \apj, 835, 265

\bibitem[{{Westmoquette} {et~al.}(2011){Westmoquette}, {Smith}, \&
  {Gallagher}}]{Westmoquette+2011}
{Westmoquette}, M.~S., {Smith}, L.~J., \& {Gallagher}, J.~S., I. 2011, \mnras,
  414, 3719

\bibitem[{{Zeng} {et~al.}(2018){Zeng}, {Jim{\'e}nez-Serra}, {Rivilla},
  {Mart{\'\i}n}, {Mart{\'\i}n-Pintado}, {Requena-Torres},
  {Armijos-Abenda{\~n}o}, {Riquelme}, \& {Aladro}}]{Zeng+2018}
{Zeng}, S., {Jim{\'e}nez-Serra}, I., {Rivilla}, V.~M., {et~al.} 2018, \mnras,
  478, 2962

\bibitem[{{Zinchenko} {et~al.}(2000){Zinchenko}, {Henkel}, \&
  {Mao}}]{Zinchenko+2000}
{Zinchenko}, I., {Henkel}, C., \& {Mao}, R.~Q. 2000, \aap, 361, 1079

\end{thebibliography}
\newpage
\appendix
\section{Spectra of molecular transitions}
\begin{figure}[b]
  \centering
  \begin{tabular}[b]{@{}p{0.43\textwidth}@{}}
    \centering\includegraphics[width=1.0\linewidth]{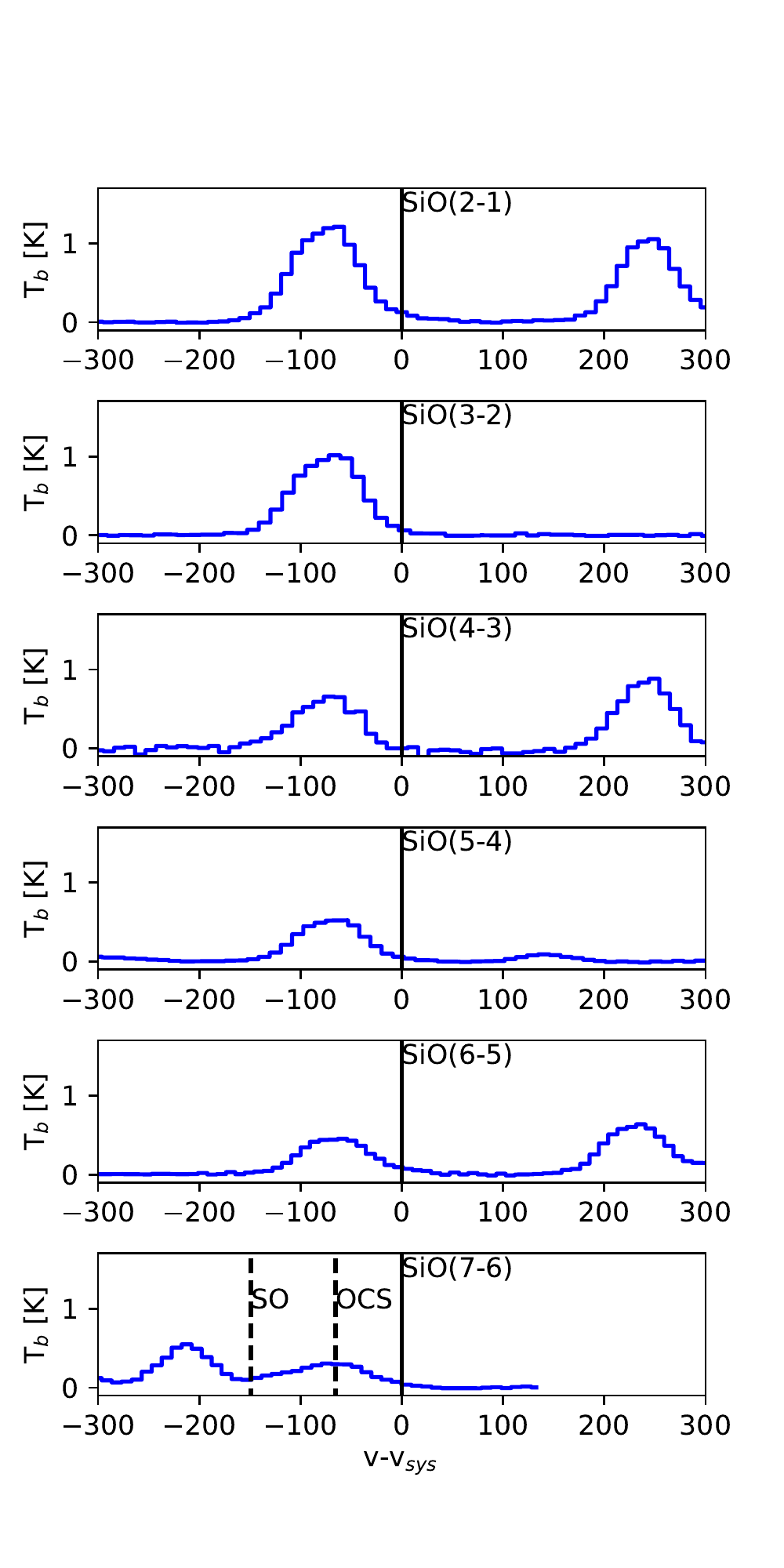} \\
  \end{tabular}
  \caption{Spectra extracted from GMC 6, for all SiO transitions. used in this work. The solid vertical line marks the rest frequency of each transition, and the dashed vertical line(s) the adjacent or blending line. }
  \label{fig:Spectra_Ss}
\end{figure}

\begin{figure*}
  \centering
  \begin{tabular}[b]{@{}p{0.43\textwidth}@{}}
    \centering\includegraphics[width=1.0\linewidth]{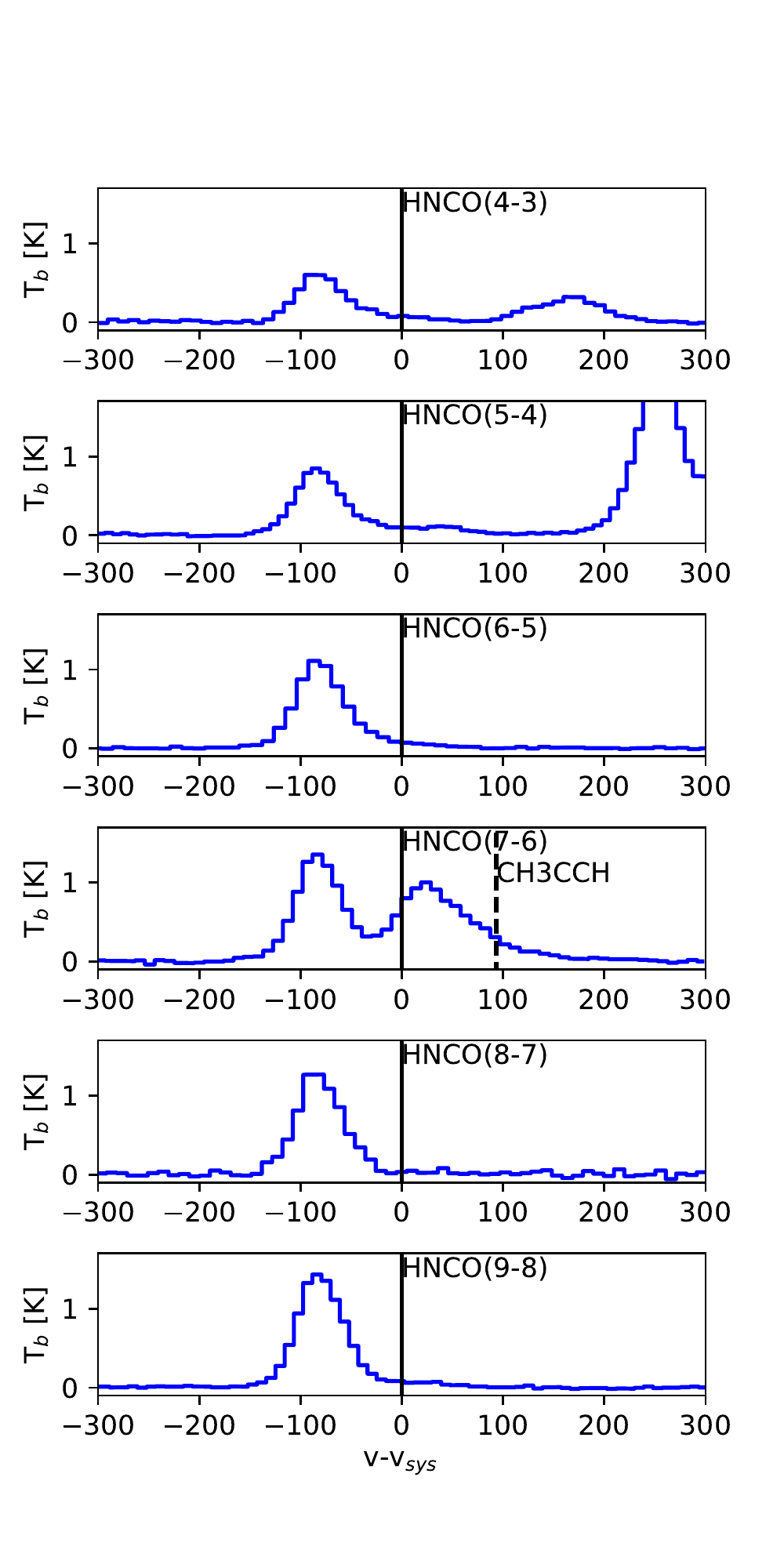} \\
  \end{tabular}
  \begin{tabular}[b]{@{}p{0.43\textwidth}@{}}
    \centering\includegraphics[width=1.0\linewidth]{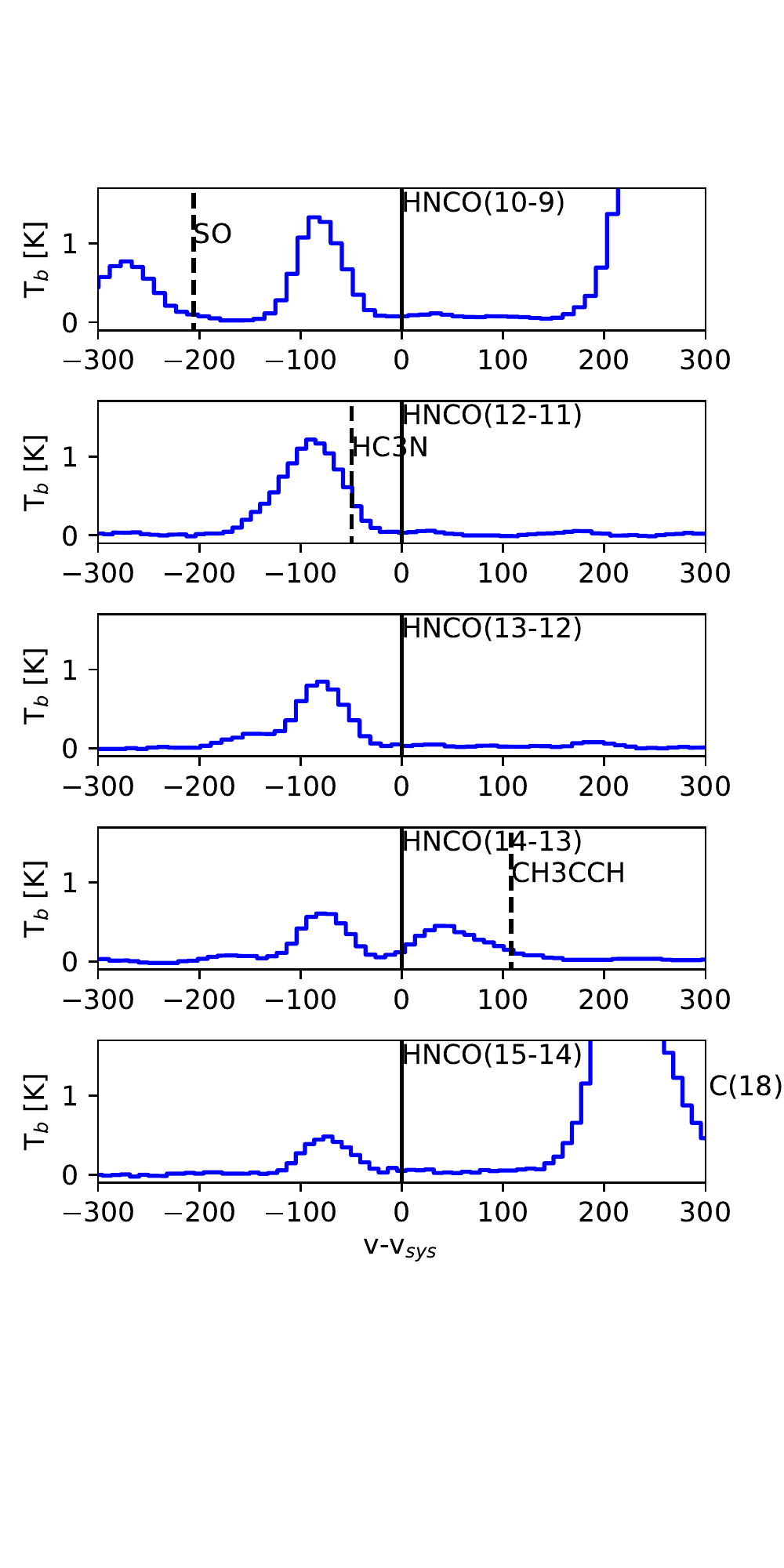} \\
  \end{tabular}
  \caption{Spectra extracted from GMC 6, for all HNCO transitions. used in this work. The solid vertical line marks reference target transition with respect to the systemic velocity of ngc 253 ($v_{sys} =258.8$ km s\textsuperscript{-1}), and the dashed vertical line the adjacent or blending line. }
  \label{fig:Spectra_Hs}
\end{figure*}

In this section we show the spectra from all the SiO (Fig. \ref{fig:Spectra_Ss}) and HNCO (Fig. \ref{fig:Spectra_Hs}) transitions used in the current work. 
The spectra are extracted from the representative region - GMC 6 - as listed in Table \ref{tab:GMC_locations}. 
\label{sec:spectra}

\section{Additional intensity maps}
In this section we display the velocity-integrated intensity maps for the remaining HNCO and SiO transition that are not shown in the main text. 
\label{sec:appen_mom0}
\begin{figure*}
  \centering
  \begin{tabular}[b]{@{}p{0.46\textwidth}@{}}
    \centering\includegraphics[width=1.0\linewidth]{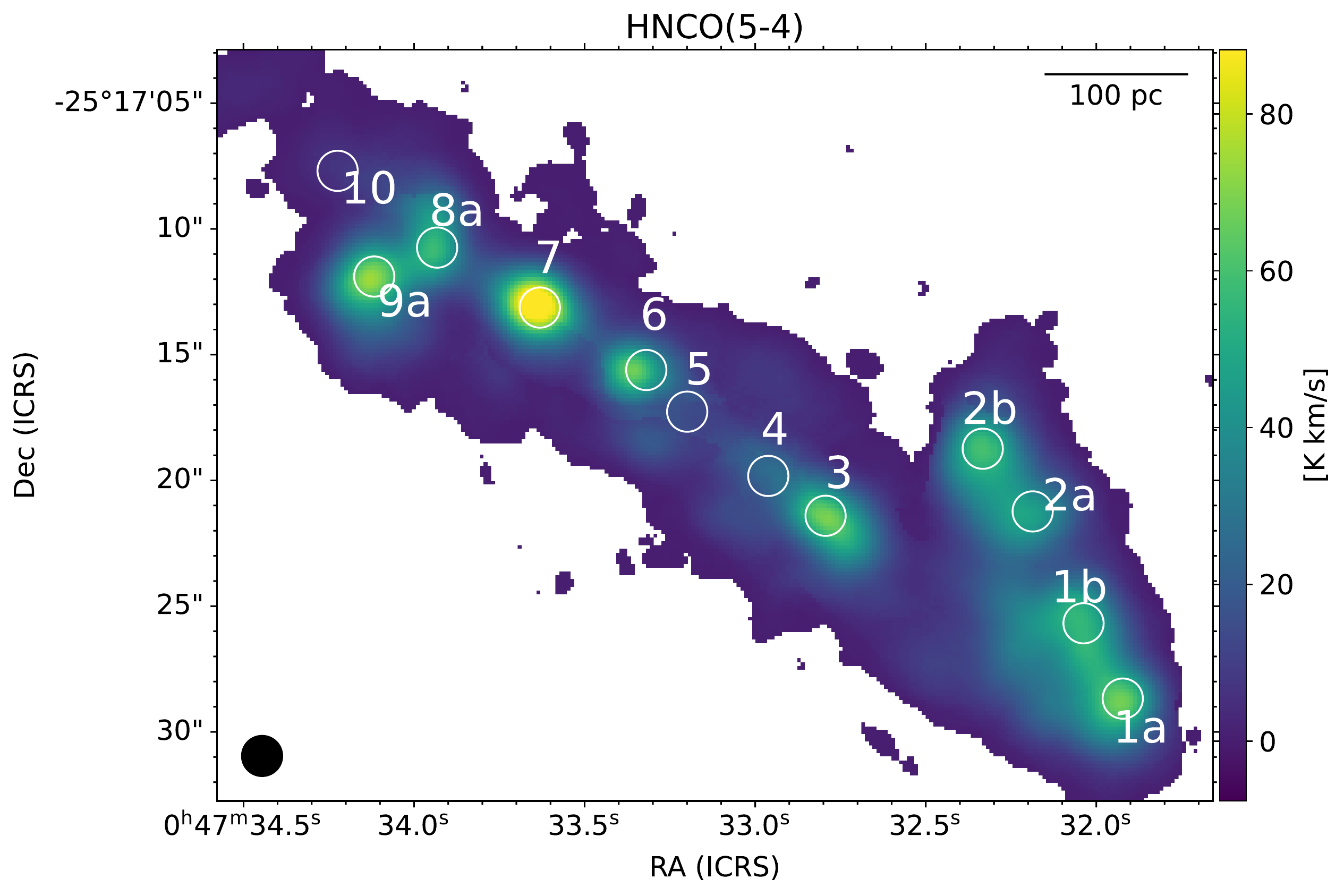} \\
    \centering\small (a) 
  \end{tabular}%
  \quad
  \begin{tabular}[b]{@{}p{0.46\textwidth}@{}}
    \centering\includegraphics[width=1.0\linewidth]{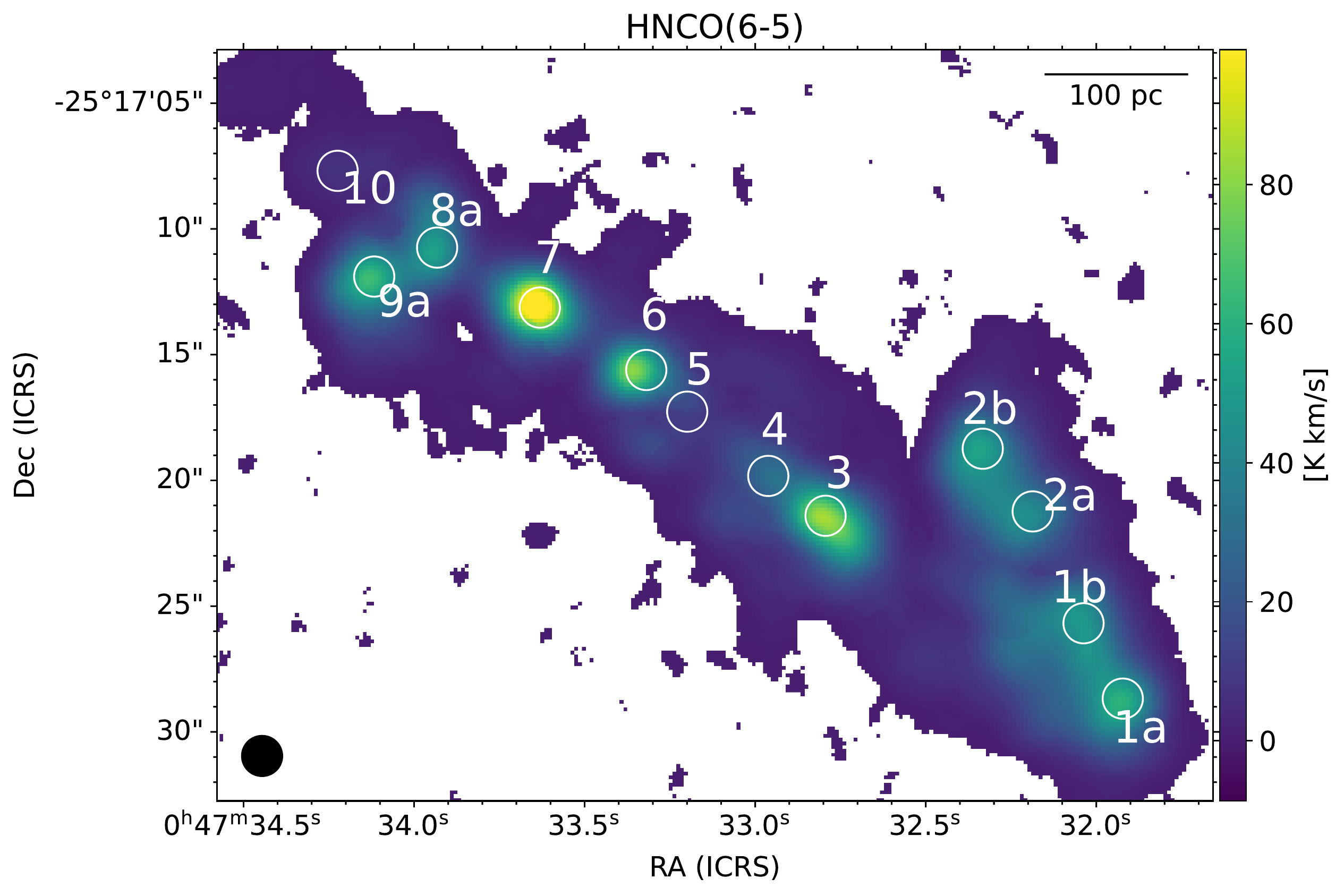} \\
    \centering\small (b) 
  \end{tabular}
  \begin{tabular}[b]{@{}p{0.46\textwidth}@{}}
    \centering\includegraphics[width=1.0\linewidth]{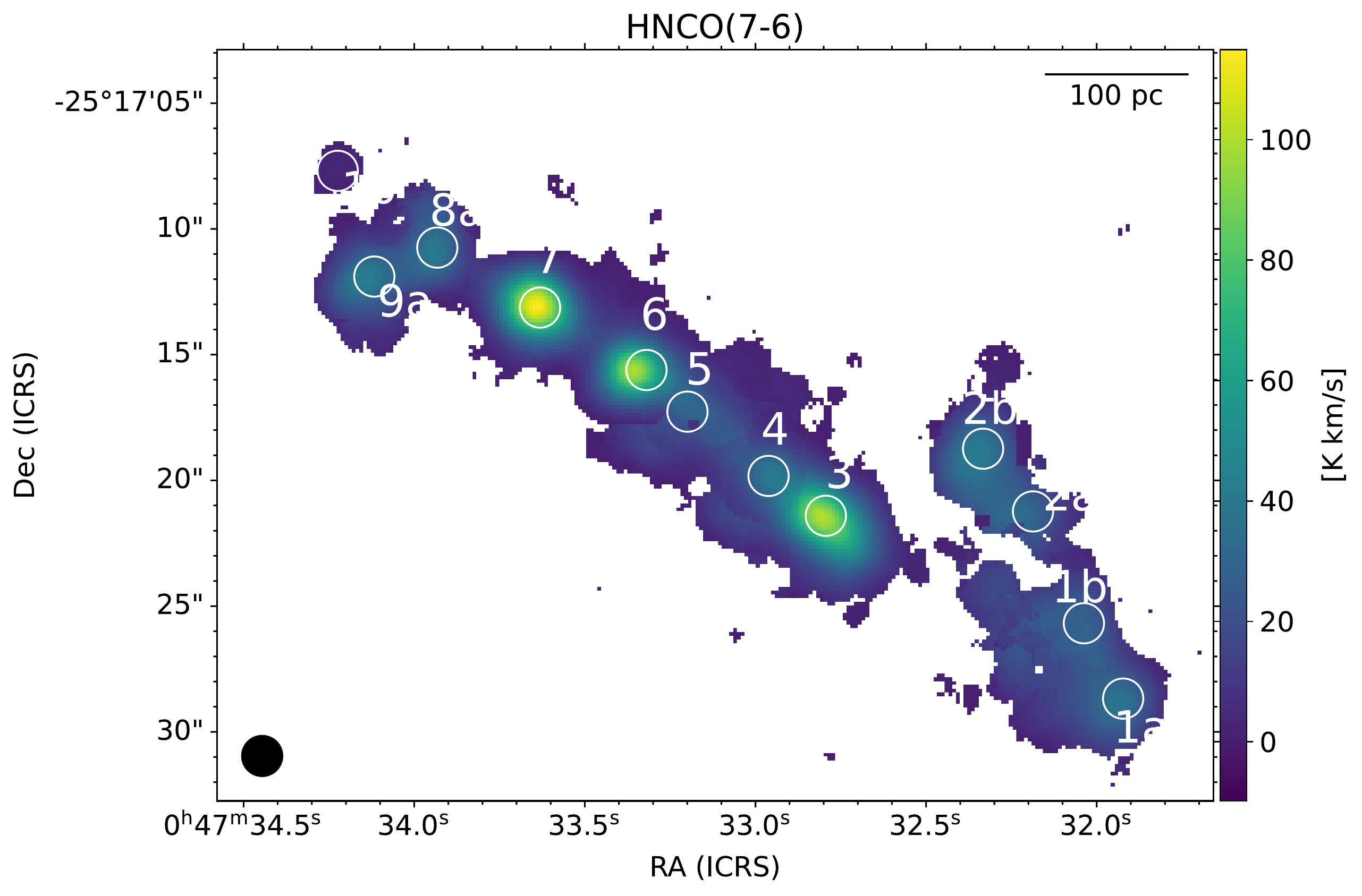} \\
    \centering\small (e)
  \end{tabular}
  \begin{tabular}[b]{@{}p{0.46\textwidth}@{}}
    \centering\includegraphics[width=1.0\linewidth]{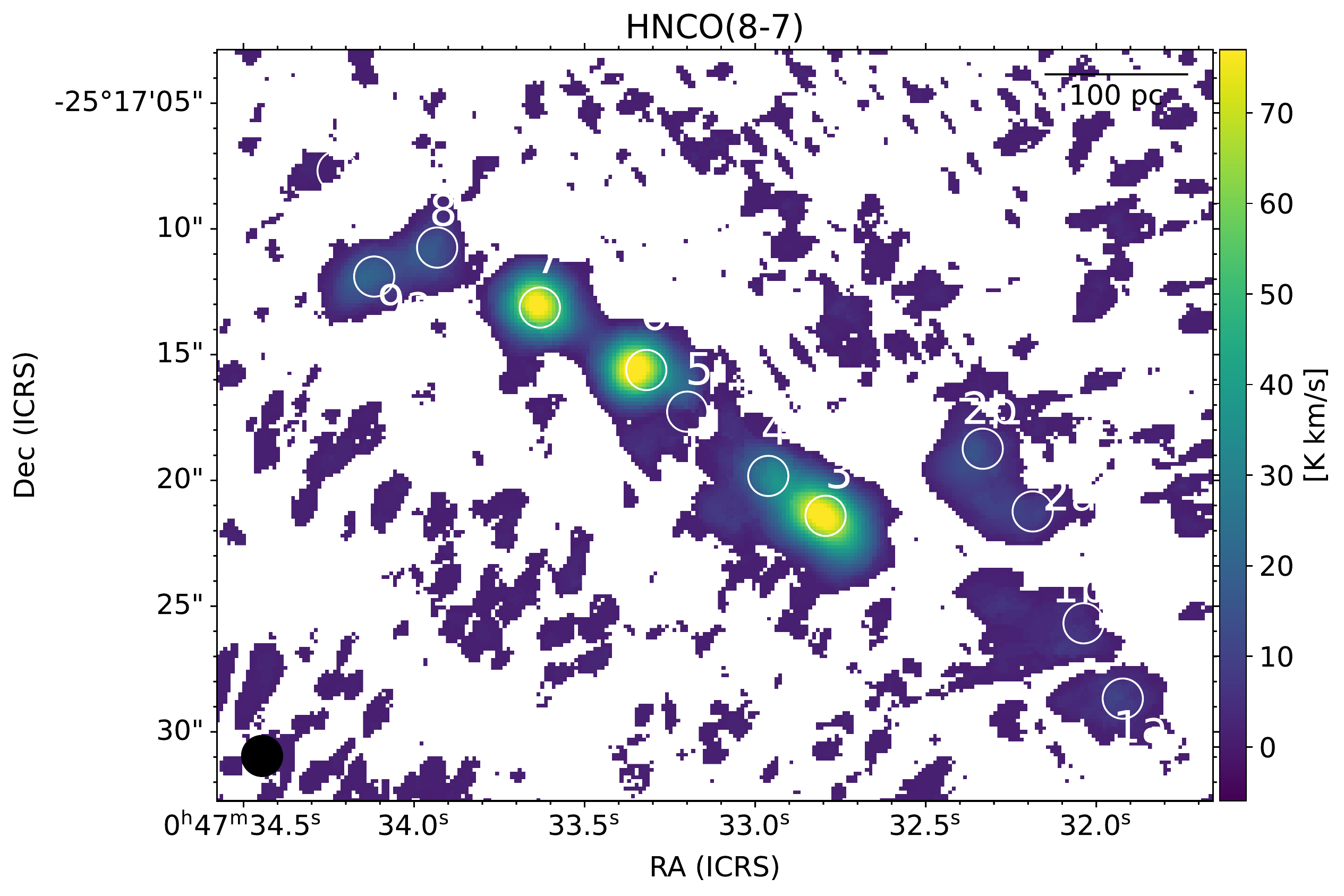} \\
    \centering\small (c) 
  \end{tabular}
  \begin{tabular}[b]{@{}p{0.46\textwidth}@{}}
    \centering\includegraphics[width=1.0\linewidth]{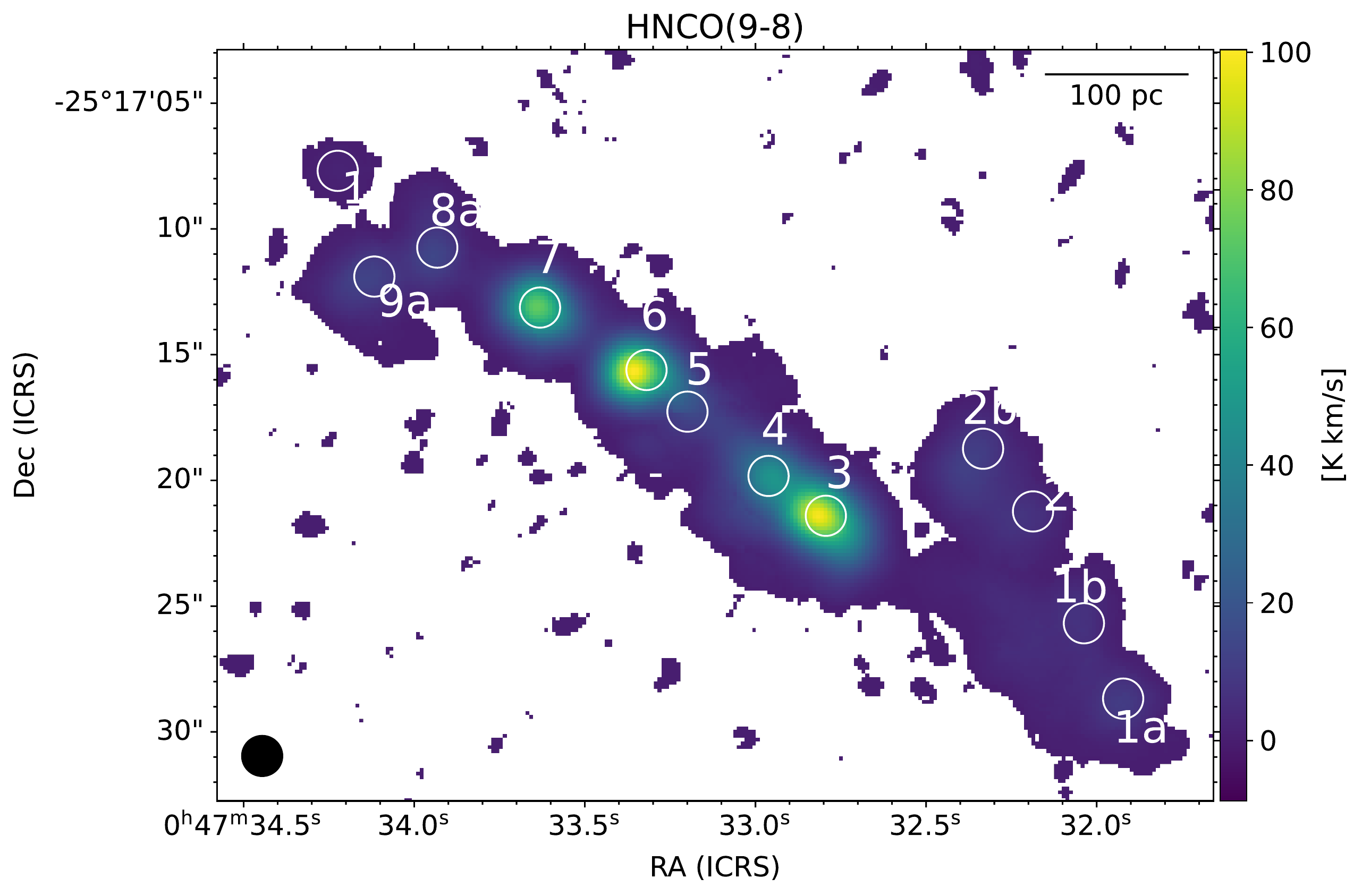} \\
    \centering\small (d)
  \end{tabular}
  \begin{tabular}[b]{@{}p{0.46\textwidth}@{}}
    \centering\includegraphics[width=1.0\linewidth]{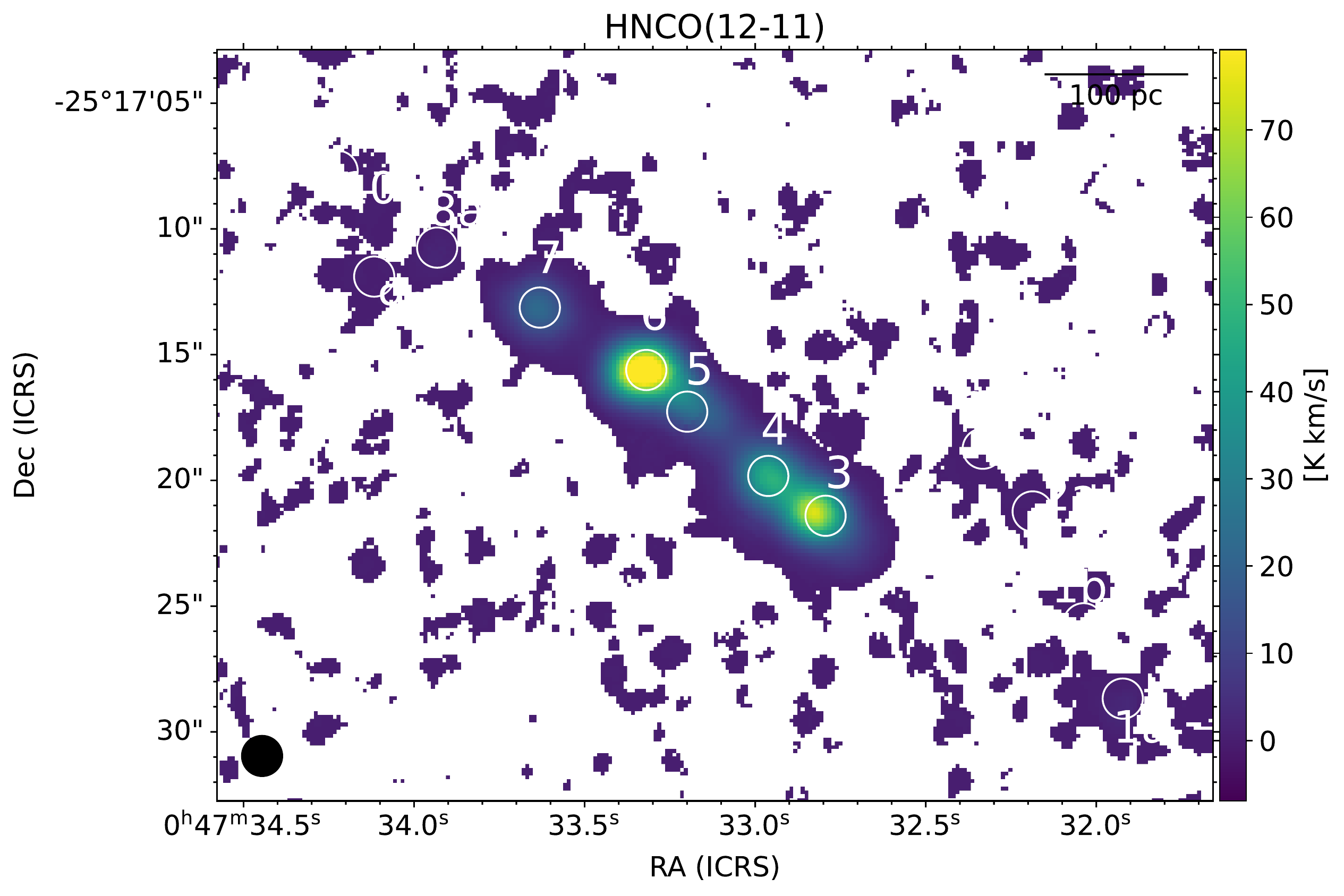} \\
    \centering\small (f) 
  \end{tabular}
  \caption{The remaining velocity-integrated line intensities in [K km s\textsuperscript{-1}] of HNCO transitions: (5-4)/(6-5)/(8-7)/(9-8)/10\textsubscript{0,10}-9\textsubscript{0,9}/(13-12), ordered accordingly from (a) to (f). Note that HNCO (8-7) transition is close to the 183 GHz water line.} 
  \label{fig:mom0_add_I}
\end{figure*}
\begin{figure*}
  \centering
  \begin{tabular}[b]{@{}p{0.46\textwidth}@{}}
    \centering\includegraphics[width=1.0\linewidth]{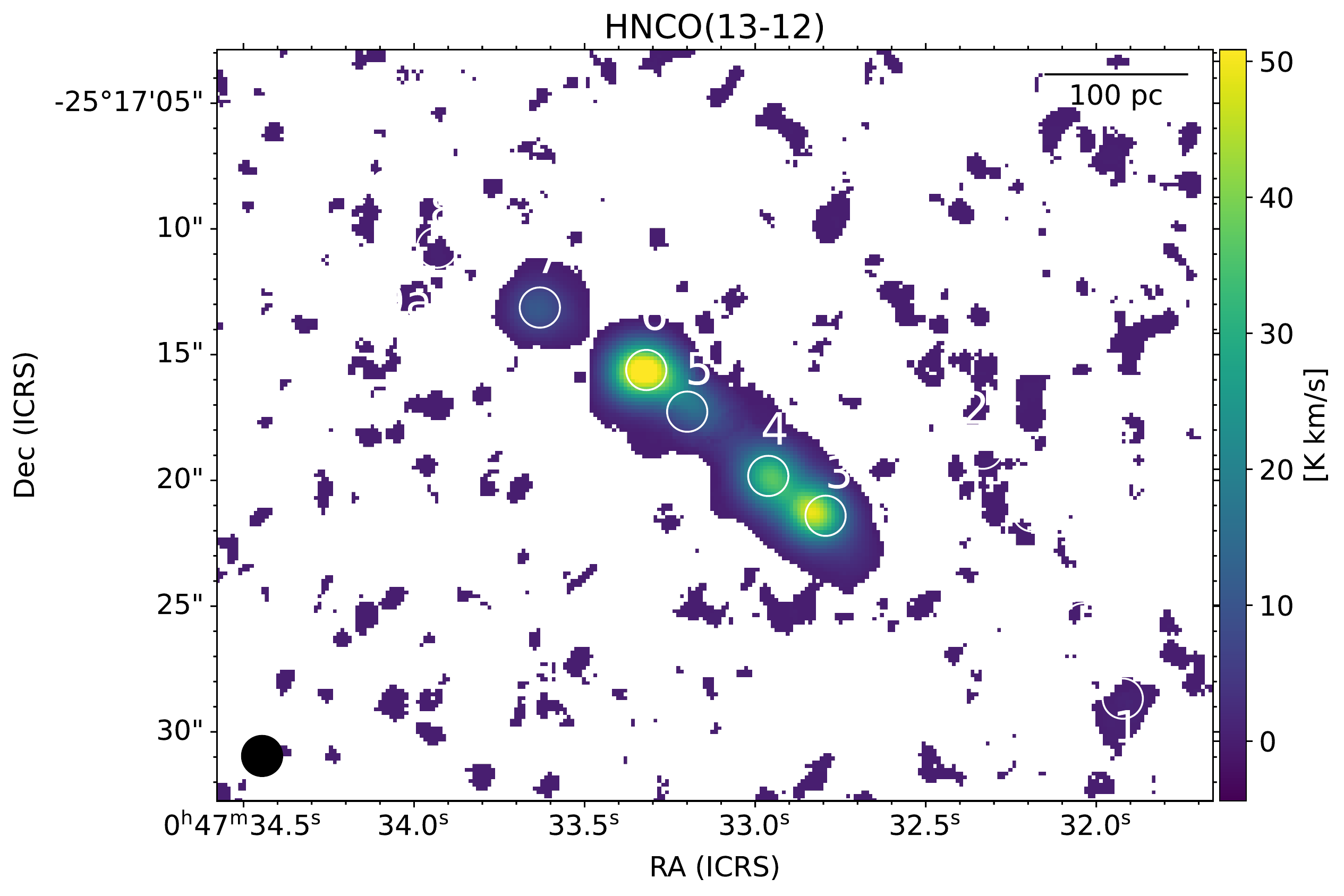} \\
    \centering\small (a)
  \end{tabular}
  \begin{tabular}[b]{@{}p{0.46\textwidth}@{}}
    \centering\includegraphics[width=1.0\linewidth]{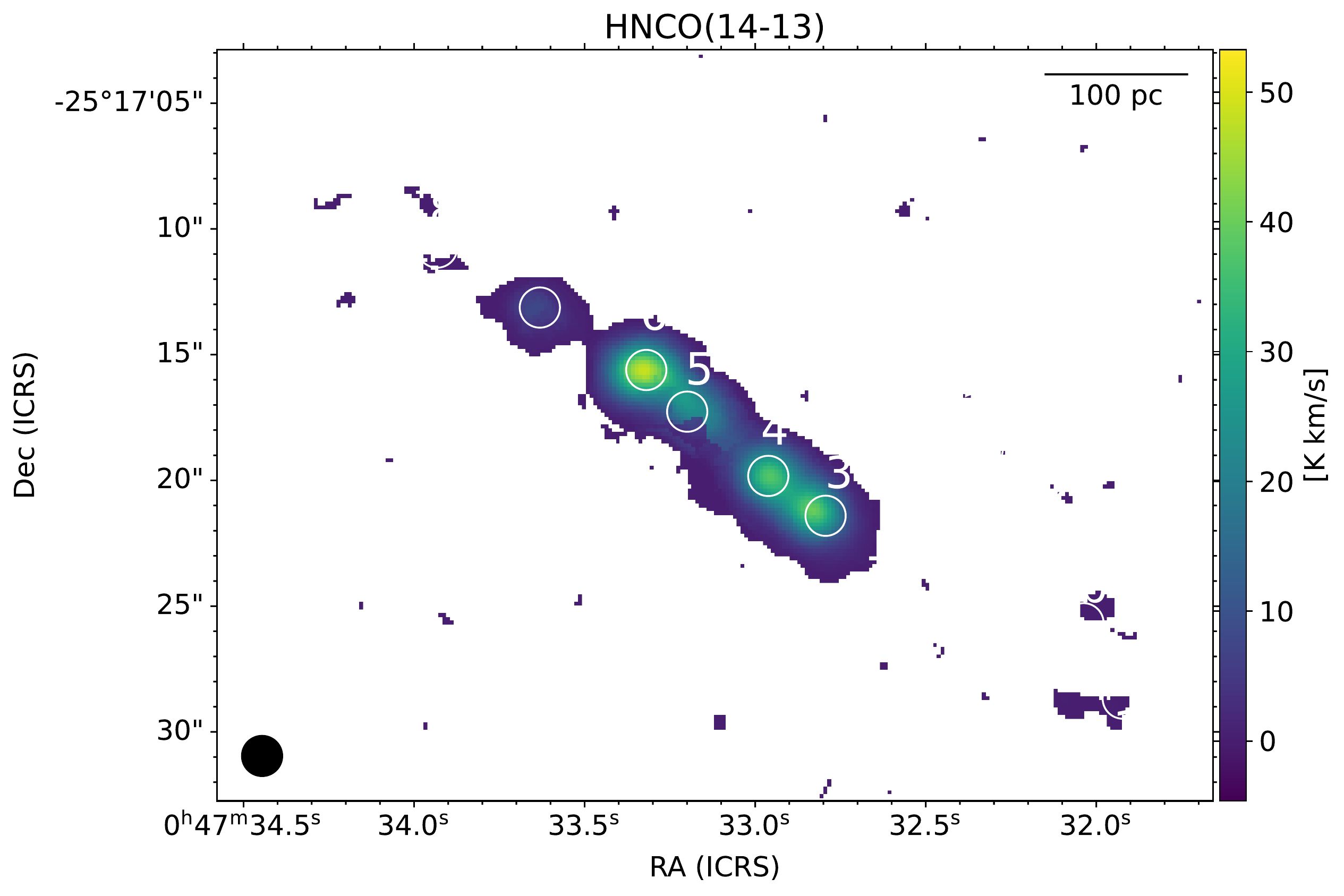} \\
    \centering\small (b) 
  \end{tabular}%
  \quad
  \begin{tabular}[b]{@{}p{0.46\textwidth}@{}}
    \centering\includegraphics[width=1.0\linewidth]{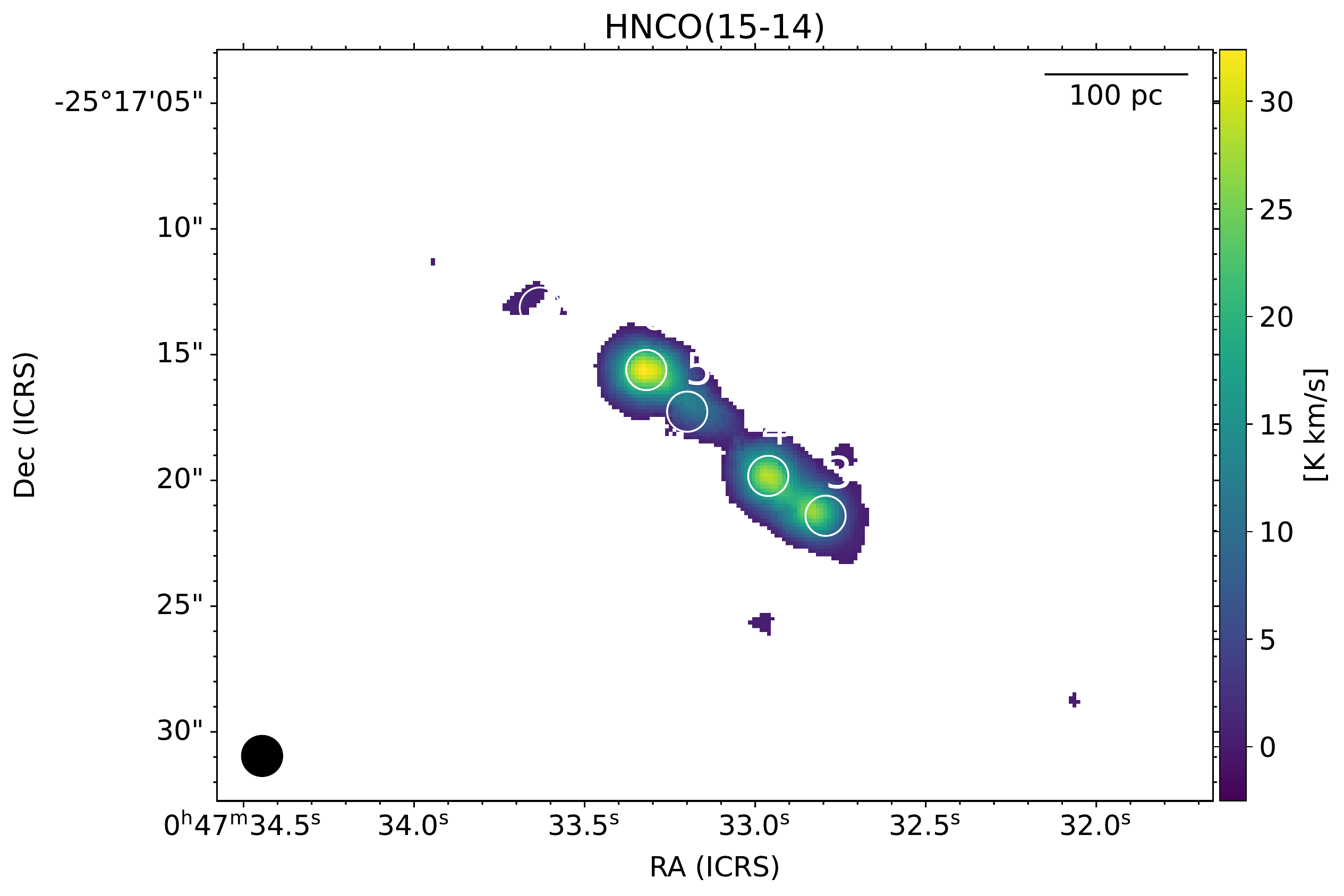} \\
    \centering\small (c) 
  \end{tabular}
  \caption{The velocity-integrated line intensities in [K km s\textsuperscript{-1}] of HNCO (13-12), (14-13), (15-14) transitions. }
  \label{fig:mom0_add_II}
\end{figure*}
\begin{figure*}
  \begin{tabular}[b]{@{}p{0.46\textwidth}@{}}
    \centering\includegraphics[width=1.0\linewidth]{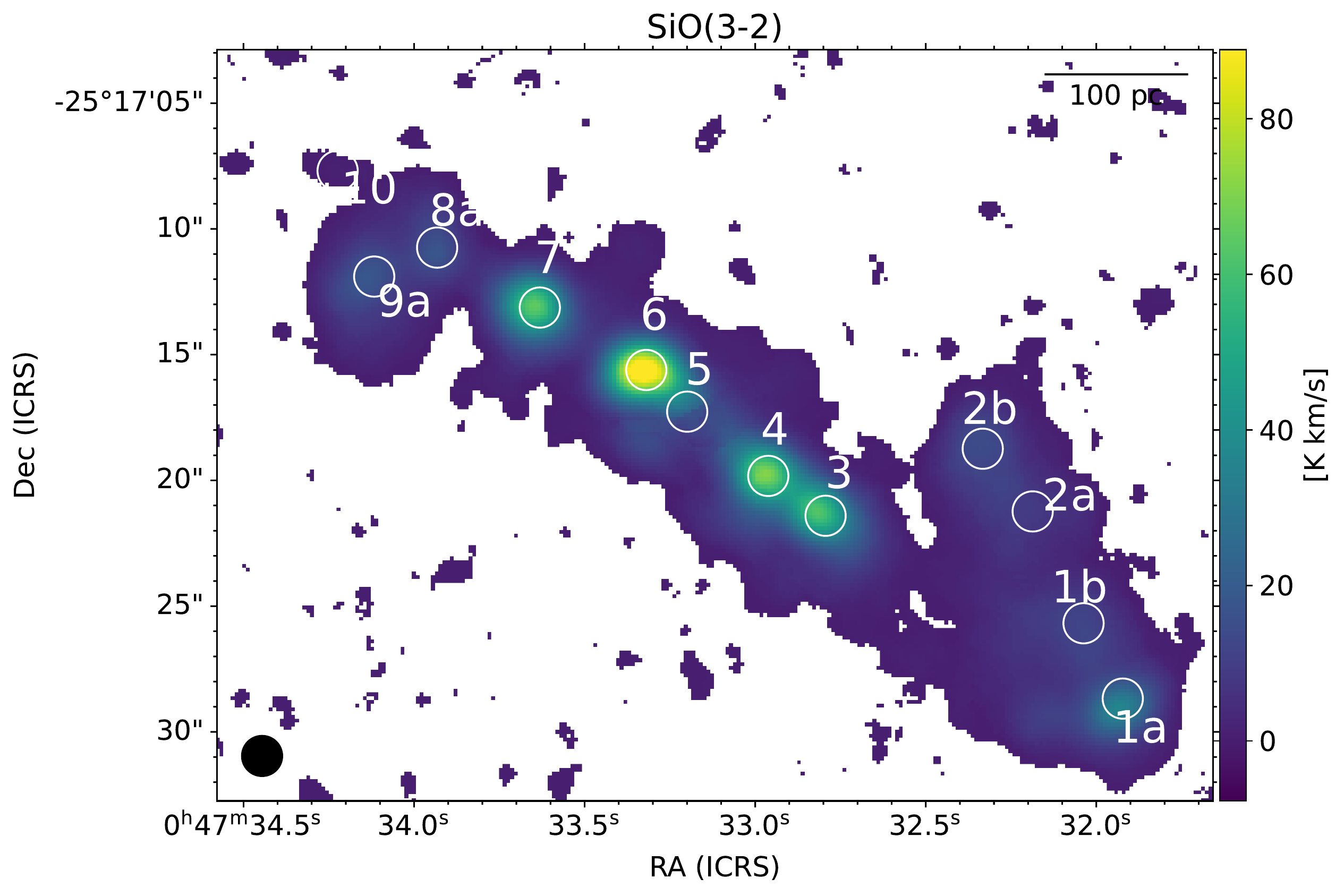} \\
    \centering\small (a) 
  \end{tabular}
  \begin{tabular}[b]{@{}p{0.46\textwidth}@{}}
    \centering\includegraphics[width=1.0\linewidth]{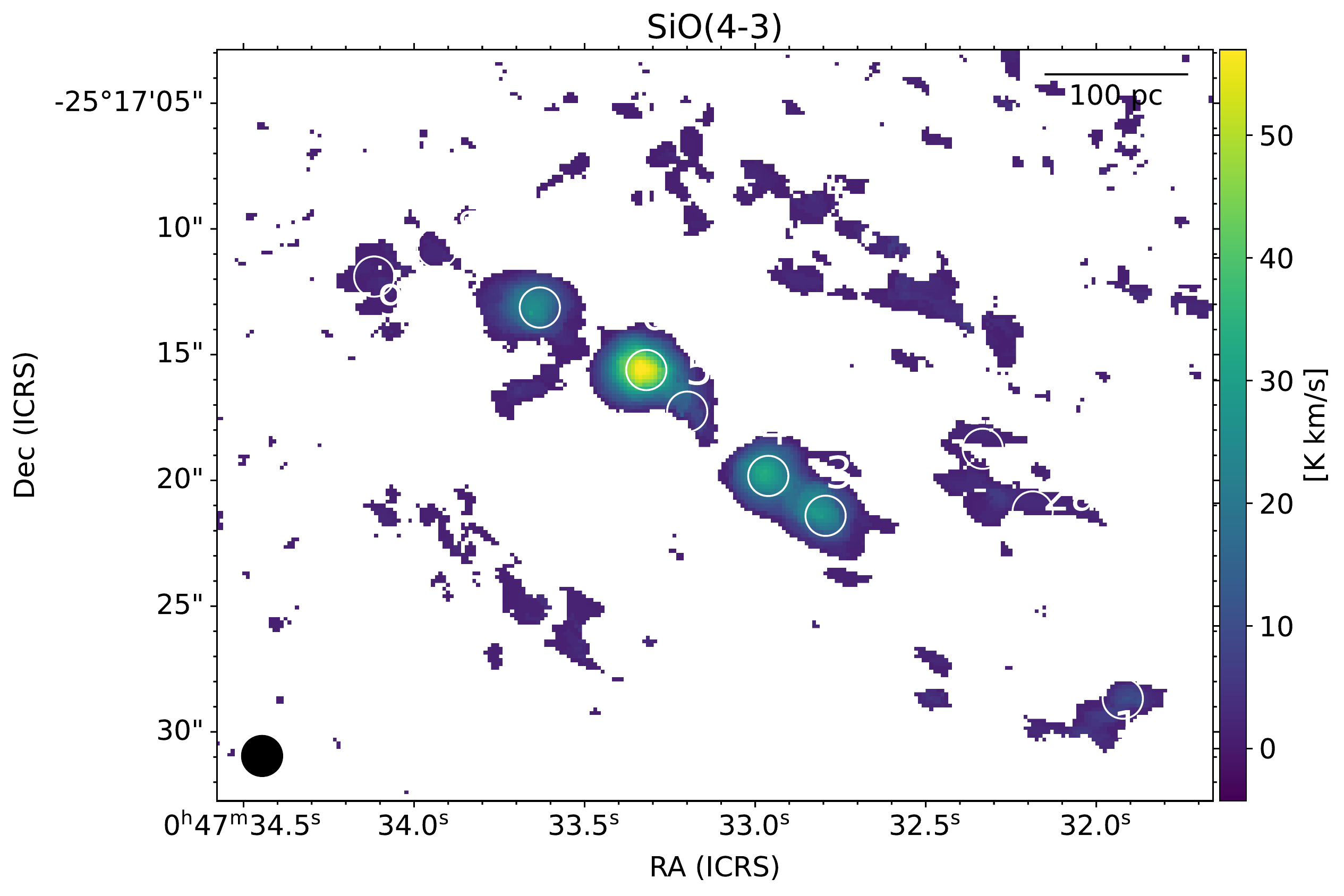} \\
    \centering\small (b)
  \end{tabular}
  \begin{tabular}[b]{@{}p{0.46\textwidth}@{}}
    \centering\includegraphics[width=1.0\linewidth]{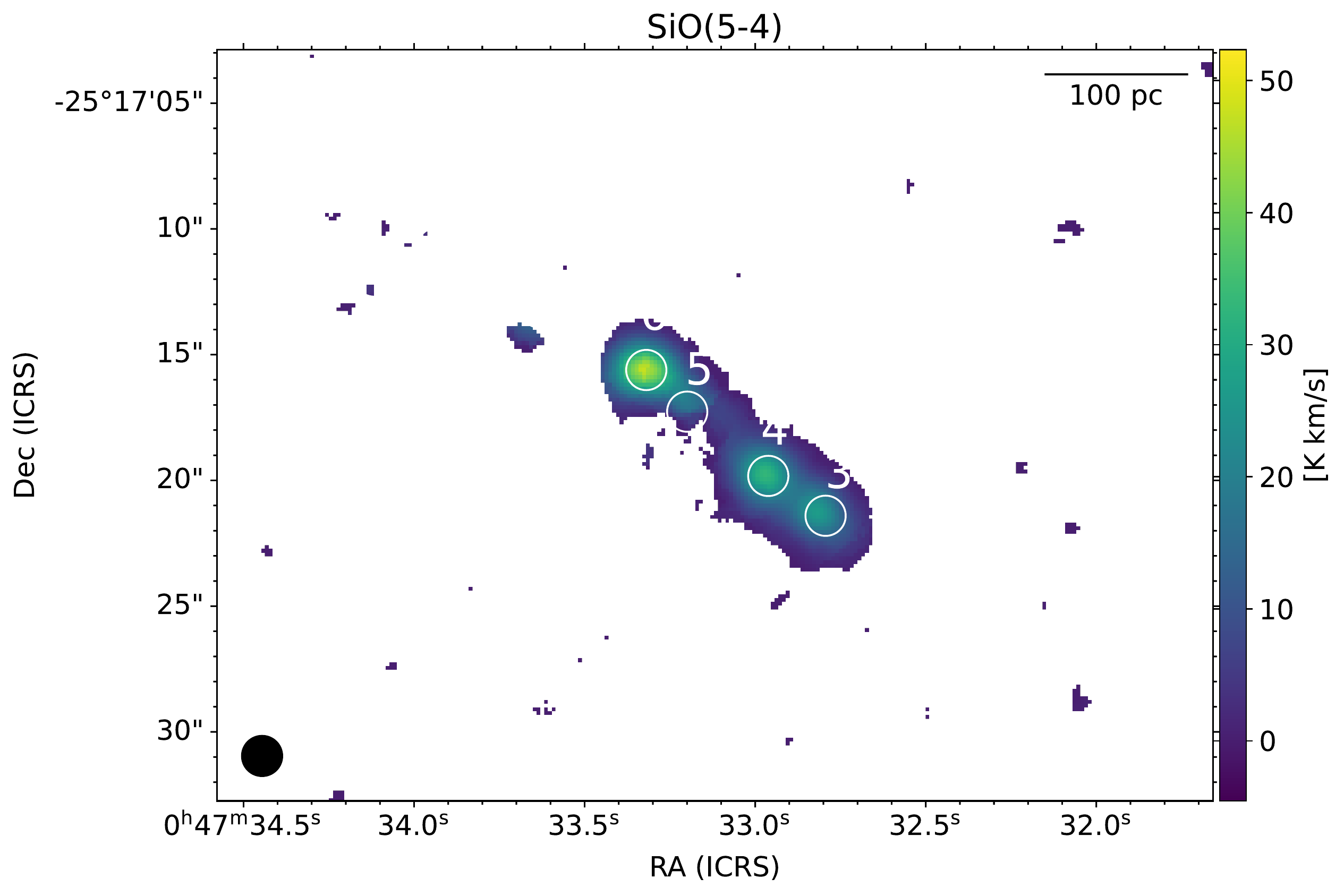} \\
    \centering\small (c) 
  \end{tabular}
  \begin{tabular}[b]{@{}p{0.46\textwidth}@{}}
    \centering\includegraphics[width=1.0\linewidth]{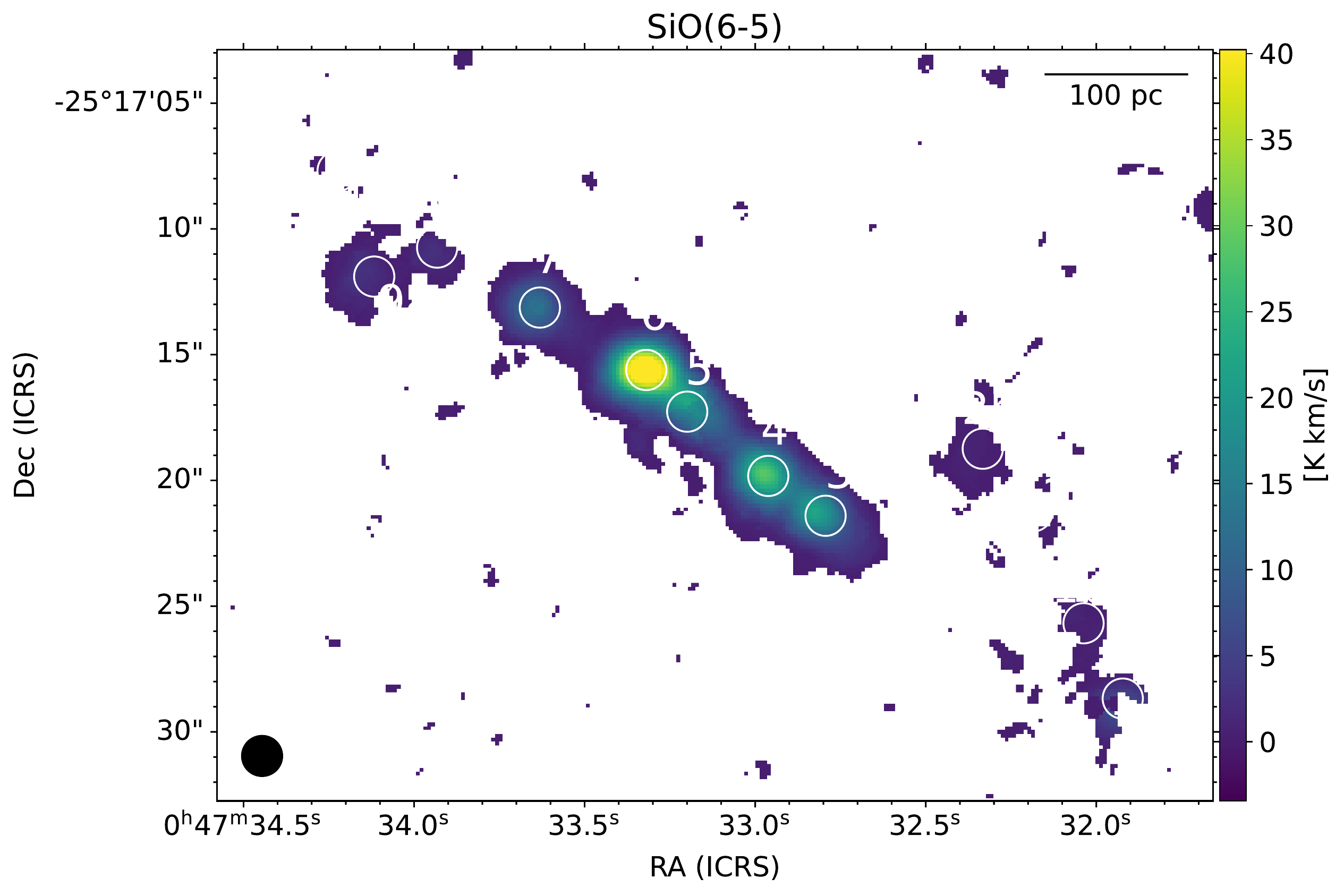} \\
    \centering\small (d)
  \end{tabular}
  \begin{tabular}[b]{@{}p{0.46\textwidth}@{}}
    \centering\includegraphics[width=1.0\linewidth]{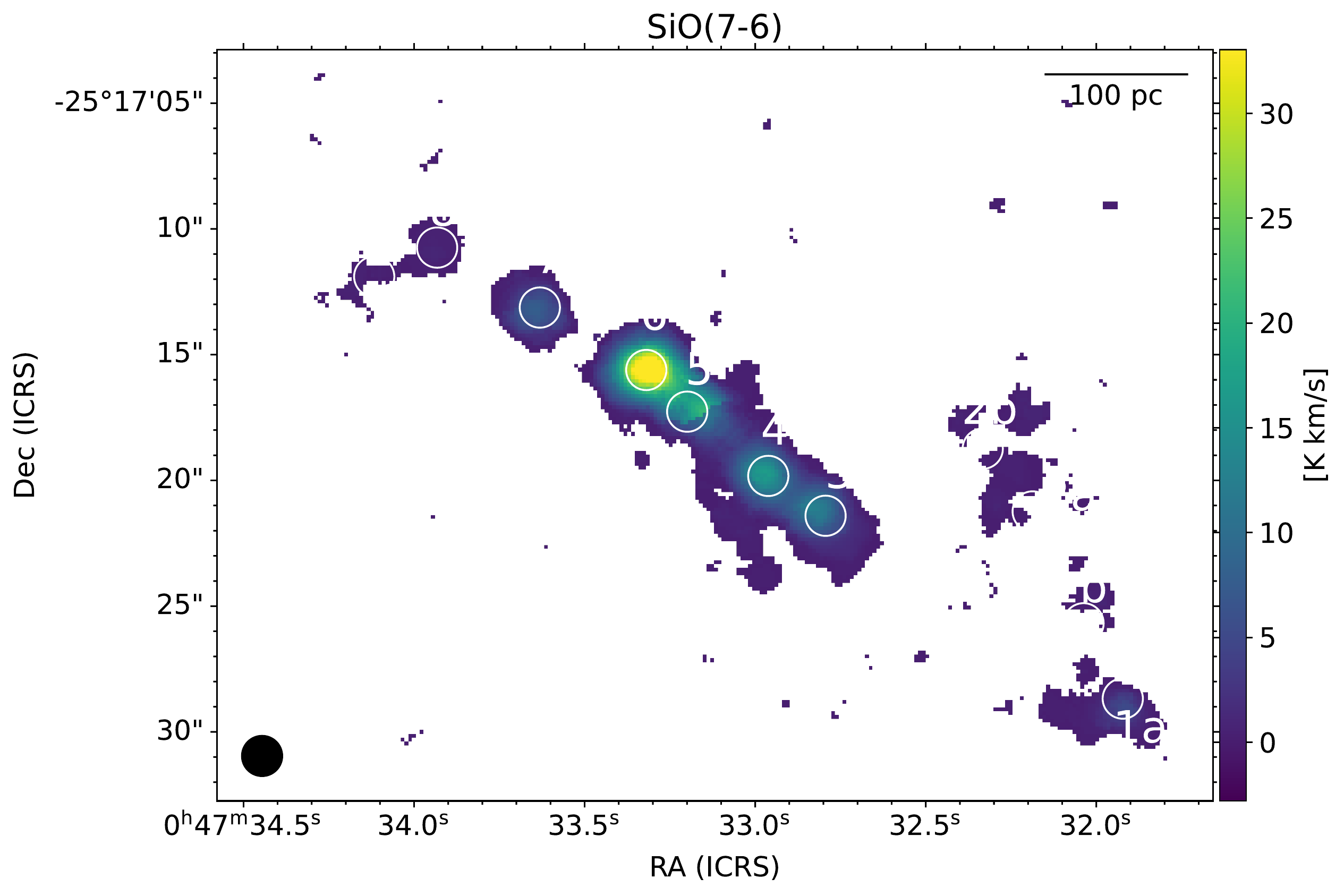} \\
    \centering\small (e)
  \end{tabular}
  \caption{The velocity-integrated line intensities in [K km s\textsuperscript{-1}] of the remaining SiO transitions: SiO (3-2) up to (7-6). }
  \label{fig:mom0_add_III}
\end{figure*}
\section{Line intensity ratio maps}
In this section we display the intensity ratio maps described in Sect. \ref{sec:intensity_ratio}. 
\label{sec:ratio_maps}
\begin{figure*}
  \centering
  \begin{tabular}[b]{@{}p{0.8\textwidth}@{}}
    \centering\includegraphics[width=1.0\linewidth]{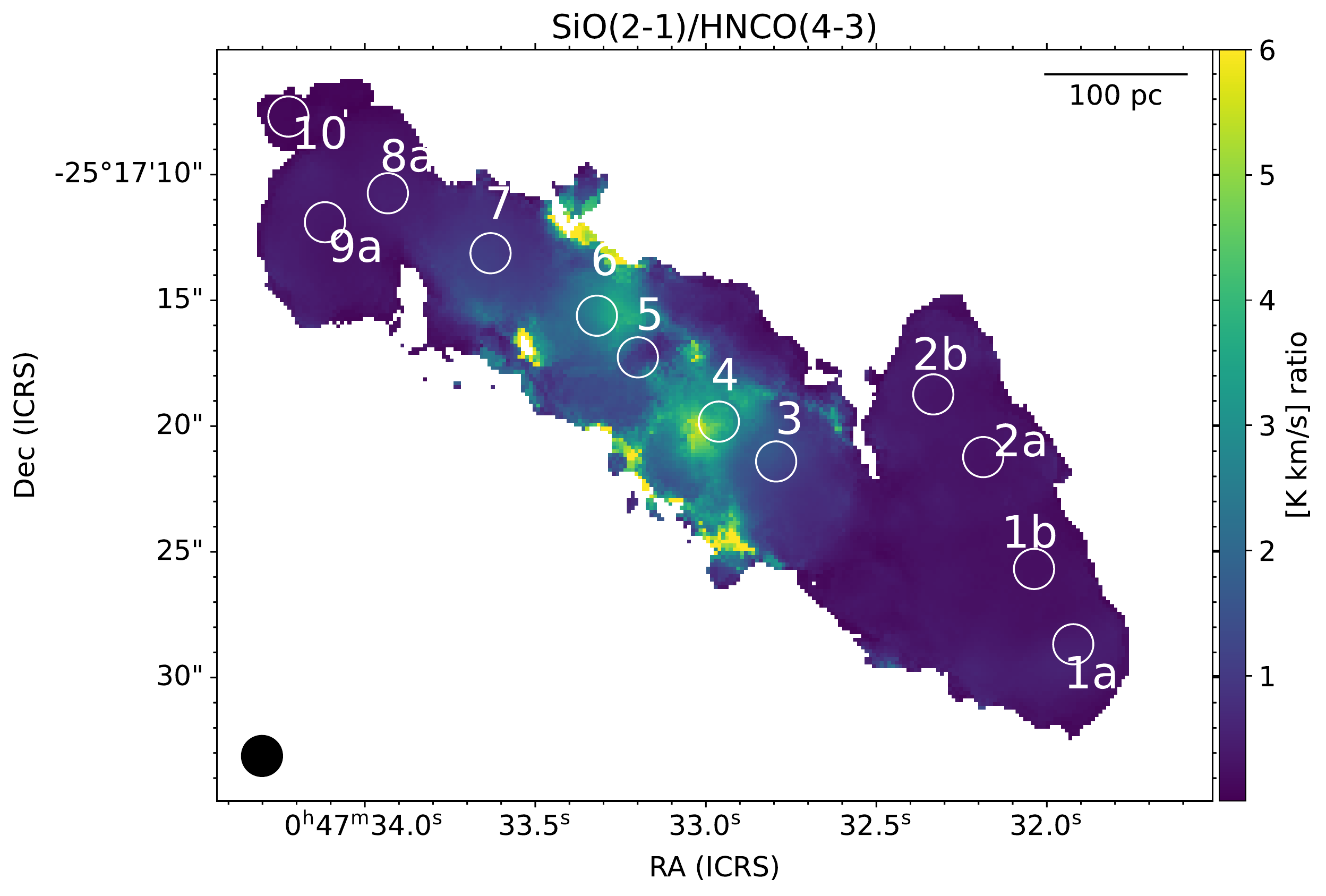} \\
    \centering\small (a) 
  \end{tabular}%
  \quad
  \begin{tabular}[b]{@{}p{0.8\textwidth}@{}}
    \centering\includegraphics[width=1.0\linewidth]{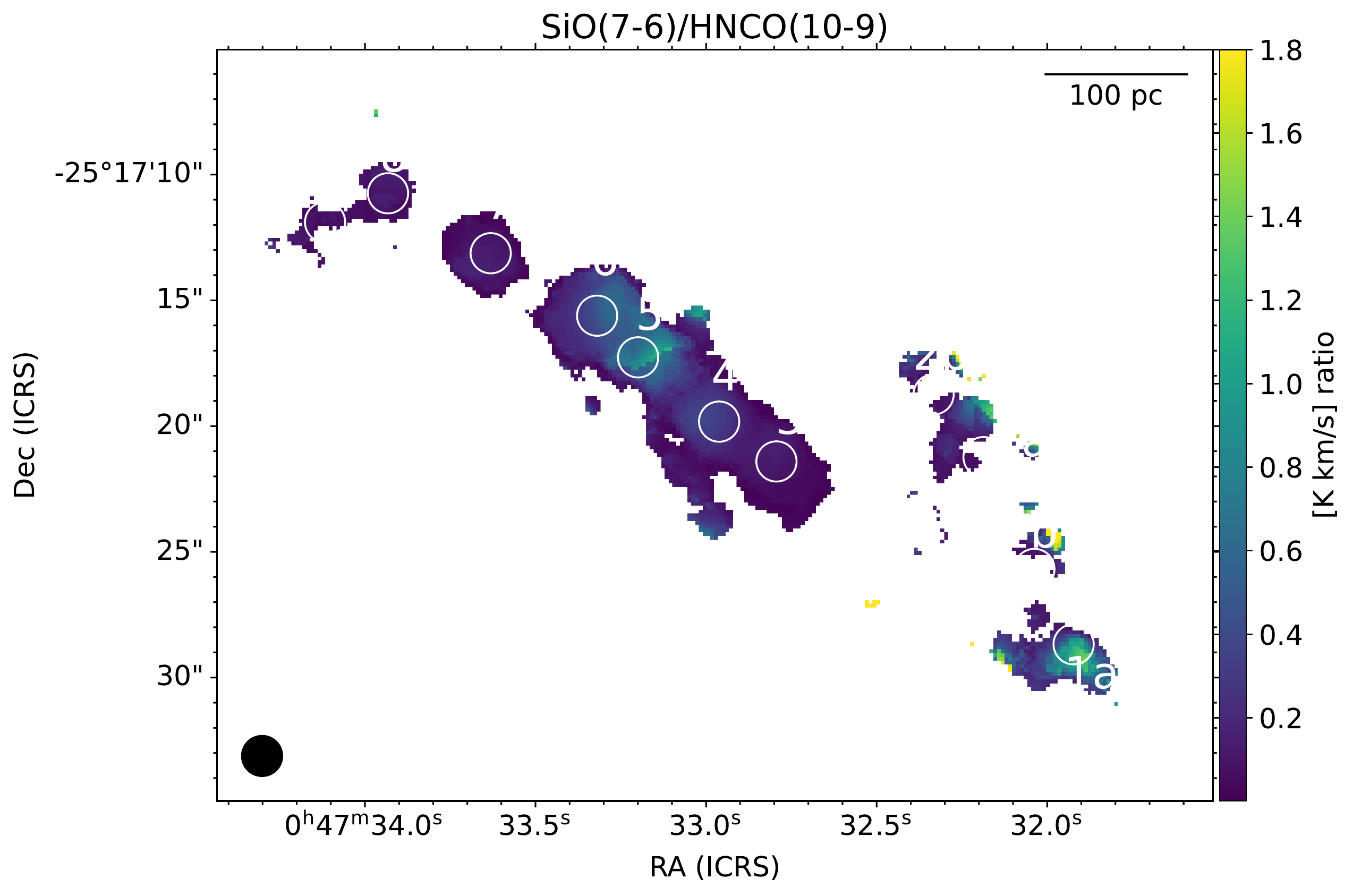} \\
    \centering\small (b) 
  \end{tabular}
  \caption{The line-intensity ratio maps of two selected pair: SiO(2-1) \& HNCO (4\textsubscript{0,4}-3\textsubscript{0,3}) and SiO(7-6) \& HNCO 10(\textsubscript{0,10}-9\textsubscript{0,9})
  }
  \label{fig:ratio_maps}
\end{figure*}
\section{Additional corner plots for GMC2b/3/5/9a/10}
The RADEX-Bayesian inference results for the rest of GMCs: GMC 2b, 3, 4, 6, and 8a. 
\label{sec:rest_corners}
\begin{figure*}
  \centering
  \includegraphics[width=16cm]{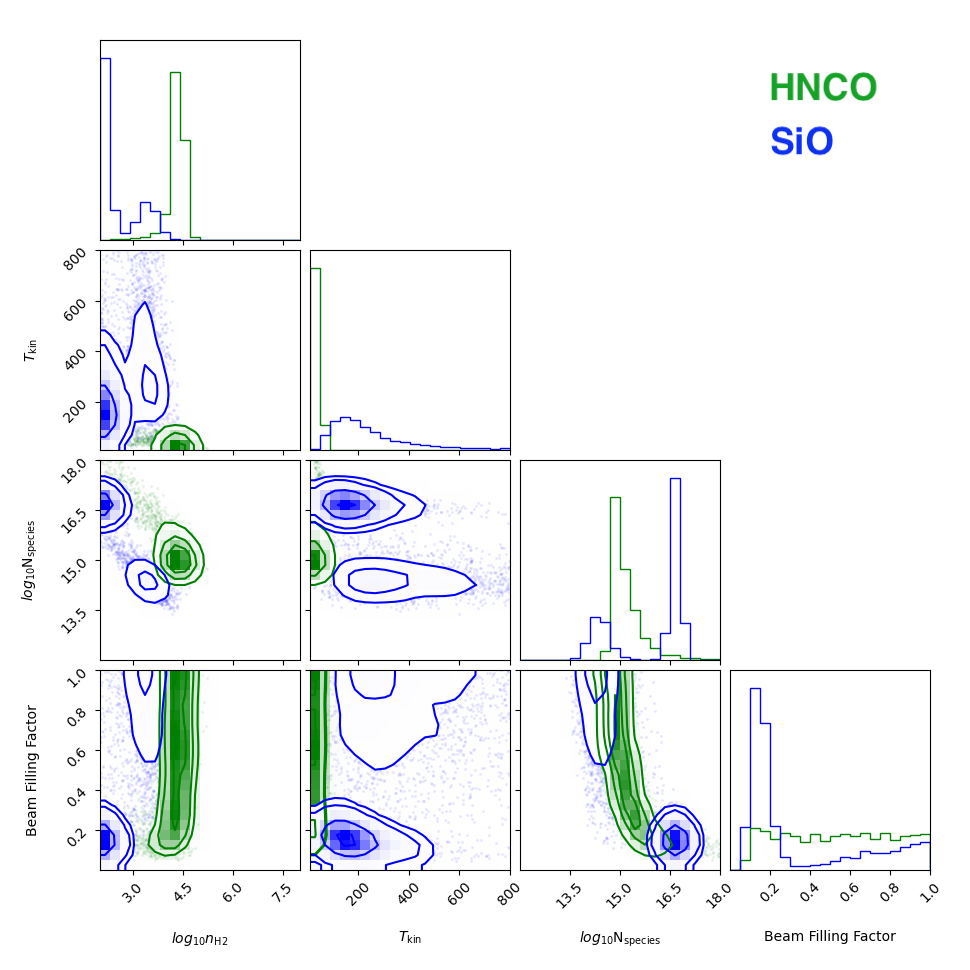}
  \caption{Same as Fig.~\ref{fig:RADEX_corner_GMC1a} but for GMC2b.
  }
  \label{fig:RADEX_corner_GMC2b}
\end{figure*}
\begin{figure*}
  \centering
  \includegraphics[width=16cm]{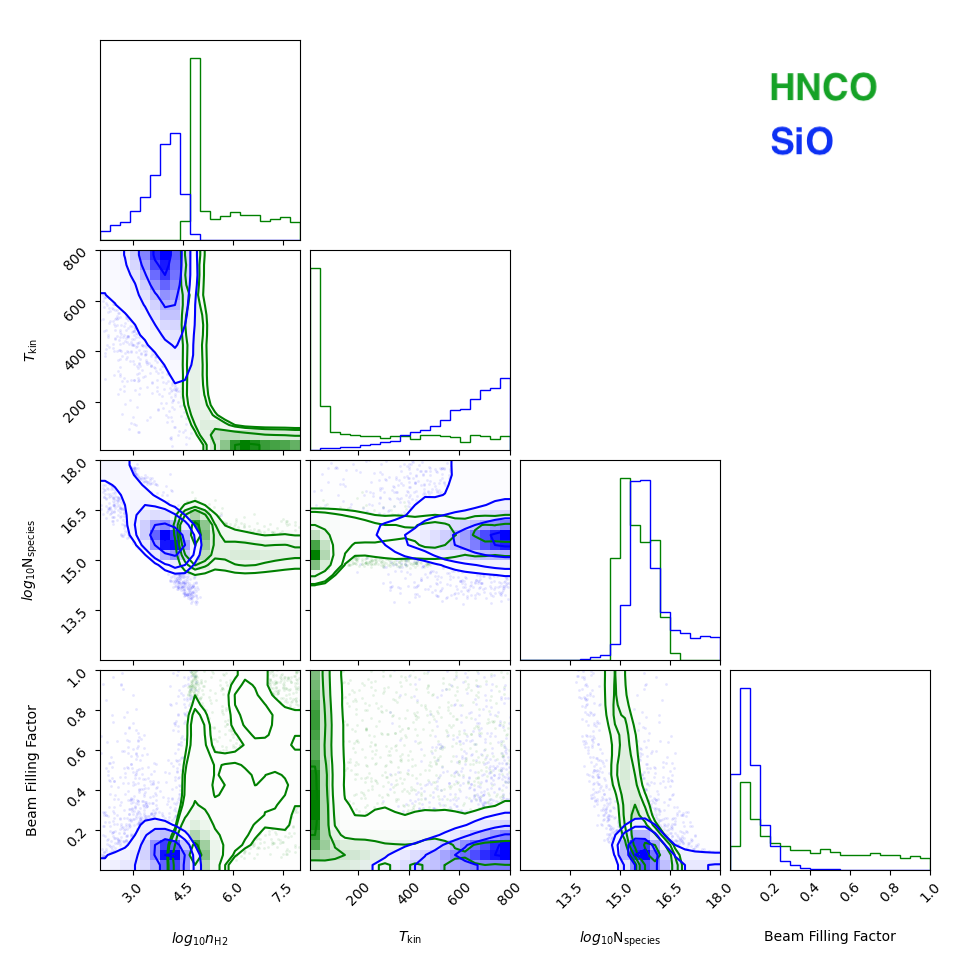}
  \caption{Same as Fig.~\ref{fig:RADEX_corner_GMC1a} but for GMC3.
  }
  \label{fig:RADEX_corner_GMC3}
\end{figure*}
\begin{figure*}
  \centering
  \includegraphics[width=16cm]{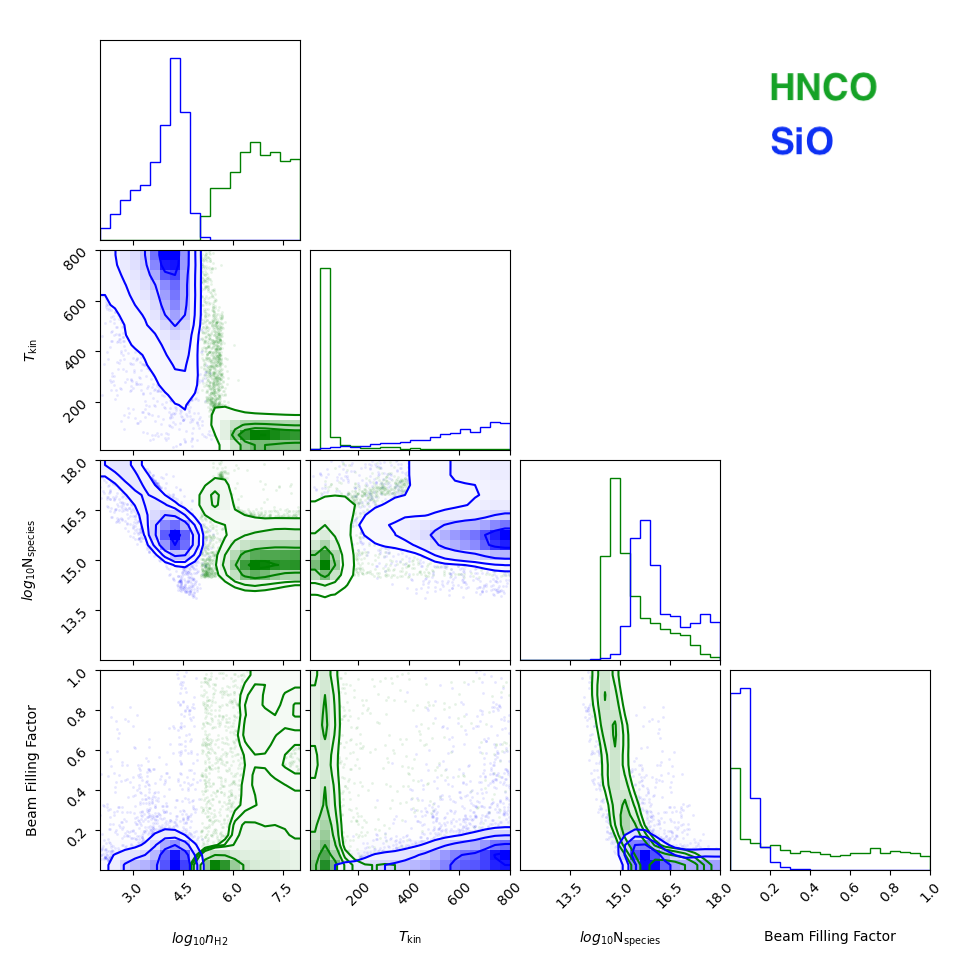}
  \caption{Same as Fig.~\ref{fig:RADEX_corner_GMC1a} but for GMC4.
  }
  \label{fig:RADEX_corner_GMC4}
\end{figure*}
\begin{figure*}
  \centering
\includegraphics[width=16cm]{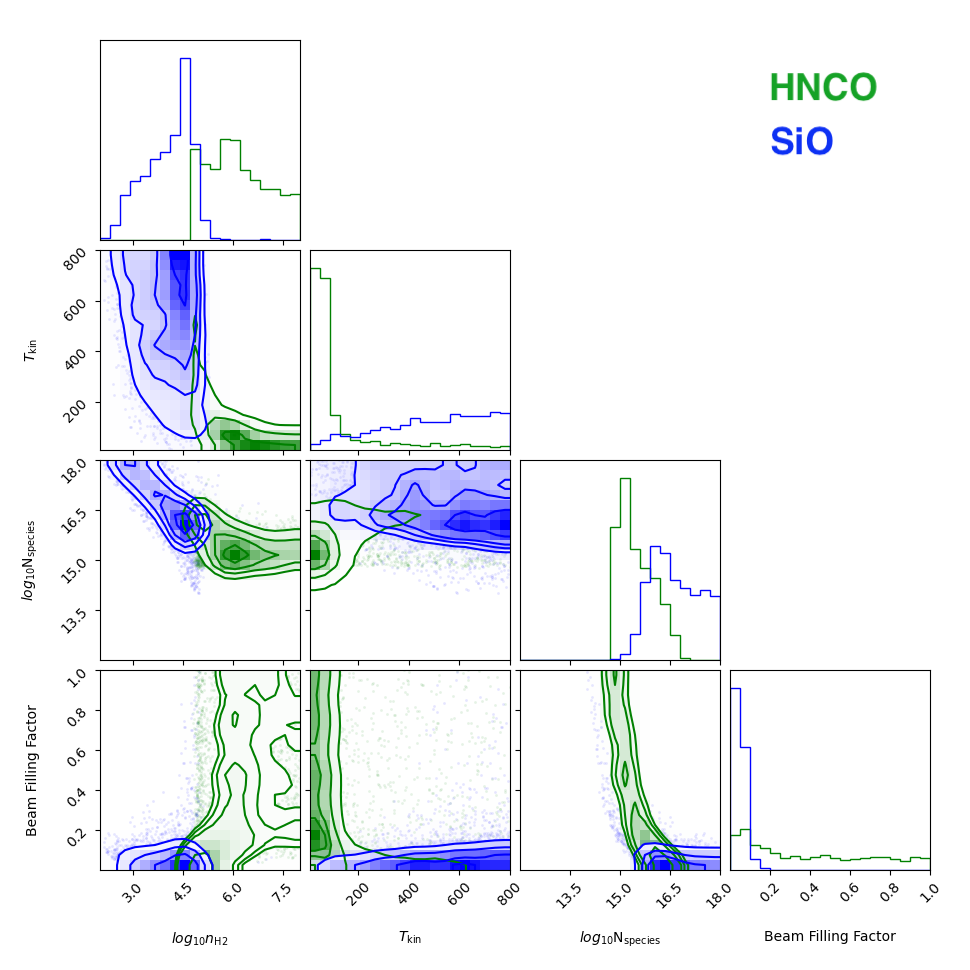}
  \caption{Same as Fig.~\ref{fig:RADEX_corner_GMC1a} but for GMC6. 
  }
  \label{fig:RADEX_corner_GMC6}
\end{figure*}
\begin{figure*}
  \centering
  \includegraphics[width=16cm]{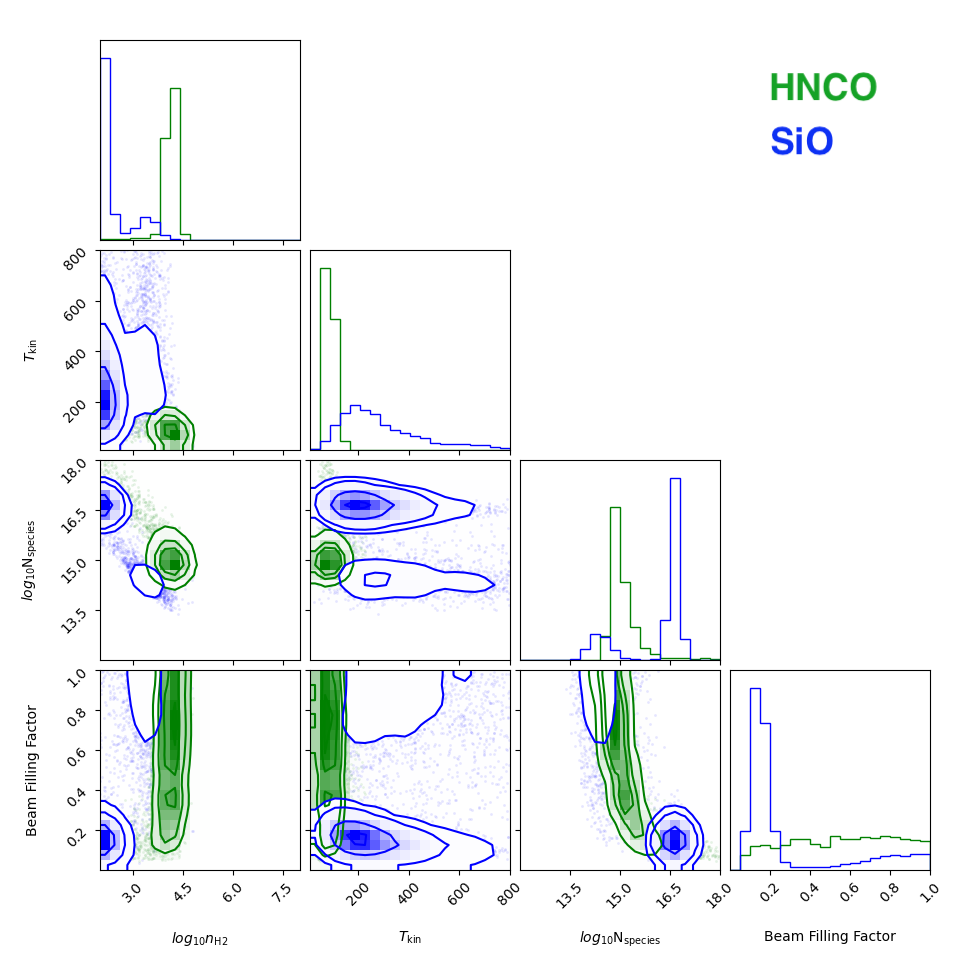}
  \caption{Same as Fig.~\ref{fig:RADEX_corner_GMC1a} but for GMC8a.
  }
  \label{fig:RADEX_corner_GMC8p}
\end{figure*}
\section{Comparison of the predicted intensity from the RADEX-Bayesian inference analysis with observed values: A posterior predictive check (PPC)}
\begin{figure*}
  \centering
  \begin{tabular}[b]{@{}p{0.45\textwidth}@{}}
    \centering\includegraphics[width=1.0\linewidth]{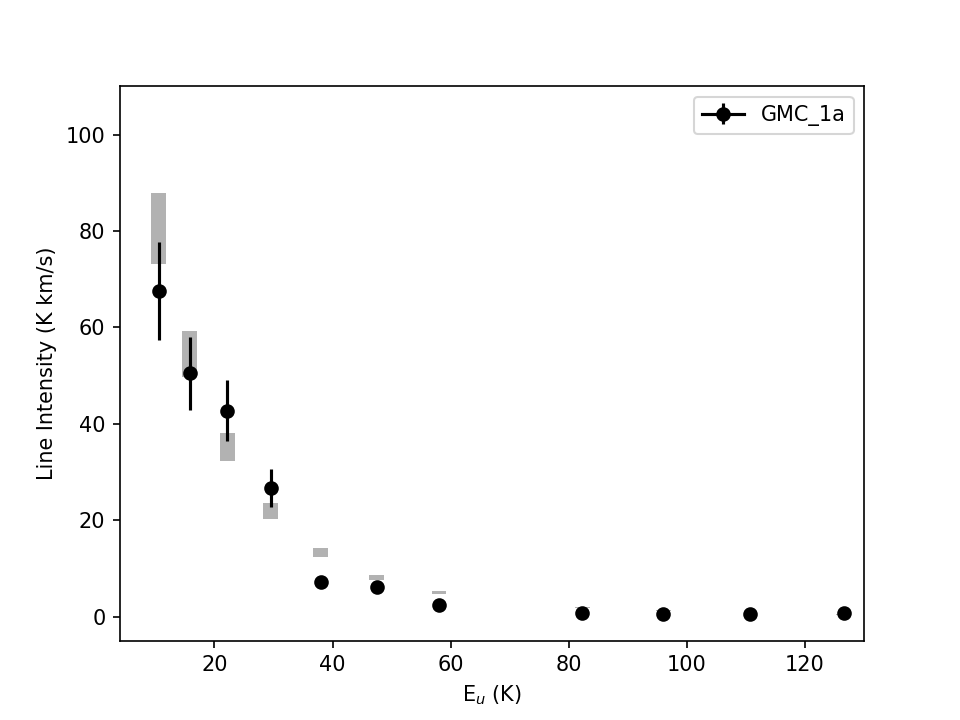} \\
    \centering\small (a) 
  \end{tabular}%
  \quad
  \begin{tabular}[b]{@{}p{0.45\textwidth}@{}}
    \centering\includegraphics[width=1.0\linewidth]{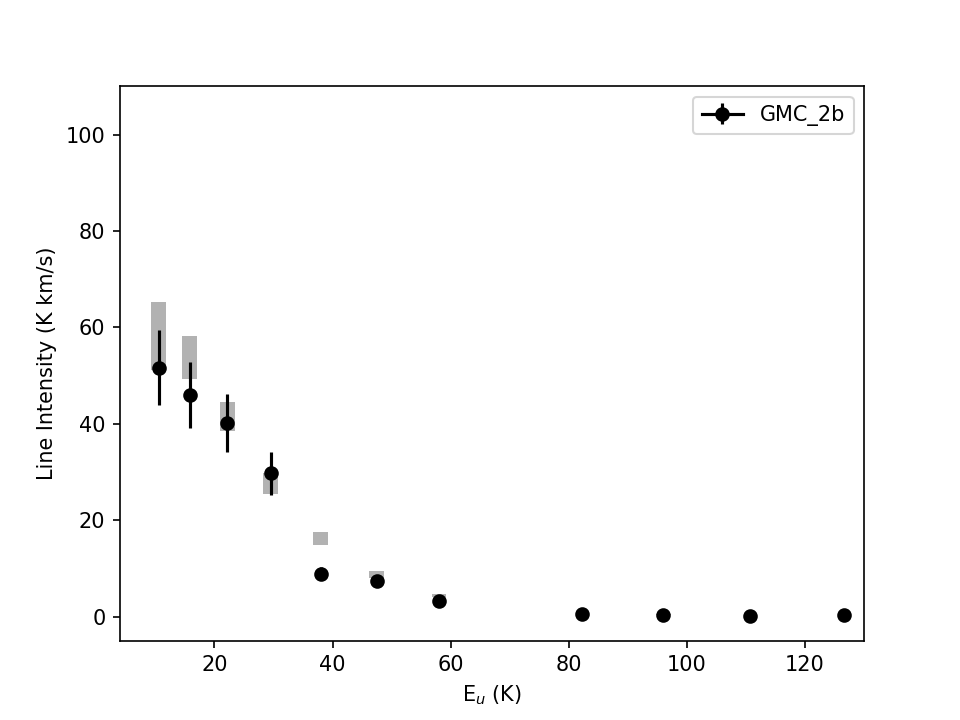} \\
    \centering\small (b) 
  \end{tabular}
  \begin{tabular}[b]{@{}p{0.45\textwidth}@{}}
    \centering\includegraphics[width=1.0\linewidth]{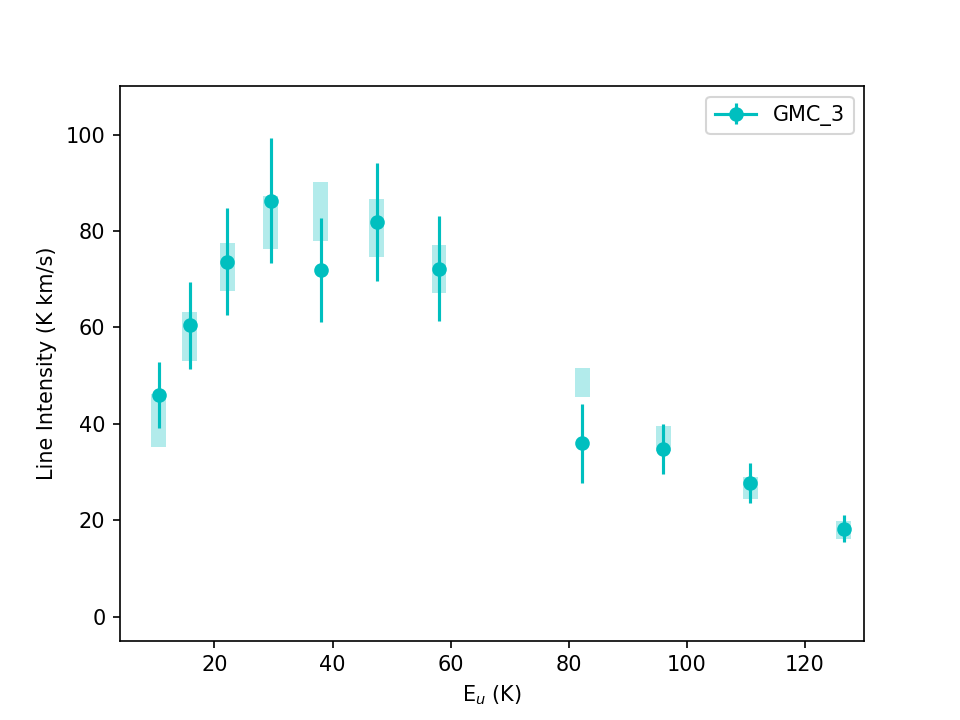} \\
    \centering\small (c) 
  \end{tabular}
  \begin{tabular}[b]{@{}p{0.45\textwidth}@{}}
    \centering\includegraphics[width=1.0\linewidth]{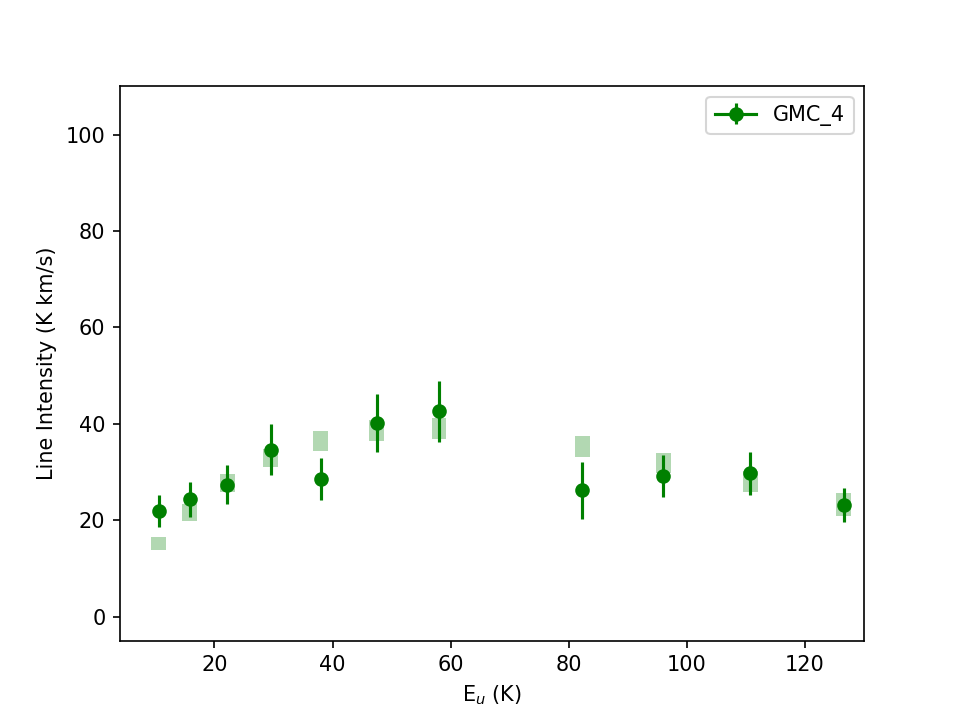} \\
    \centering\small (d) 
  \end{tabular}
  \begin{tabular}[b]{@{}p{0.45\textwidth}@{}}
    \centering\includegraphics[width=1.0\linewidth]{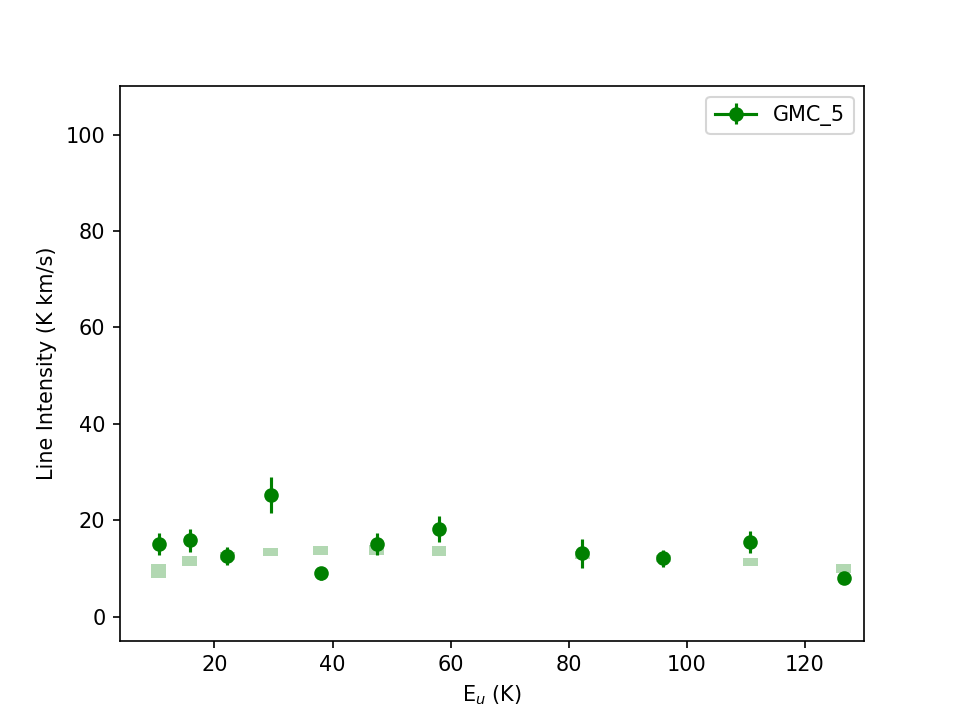} \\
    \centering\small (e) 
  \end{tabular}
  \begin{tabular}[b]{@{}p{0.45\textwidth}@{}}
    \centering\includegraphics[width=1.0\linewidth]{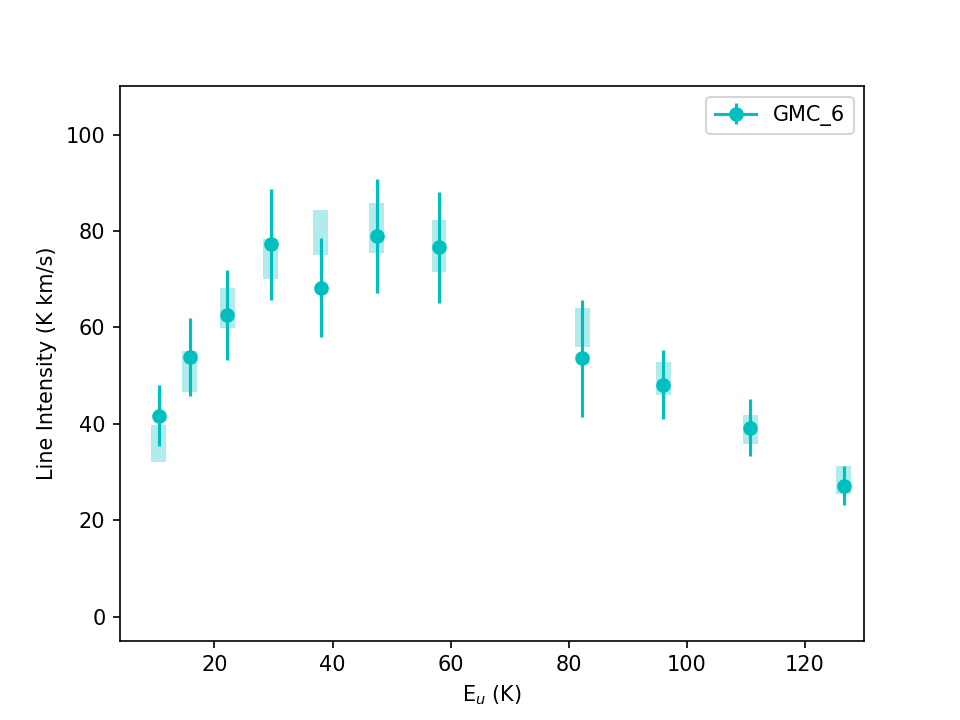} \\
    \centering\small (f) 
  \end{tabular}
  \caption{The posterior predictive checks (PPCs) of all HNCO transitions, GMC 1a-6, ordered accordingly from (a) to (f). The observed line intensity is in marker with uncertainty in line segment. The predicted line intensity is in colored band overlaid in the background. }
  \label{fig:PPC_HNCO_I}
\end{figure*}

\begin{figure*}
  \centering
  \begin{tabular}[b]{@{}p{0.45\textwidth}@{}}
    \centering\includegraphics[width=1.0\linewidth]{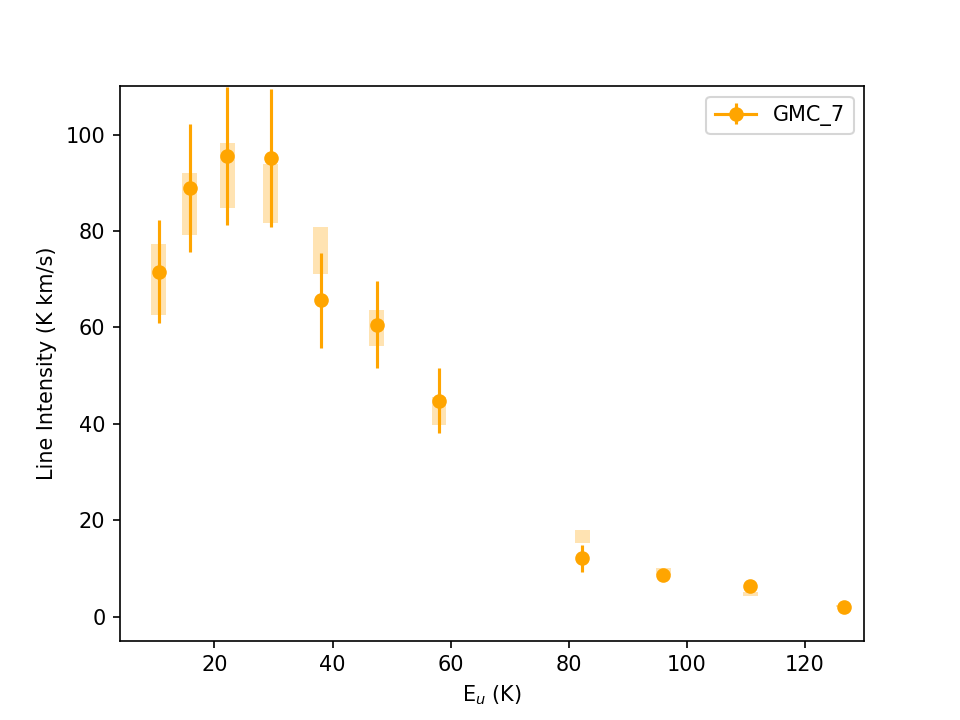} \\
    \centering\small (a) 
  \end{tabular}%
  \quad
  \begin{tabular}[b]{@{}p{0.45\textwidth}@{}}
    \centering\includegraphics[width=1.0\linewidth]{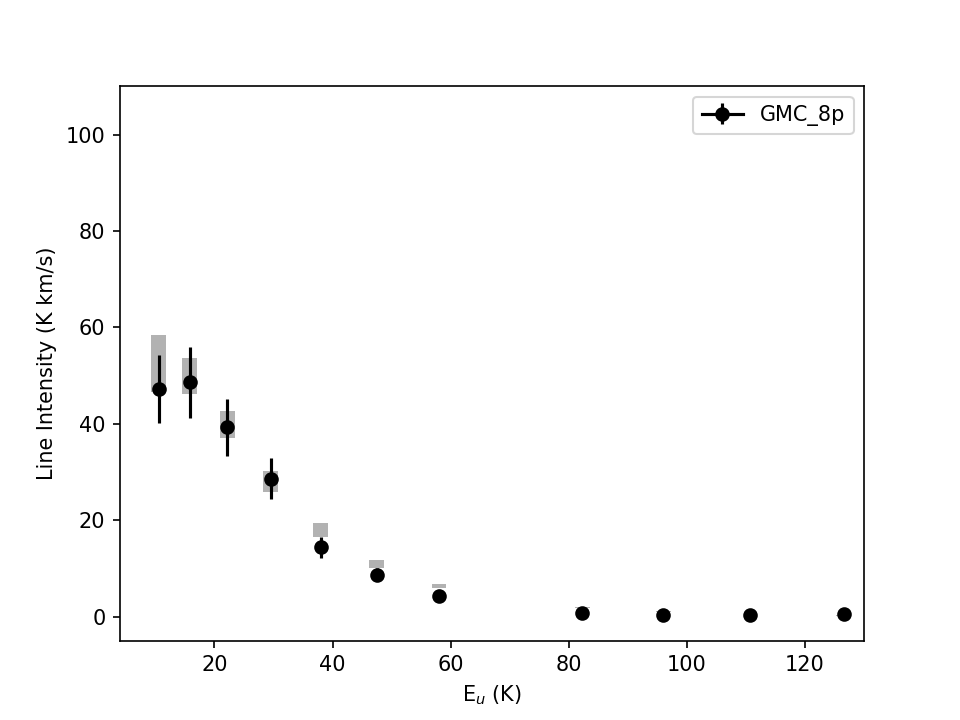} \\
    \centering\small (b) 
  \end{tabular}
  \begin{tabular}[b]{@{}p{0.45\textwidth}@{}}
    \centering\includegraphics[width=1.0\linewidth]{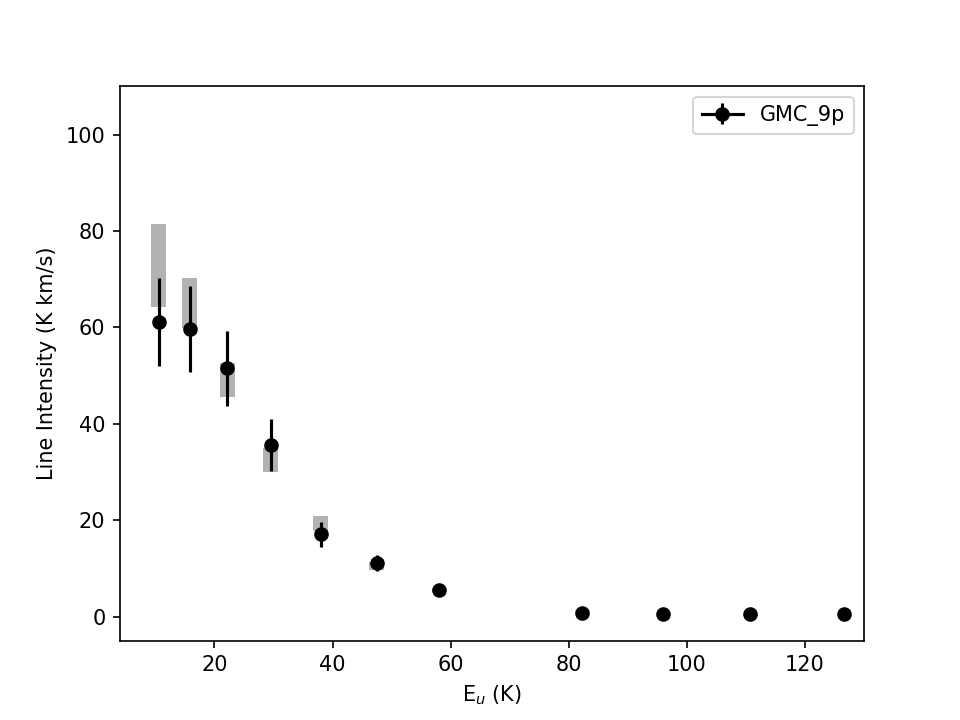} \\
    \centering\small (c) 
  \end{tabular}
  \begin{tabular}[b]{@{}p{0.45\textwidth}@{}}
    \centering\includegraphics[width=1.0\linewidth]{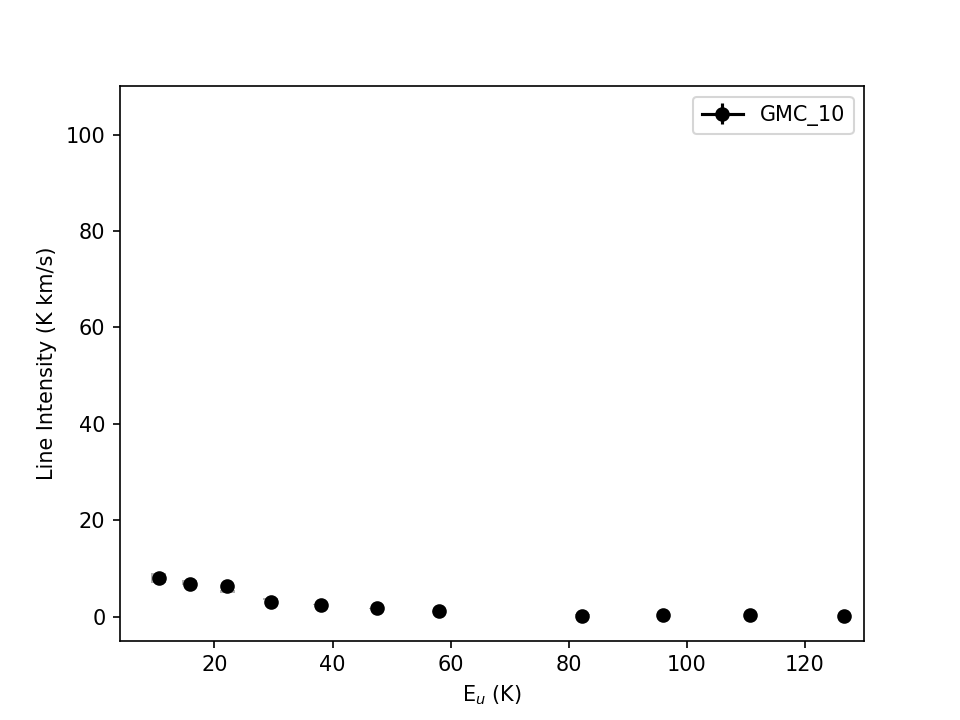} \\
    \centering\small (d) 
  \end{tabular}
  \caption{The posterior predictive checks (PPCs) of all HNCO transitions, GMC 7-10, ordered accordingly from (a) to (d). The observed line intensity is in marker with uncertainty in line segment. The predicted line intensity from RADEX-Bayesian analysis is in colored band overlaid in the background. }
  \label{fig:PPC_HNCO_II}
\end{figure*}

\begin{figure*}
  \centering
  \begin{tabular}[b]{@{}p{0.45\textwidth}@{}}
    \centering\includegraphics[width=1.0\linewidth]{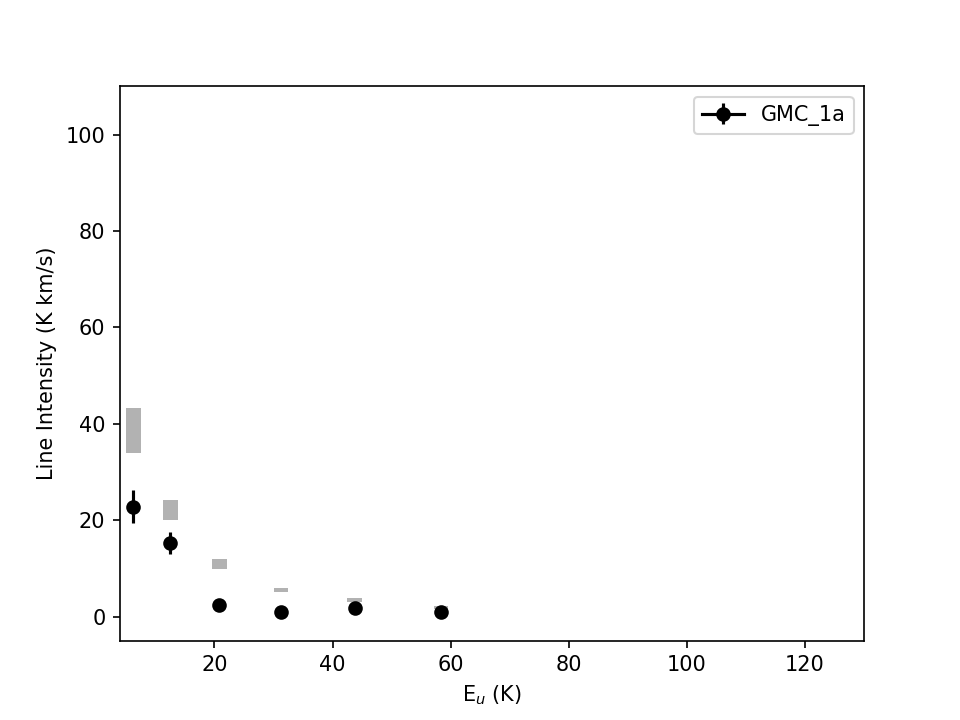} \\
    \centering\small (a) 
  \end{tabular}%
  \quad
  \begin{tabular}[b]{@{}p{0.45\textwidth}@{}}
    \centering\includegraphics[width=1.0\linewidth]{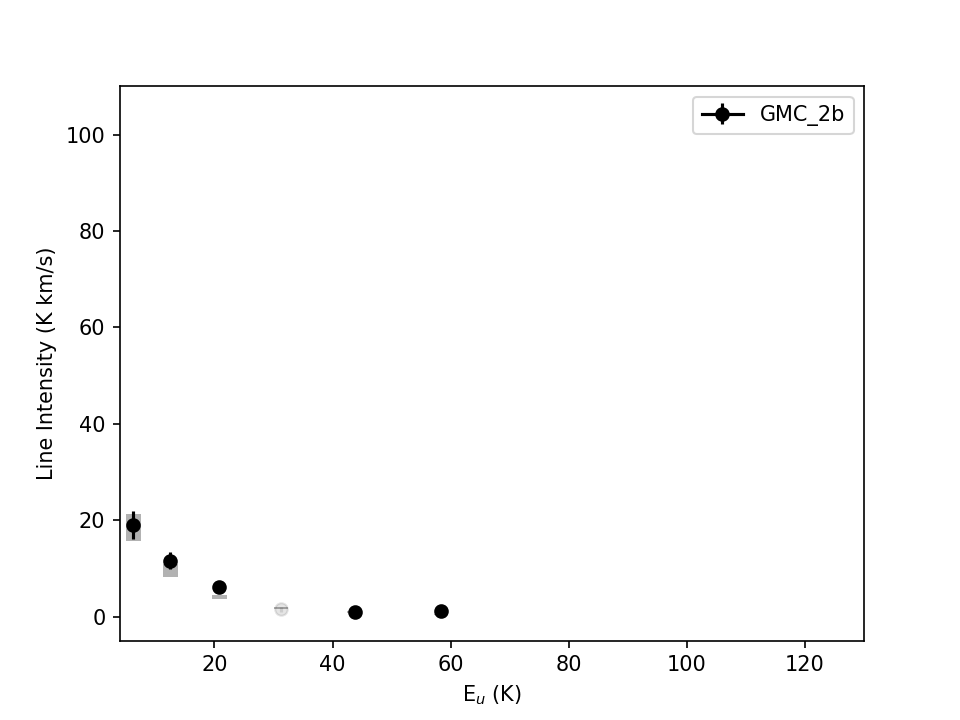} \\
    \centering\small (b) 
  \end{tabular}
  \begin{tabular}[b]{@{}p{0.45\textwidth}@{}}
    \centering\includegraphics[width=1.0\linewidth]{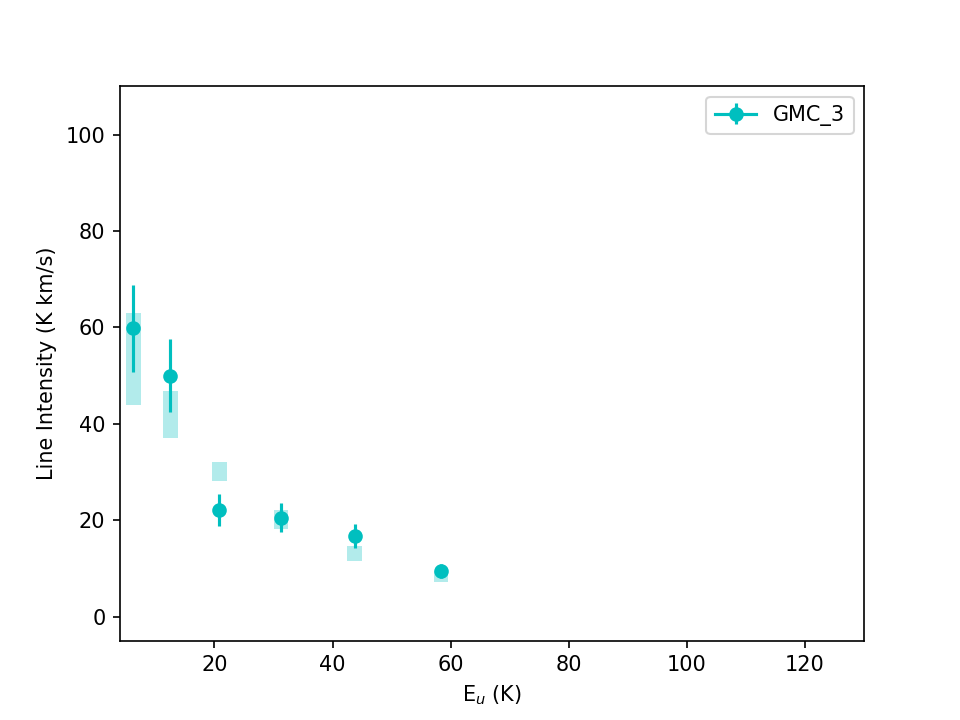} \\
    \centering\small (c) 
  \end{tabular}
  \begin{tabular}[b]{@{}p{0.45\textwidth}@{}}
    \centering\includegraphics[width=1.0\linewidth]{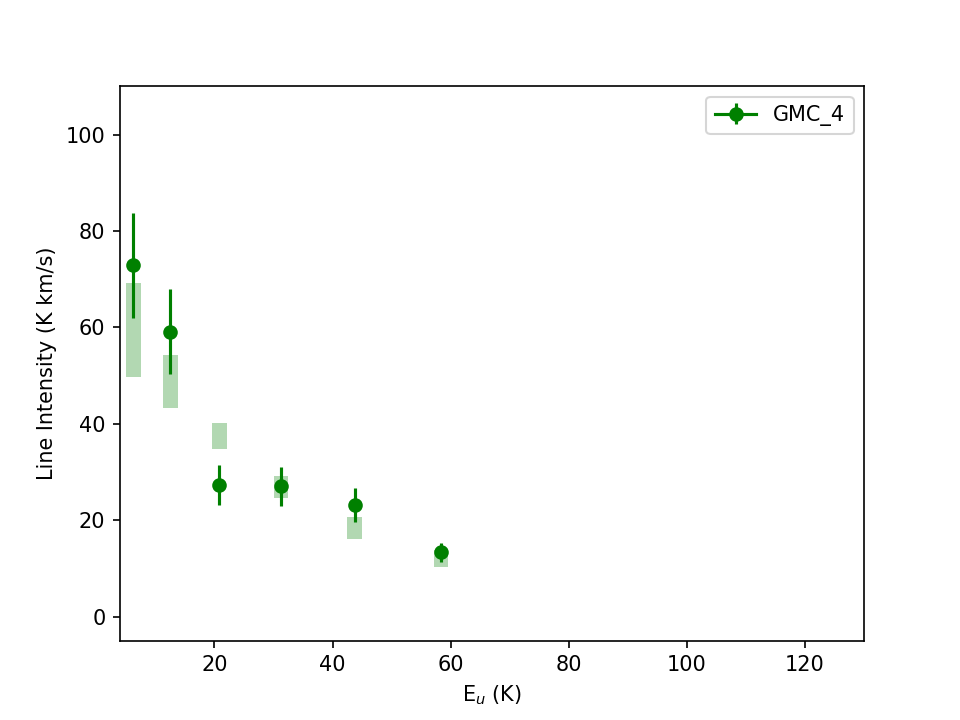} \\
    \centering\small (d) 
  \end{tabular}
  \begin{tabular}[b]{@{}p{0.45\textwidth}@{}}
    \centering\includegraphics[width=1.0\linewidth]{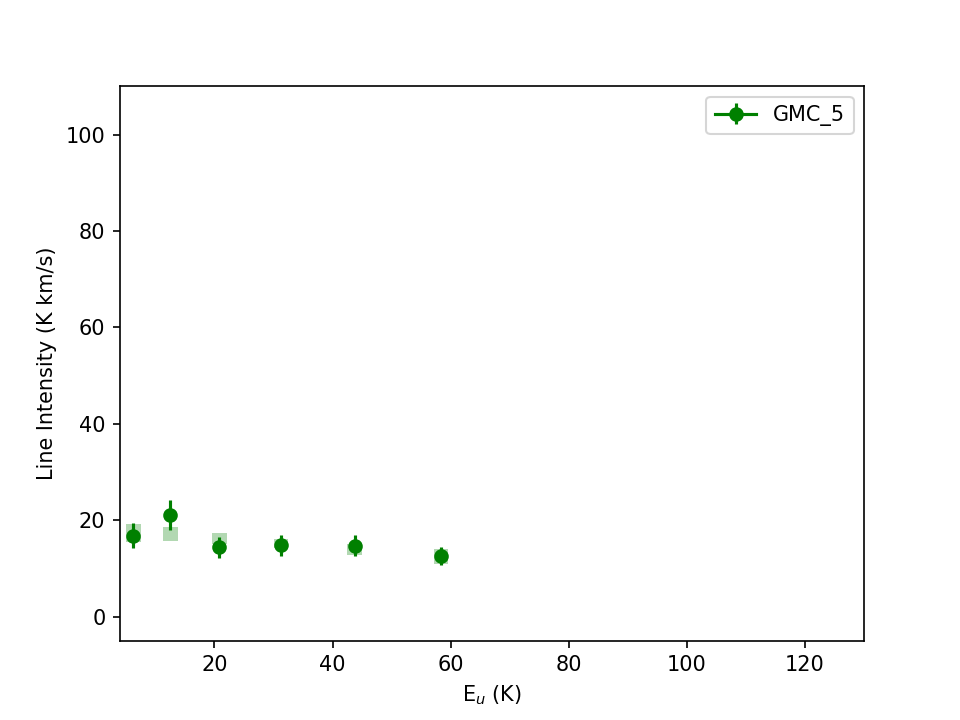} \\
    \centering\small (e) 
  \end{tabular}
  \begin{tabular}[b]{@{}p{0.45\textwidth}@{}}
    \centering\includegraphics[width=1.0\linewidth]{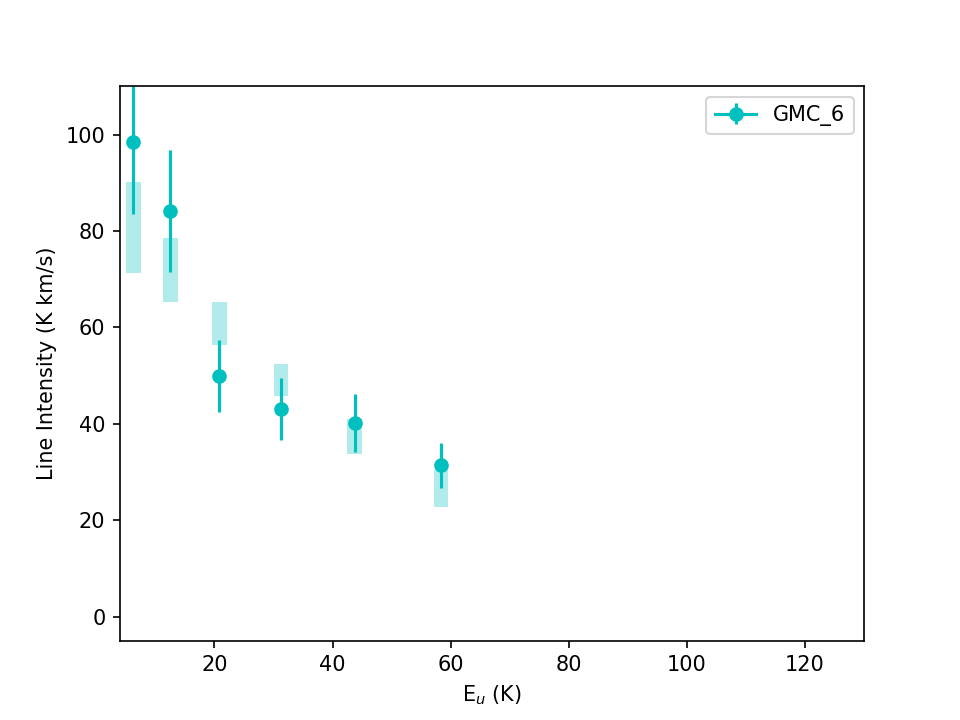} \\
    \centering\small (f) 
  \end{tabular}
  \caption{The posterior predictive checks (PPCs) of all SiO transitions, GMC 1a-6, ordered accordingly from (a) to (f). The observed line intensity is in marker with uncertainty in line segment. The predicted line intensity from RADEX-Bayesian analysis is in colored band overlaid in the background. }
  \label{fig:PPC_SiO_I}
\end{figure*}

\begin{figure*}
  \centering
  \begin{tabular}[b]{@{}p{0.45\textwidth}@{}}
    \centering\includegraphics[width=1.0\linewidth]{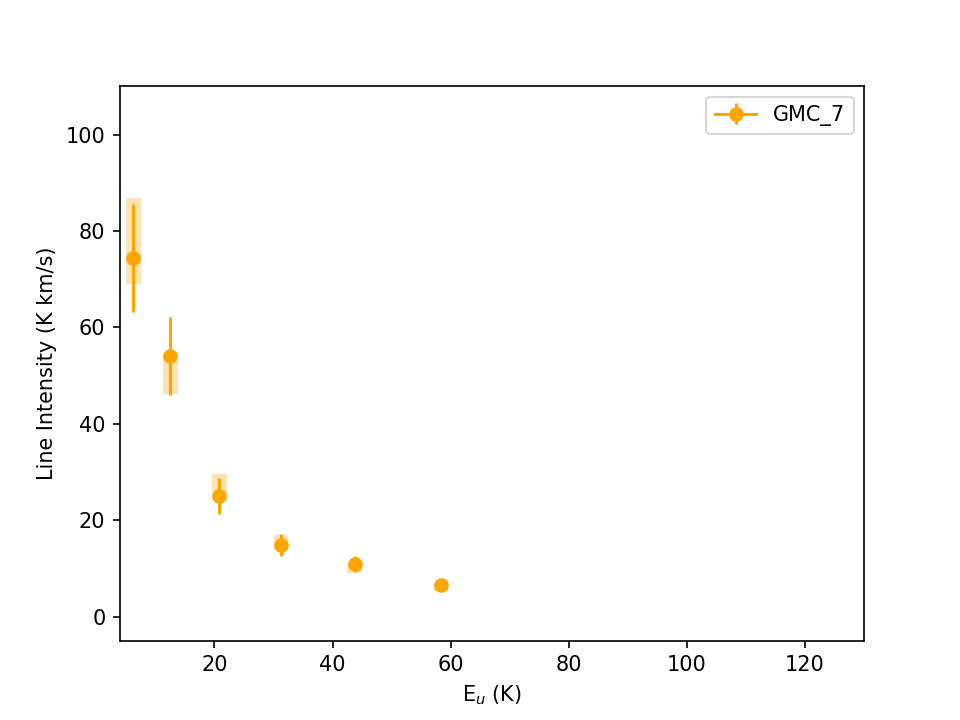} \\
    \centering\small (a) 
  \end{tabular}%
  \quad
  \begin{tabular}[b]{@{}p{0.45\textwidth}@{}}
    \centering\includegraphics[width=1.0\linewidth]{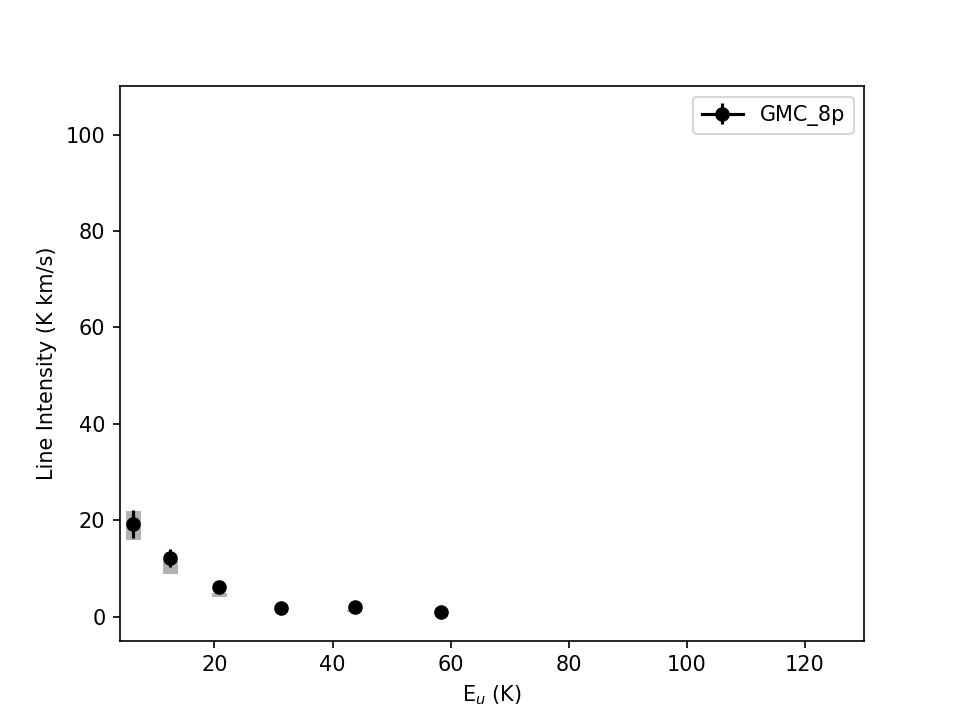} \\
    \centering\small (b) 
  \end{tabular}
  \begin{tabular}[b]{@{}p{0.45\textwidth}@{}}
    \centering\includegraphics[width=1.0\linewidth]{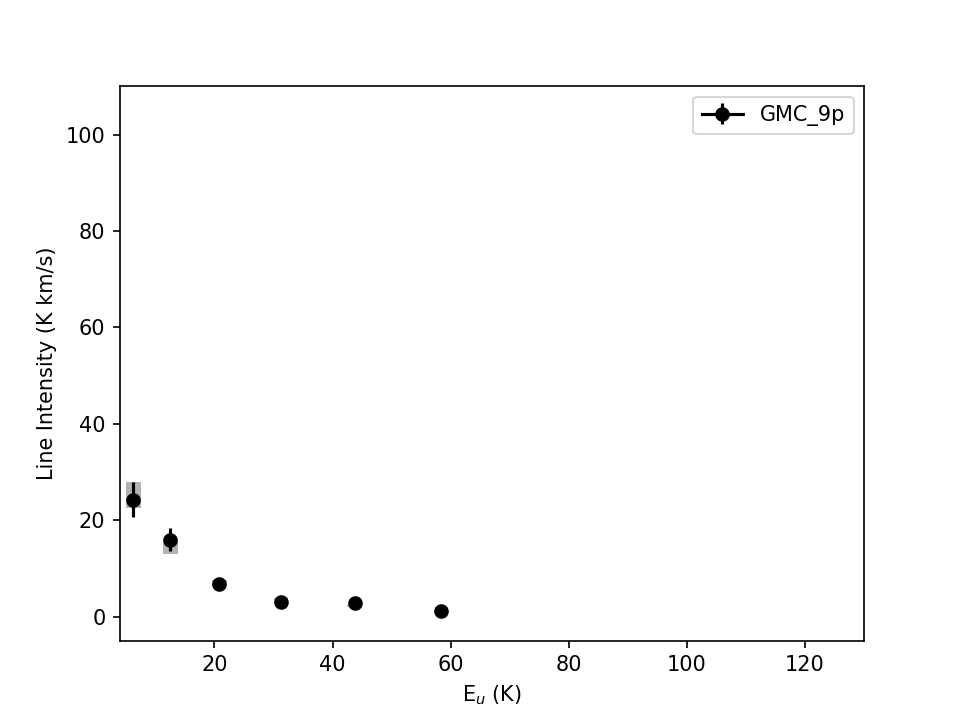} \\
    \centering\small (c) 
  \end{tabular}
  \begin{tabular}[b]{@{}p{0.45\textwidth}@{}}
    \centering\includegraphics[width=1.0\linewidth]{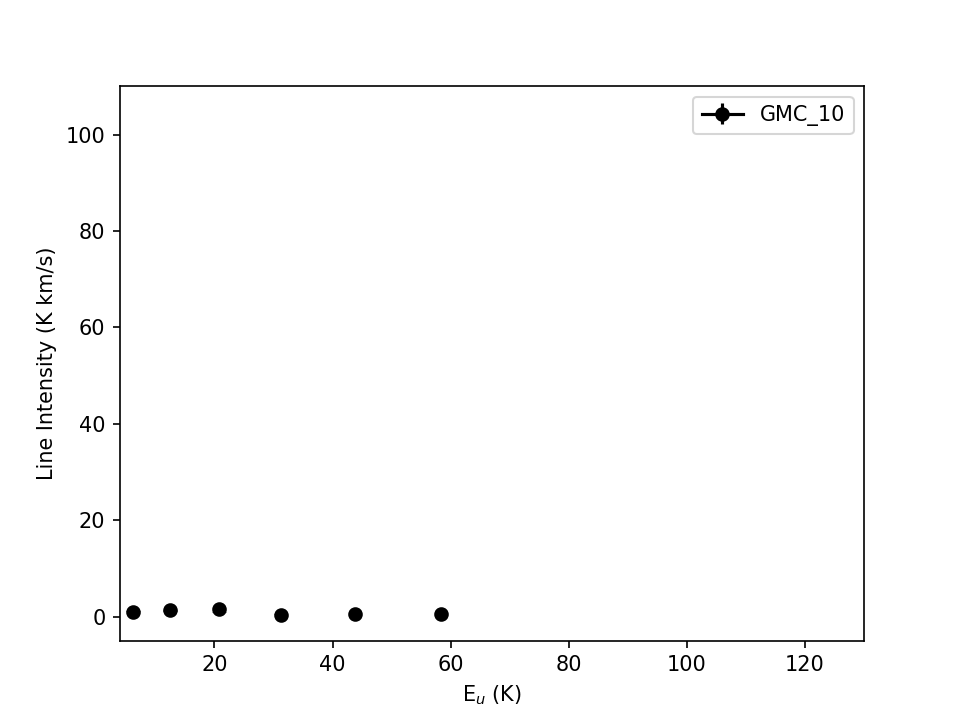} \\
    \centering\small (d) 
  \end{tabular}
  \caption{The posterior predictive checks (PPCs) of all SiO transitions, GMC 7-10, ordered accordingly from (a) to (d). The observed line intensity is in marker with uncertainty in line segment. The predicted line intensity from RADEX-Bayesian analysis is in colored band overlaid in the background. }
  \label{fig:PPC_SiO_II}
\end{figure*}

We perform a posterior predictive check for the inferred gas properties in Sect. \ref{sec:radex_tech} from the coupled RADEX and Bayesian inference process. 
This is to verify our posterior distribution produces a distribution for the data that is consistent with the actual data, which is the velocity-integrated intensity in our case. 
We sample the predicted line intensities from our posterior between 16-84 percentile, and plot against the observed line intensities. 
The comparisons are shown in Fig. \ref{fig:PPC_HNCO_I}-\ref{fig:PPC_SiO_II}. 
\section{Additional cases explored in chemical modelling}

\begin{figure*}
  \centering
  \begin{tabular}[b]{@{}p{0.43\textwidth}@{}}
    \centering\includegraphics[width=1.0\linewidth]{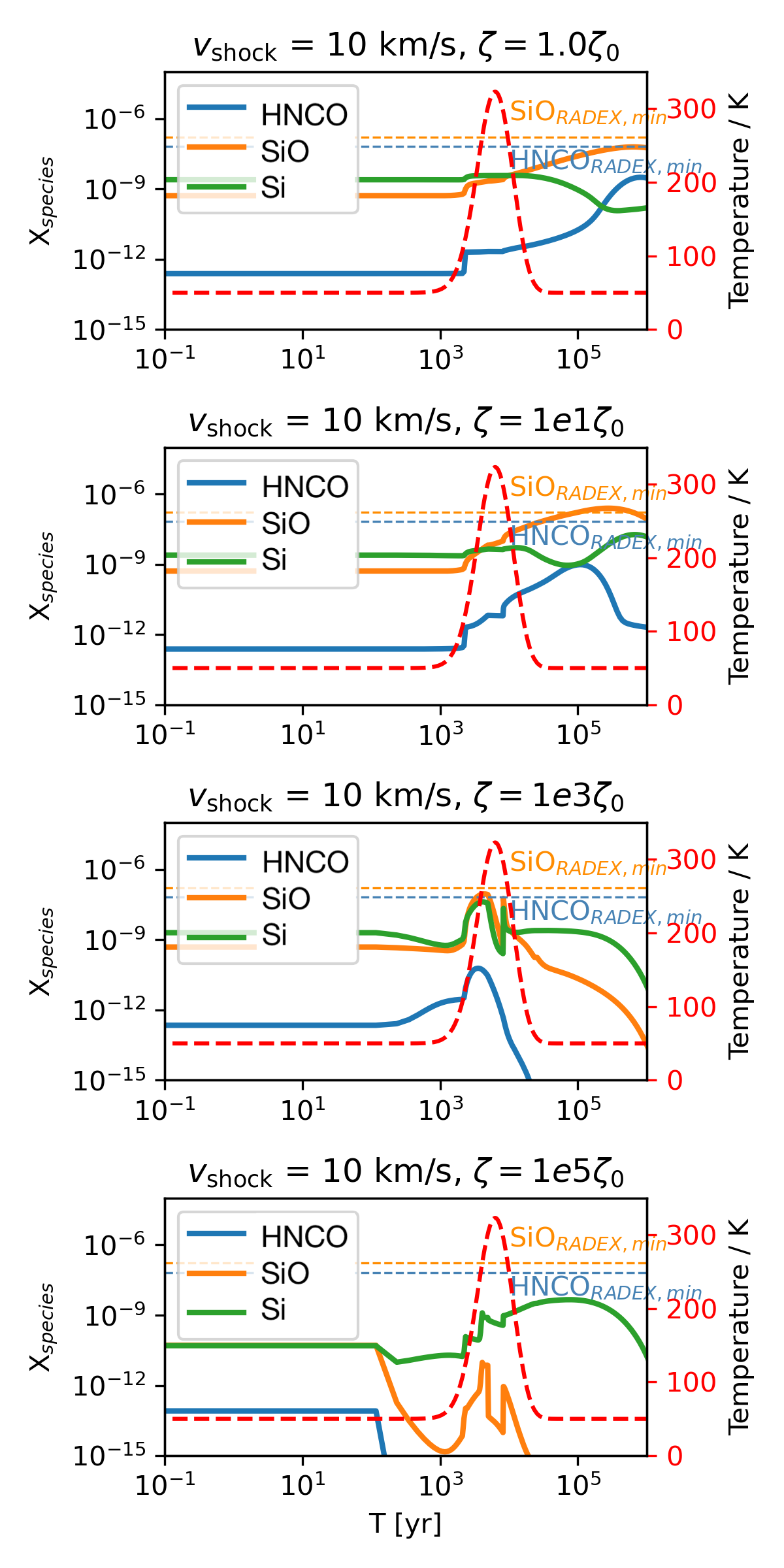} \\
  \end{tabular}\\
  \caption{Alternative cases with pre-shock gas density $n_{\rm H2}=10^{3}$ cm\textsuperscript{-3} in a slow shock scenario ($v_{s}=10$ km/s, where we compare cases of CRIR from $1\zeta_{0}$ (top) to $10^{5}\zeta_{0}$ (bottom). It is clear that an higher CRIR could further suppress the HNCO abundance enhanced via slow shocks. }
  \label{fig:append_den03Cshock_varZeta}
\end{figure*}

Fig. \ref{fig:append_den03Cshock_varZeta} shows the outputs of extra chemical models we have performed, where we varied further the cosmic ray ionization rate. 
With a standard galactic cosmic ray ionization rate ($\zeta=\zeta_{0}$, top panel in Fig. \ref{fig:append_den03Cshock_varZeta}) in the slow shock scenario at the low density of $n=10^{3}$ cm\textsuperscript{-3}) case, it is possible to enhance the HNCO abundance to higher level, which is in contrast with other higher CRIR conditions (lower panels in Fig. \ref{fig:append_den03Cshock_varZeta}). 
We note, however, this enhancement is still insufficient in achieving the lower limit imposed by RADEX result (blue dashed horizontal line). 

\begin{figure*}
  \centering
  \begin{tabular}[b]{@{}p{0.43\textwidth}@{}}
    \centering\includegraphics[width=1.0\linewidth]{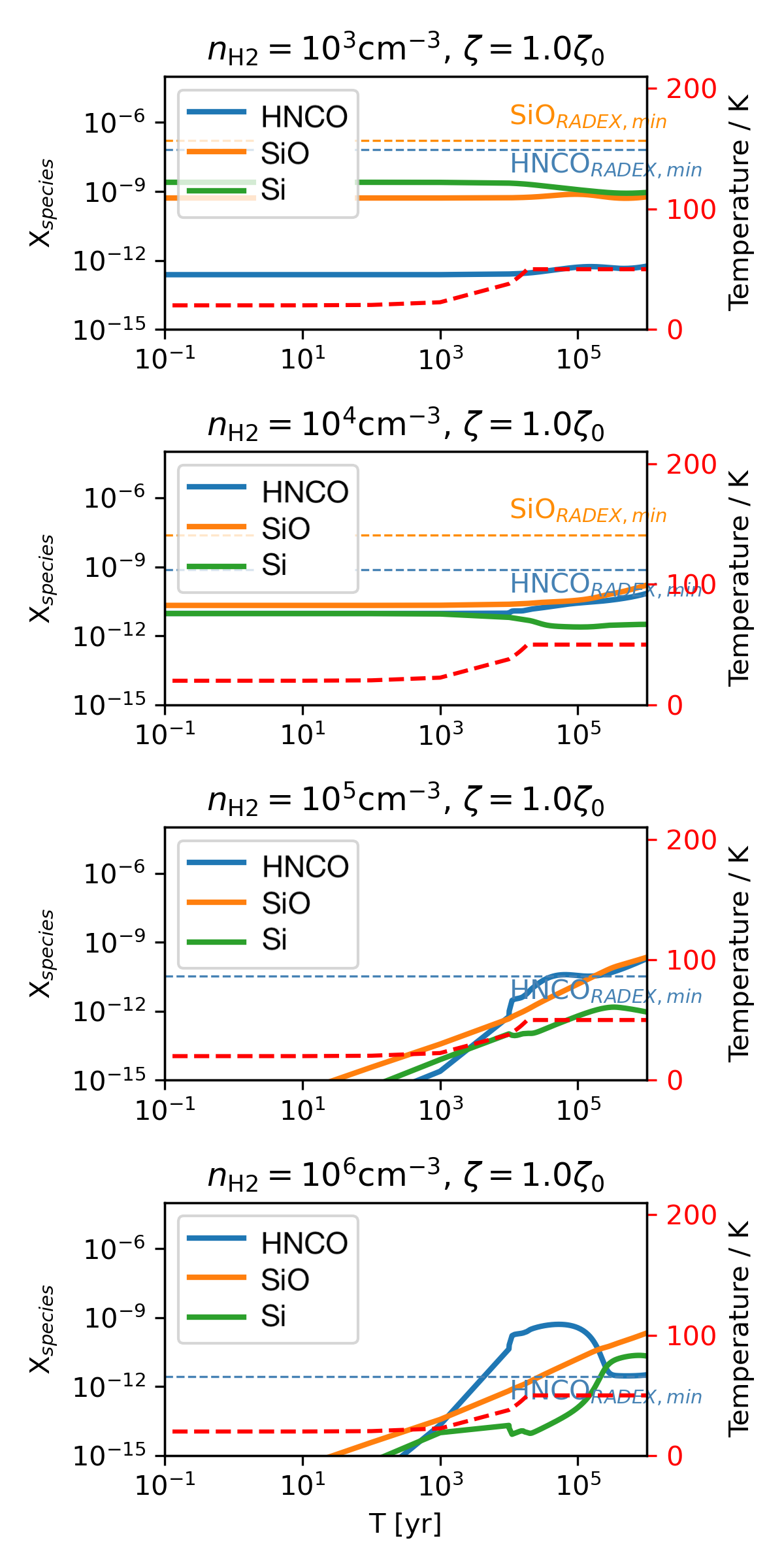} \\
    \centering\small (a) 
  \end{tabular}
  \begin{tabular}[b]{@{}p{0.43\textwidth}@{}}
    \centering\includegraphics[width=1.0\linewidth]{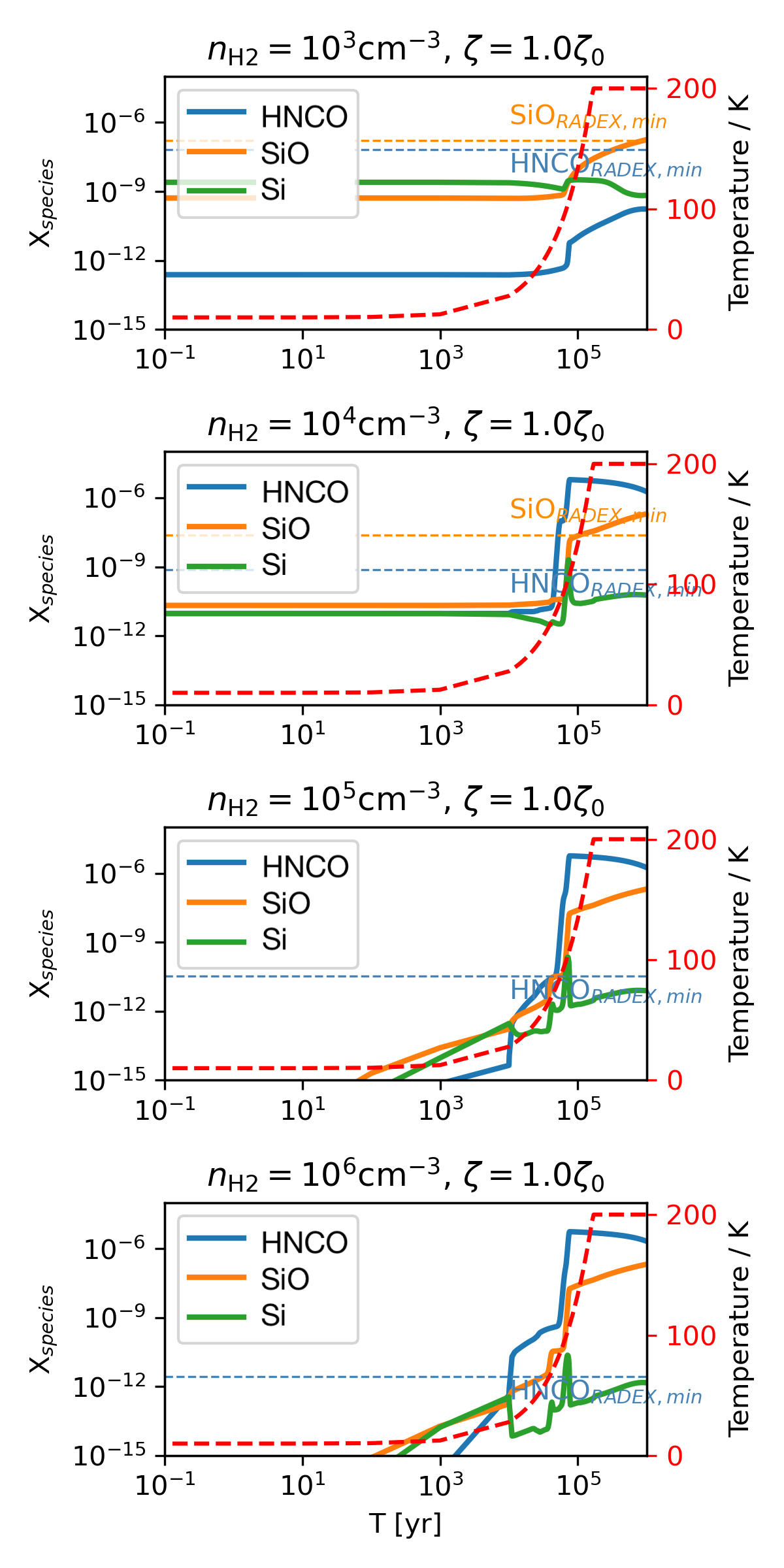} \\
    \centering\small (b) 
  \end{tabular}\\
  \caption{Alternative chemical models  without shocks with temperature = 50 K (left panel) and with temperature = 200 K (right panel) using a low CRIR of $\zeta=1\zeta_{0}$. At this cooler temperature, even with low CRIR, the HNCO abundance cannot be enhanced to a reasonable level to be matched with observational results. }
  \label{fig:append_hotcore_Zeta1_50K200K}
\end{figure*}

Fig. \ref{fig:append_hotcore_Zeta1_50K200K} show the non-shock models with $\zeta=\zeta_{0}$ for temperatures of T=50 K (left panel) and T=200 K (right panel) respectively. 
At higher gas density ($n\geq 10^{4}$ cm\textsuperscript{-3}) it shows HNCO abundance can also be enhanced in the absence of shocks with low CRIR, $\zeta=\zeta_{0}$, and high temperature T=200 K. 
\label{sec:append_varZeta}
\end{document}